\documentclass[nofootinbib,eqsecnum,tightenlines,superscriptaddress,10pt]{revtex4}

\usepackage{graphicx}
\usepackage{amsmath,amssymb,amsfonts,amsthm,stmaryrd,mathtools,bm}
\usepackage{mathrsfs}
\usepackage{color}
\usepackage{multirow,bigdelim}
\usepackage{hyperref}
\usepackage{dsfont}
\usepackage{booktabs}
\allowdisplaybreaks[0]

\usepackage{pstricks}
\usepackage{tikz}
\usepackage{ulem}

\usepackage{caption}
\usepackage{subcaption}

\newtheorem{theorem}{Theorem}[section]
\newtheorem{lemma}{Lemma}[section]
\newtheorem{definition}{Definition}[section]

\def\be#1\ee{\begin{align}#1\end{align}}

\def\ba{\begin{eqnarray}}
\def\ea{\end{eqnarray}}
\def\nn{\nonumber}
\def\q{\quad}

\begin{document}

\title{Complex  actions and causality violations:\\ Applications to Lorentzian quantum cosmology}

\author{Seth K. Asante}
\affiliation{Theoretisch-Physikalisches Institut,
Friedrich-Schiller-Universit\"at Jena,
Max-Wien-Platz 1,
07743 Jena, Germany}
\affiliation{Perimeter Institute, 31 Caroline Street North, Waterloo, ON, N2L 2Y5, Canada}
\author{Bianca Dittrich}
\affiliation{Perimeter Institute, 31 Caroline Street North, Waterloo, ON, N2L 2Y5, Canada}
\affiliation{Institute for Mathematics, Astrophysics and Particle Physics,
Radboud University, Heyendaalseweg 135, 6525 AJ Nijmegen, The Netherlands}
\author{Jos\'e Padua-Arg\"uelles}
\affiliation{Perimeter Institute, 31 Caroline Street North, Waterloo, ON, N2L 2Y5, Canada}
\affiliation{Department of Physics and  Astronomy, University of Waterloo, 200 University Avenue West, Waterloo, ON, N2L 3G1, Canada}

\begin{abstract}
For the construction of the Lorentzian path integral for gravity one faces two main questions: Firstly, what configurations to include, in particular whether to allow Lorentzian metrics that violate causality conditions. And secondly, how to evaluate a highly oscillatory path integral over unbounded domains. Relying on Picard-Lefschetz theory to address the second question for discrete Regge gravity, we will illustrate that it can also answer the first question. To this end we will define  the Regge action for complexified variables and study its analytical continuation. Although there have been previously two different versions defined for the Lorentzian Regge action, we will show that the complex action is unique. More precisely, starting from the different definitions for the action one arrives at equivalent analytical extensions.  The difference between the two  Lorentzian versions is only realized along  branch cuts which arise for a certain class of causality violating configurations.

As an application we discuss the path integral  describing a finite evolution step of the discretized deSitter universe. We will in particular consider an evolution from vanishing to finite scale factor, for which the path integral defines the no-boundary wave function.
\end{abstract}

\maketitle

\section{Introduction}

Recently, there has been increased activity in making the Lorentzian or real time path integral computable \cite{Witten1,RelTime,QCDReview}. Lorentzian path integrals are challenging as the integrands are typically highly oscillating and the integrations are over unbounded domains, leading to integrals which are not absolutely converging. An appropriate deformation of the integration contour into the complex plane, based on Picard-Lefschetz theory \cite{PLTheory}, can however turn the integrand into a quickly decaying function, and allow for numerical techniques, such as Monte-Carlo simulations \cite{QCDReview, HanLefschetz}. 

The Lorentzian path integral is in particular important for quantum gravity \cite{CDT,TurokEtAl,EPRL-FK,SFReview,EffSF3,Ding2021}. Euclidean quantum gravity approaches use a  formal Wick rotation, to justify a path integral based on the Euclidean gravity action, over Euclidean metrics. Such Euclidean approaches suffer however from two main draw-backs: firstly the conformal factor problem \cite{ConfFac}. That is, the kinetic term for the conformal factor comes with the `wrong sign', rendering the Euclidean gravity action unbounded from below. Secondly, the space of Euclidean metrics is quite different from the space of Lorentzian metrics, making the notion of (inverse) Wick rotation in general  ill-defined \cite{CDT}. 

\begin{figure}[ht!]
\begin{picture}(500,150)
\put(90,7){ \includegraphics[width=4cm,height=4cm]{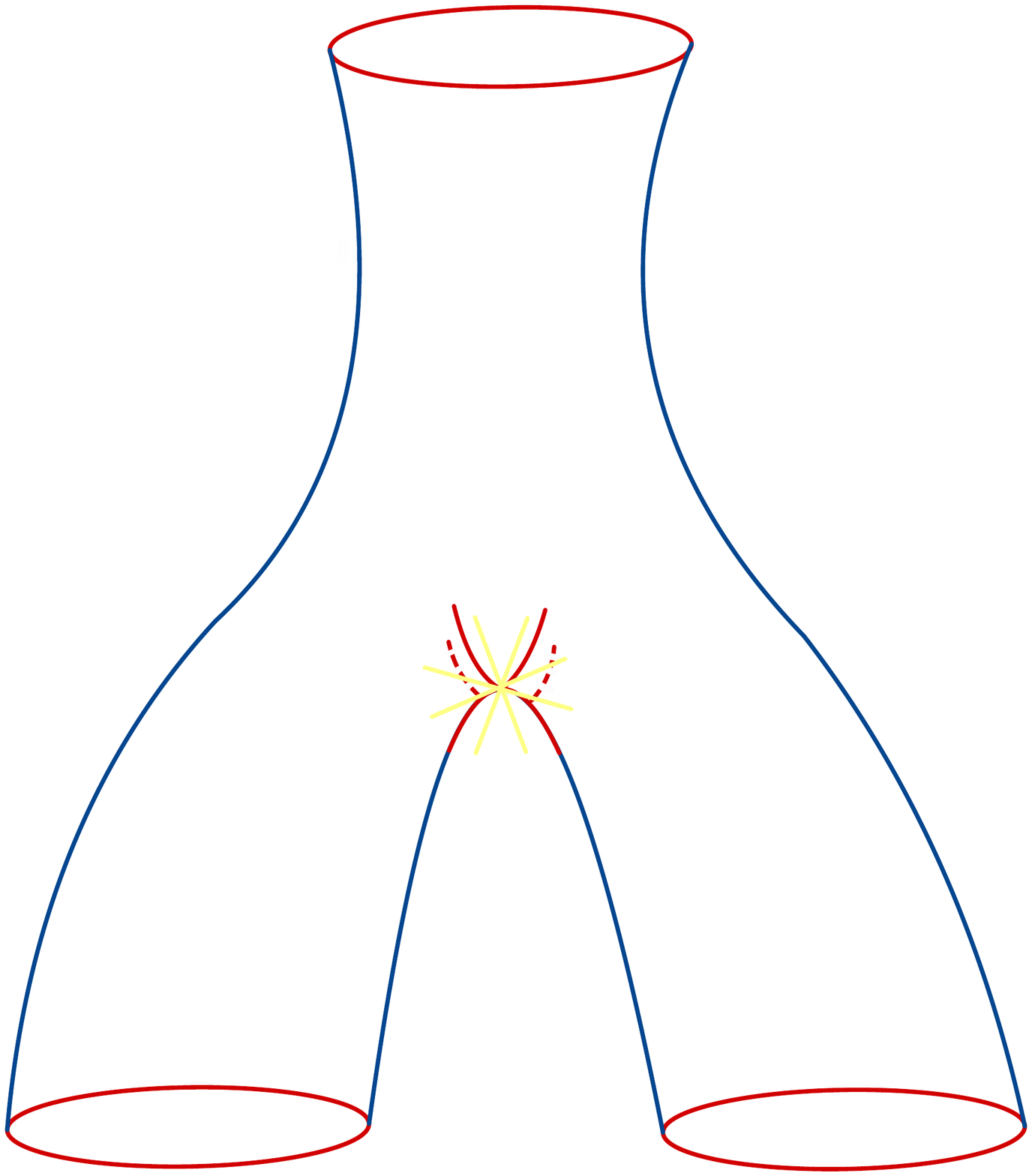} }
\put(300,7){ \includegraphics[width=4cm,height=4cm]{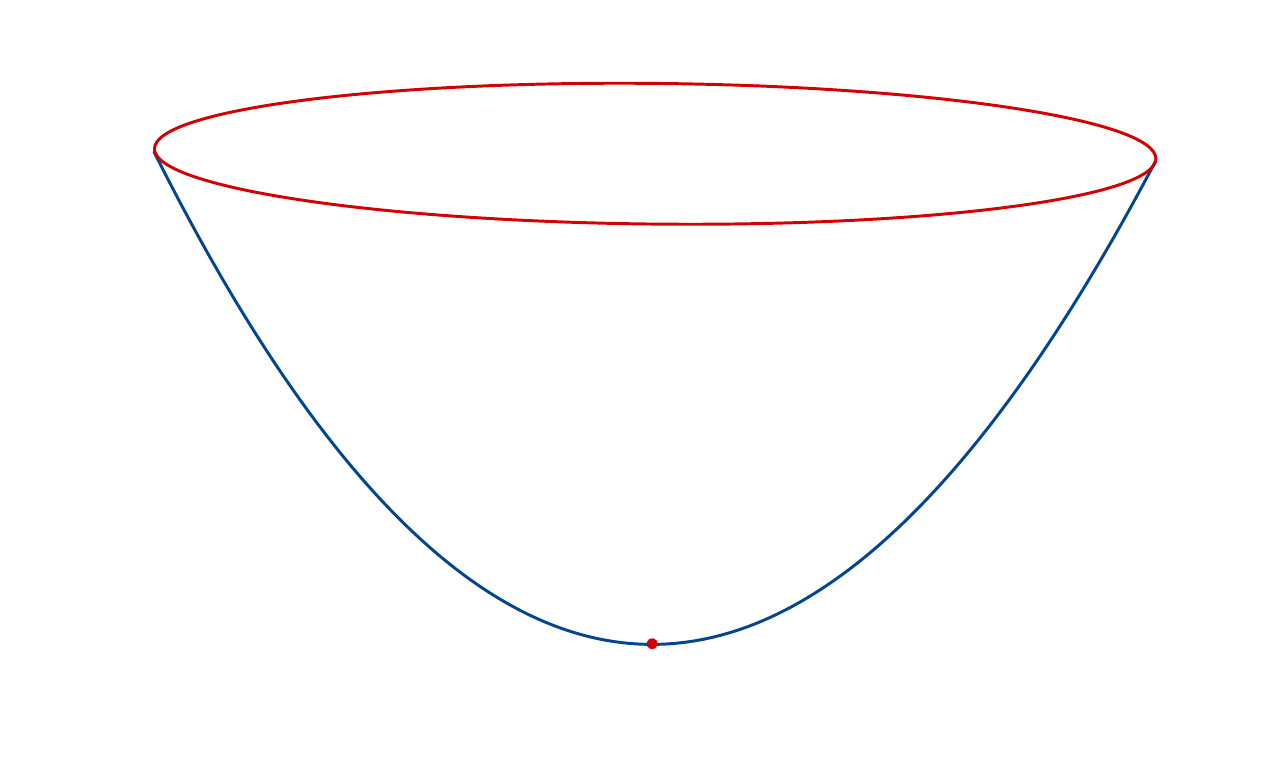} }
\end{picture}
\caption{The two figures illustrate a change of topology for (one-dimensional) spatial hypersurfaces during time evolution. The left figure shows a trouser, where at the crotch one has an irregular light cone structure: instead of one future and one past light cone one has overall four light cones. The right figure shows a so-called yarmulke configuration where at the lower tip there are no light cones.}
\label{fig:topology_change}
\end{figure}

In particular, Lorentzian geometries carry a notion of causality, whereas Euclidean configurations do not. We might have  configurations, where certain causality conditions\footnote{\textit{E.g.} that each point has one well-defined future light cone and one well-defined past light cone.} are violated. Such causality violating configurations might arise in geometries describing topology change during time evolution \cite{LoukoSorkin,TopChange}, see Figure \ref{fig:topology_change}, but as we will see here, at least for discrete gravity also arise independently from topology change.

This leads to the question of how to define the Lorentzian path integral for quantum gravity, see also the recent discussions in \cite{Witten2021,Lehners2021,Visser2021}. In particular whether to allow or to forbid causality violating configurations in the Lorentzian path integral. In Causal Dynamical Triangulations \cite{CDT,LollJordan} such configurations are forbidden. This has led to a far better behaved continuum limit than for (Euclidean) Dynamical Triangulations, and in particular a smooth deSitter universe emerging in the continuum limit for four-dimensional Causal Dynamical Triangulations \cite{CDT4DResult}. 

Another path integral approach based on discrete geometries is quantum Regge calculus \cite{Regge}. It plays also an important role for the spin foam approach, \textit{e.g.} \cite{SFReview,AreaRegge,AreaAngle, EffSF}. Sorkin \cite{Sorkin1974,Sorkin2019} constructed a Regge action for Lorentzian discrete geometries, and noted that certain type of causality violations lead to imaginary terms in the Regge action and thus either to suppressed or enhanced amplitudes \cite{Sorkin2019}.  However which causality violating configurations are enhanced and which are suppressed seems to be dependent on a convention: a (sign) choice for the definition of Minkowskian angles determines whether trouser-like configurations are enhanced and yarmulke configurations (see Figure \ref{fig:topology_change}) are suppressed or vice versa.  Based on an investigation of topology change in two-dimensional continuum gravity \cite{LoukoSorkin}, where one obtains also imaginary terms in the action for causality violating configurations, Sorkin argued for the choice that enhances yarmulkes and suppresses trousers \cite{Sorkin2019}.  

But causality violations that are similar in their behavior to yarmulkes appear quite generically in discrete four-dimensional approaches to (quantum) gravity, \textit{e.g.} in Regge gravity, even for configurations describing cosmological space times \cite{DGS} and in spin foams \cite{EffSF3}. Initial investigations based on four-dimensional effective spin foams \cite{EffSF} showed that an enhancing behavior for such configurations does not lead to physically sensible results.  

In this work we will show that the question of how to compute the Lorentzian path integral can also inform on how to treat causality violating configurations. In particular, we will show that a deformation of the  integration contour into the complexified configuration space, informed by Picard-Lefschetz theory \cite{PLTheory},  does also determine whether causality violating amplitudes are suppressed or enhanced. We will consider examples with yarmulke like causality violations for which the deformation of the contour leads to a suppression. We expect that this will be generically the case.

To this end we will revisit the definition of the Regge action for Lorentzian geometries, and more generally for discrete geometries described by complexified length variables. The complex Regge action does seem to depend on a sign choice, which is related to the choice for the Minkowskian angles mentioned above. We will show that this choice only matters for configurations with a certain class of causality violations: such configurations appear along  branch cuts for the  complex Regge action. The Regge action differs across the branch cuts exactly by this sign choice. 

The definition of the Regge action for complexified length variables does then allow to deform the integration contour into the complex plane. The choice of a converging deformation given by a Lefschetz thimble or combinations thereof, does also determine how to navigate the branch cuts, and those with which amplitudes the causally irregular configurations will contribute. 

Hence the question which class of causality violations are suppressed and which are enhanced is not decided at the outset but by the properties of the complex Regge action.

To illustrate this mechanism we will consider examples that describe a finite time evolution step for the  discretized deSitter universe \cite{DGS}. Homogeneity and isotropy reduce the dynamical variables in these examples to the conformal factor.  The first example considers an evolution step from vanishing scale factor to finite scale factor, and the corresponding partition function can be understood as a no-boundary wave function. Here we will have saddle points for Euclidean data. The second example considers an evolution step from finite scale factor to finite scale factor, and we will consider boundary values for which we obtain saddle points for Lorentzian data. For both examples we will find that a deformation of the contour as suggested by the Lefschetz thimble, leads to a suppression of the causality violating configurations.

~\\
This paper is organized as follows: In the next section, we will construct a Regge action for complexified edge lengths. This includes the construction of  complex angle functions $\theta^\pm$  for complexified edge lengths, in Section \ref{CDA}. We will explore the properties of the analytical extension of these angle functions, using a particular complexification of the edge lengths, induced by a Wick rotation. This will allow us to see how Euclidean and Lorentzian angles are unified into the complex angles. The complex angle function is singular if one of the edges becomes null. In Section \ref{SecNull} we will consider paths that go around these singular configurations, and illustrate that the Riemann surface associated to the angle functions is rather involved. 

We then move on to the construction of deficit angles, in Section \ref{CEps},  and finally the Regge action, in Section \ref{CRA}, as functions of complexified edge lengths. Assuming a notion of a global Wick rotation exists for a given triangulation, we then discuss some properties of the analytical continuation of the complex Regge action, and show how it reproduces two copies of Lorentzian and Euclidean Regge actions, with different global signs.

We will see that certain kind of causality violations will lead to imaginary terms for the Lorentzian Regge action. We will therefore explain in Section \ref{CCond} a class of local causality conditions for piecewise flat manifolds.

We then detail explicit example triangulations, which describes a discrete evolution step of the deSitter universe in Section \ref{SecEx}, and study the analytical extension of its Regge action in Section \ref{ssec:ball_extension}. This will show that violations of so-called hinge causality along Lorentzian data will lead to branch cuts for the complex Regge action, which result from the singularities of the complex angles for null edges, mentioned above. 

We will then compute the Regge path integral for these example triangulations in Section \ref{Sec:PLIntegral}. To this end we will use a deformation of the contour as suggested by Picard-Lefschetz theory. This will lead, on the one hand, to a converging path integral. On the other hand, these examples will illustrate how the suggested deformations lead to a choice of contour in the Riemann surface for the complex Regge action, that leads to a suppression for the causally irregular configurations. 
We end with a summary and discussion in Section \ref{discussion}.

The appendix contains more technical material referred to in the main text:  Appendix \ref{AppA} details the analytical continuations of the complex dihedral angle in the Wick rotation angle $\phi$. Appendix \ref{AppProj} provides formulas needed to compute the complex dihedral angle for higher dimensional simplices. Appendix \ref{AppC} and \ref{AppD} detail the complex Regge action, and its analytical continuation, for the example triangulations of section \ref{SecEx}.

\section{The complex Regge action}\label{Sec:CRA}

In this section we will construct a Regge action for complexified length variables and discuss its analytical continuations. The Regge action does rely heavily on the notion of (curvature) angles, we will therefore start with recalling the notion of angles in the Minkowskian plane in Section \ref{MinkA}. The definition of these Minkowskian angles comes with an ambiguity. We will however see that both choices lead to an equivalent definition, if one considers an analytical continuation for both choices, as we will do in Section \ref{CDA}.  Having constructed angles for complexified length variables, we will define the complex Regge action in Section \ref{CRA}.

\subsection{On the definition of angles in the Minkowskian plane}\label{MinkA}

On the Euclidean plane the angle between two vectors attached to the origin can be understood as a measure of distance between two points on the unit circle.  These two points arise as intersections of the two vectors with the unit circle.  Assuming that we consider always the angle of the convex wedge enclosed by the two vectors  this angle is valued in $\psi_E\in [0,\pi]$.

 For the Minkowskian plane the unit circle is replaced with four disconnected hyperbolae branches. Defining a distance measure for two points that may lie  on different hyperbolea is more subtle, but has been discussed in detail in e.g. \cite{Sorkin1974,Sorkin2019}.   If the two points are on neighbouring hyperbolae, the angle is specified (in the convention of \cite{Sorkin2019})  by a real part ${\rm Re}(\psi_{L\pm}) \in {\mathbb R}$ and an imaginary contribution  of either ${\rm Im}(\psi_{L+})=-\pi/2$ (a choice advertised in \cite{Sorkin2019}) or ${\rm Im}(\psi_{L-})=+\pi/2$ (which was used in \cite{Ding2021}). As discussed in the introduction,  the choice of sign determines which kind of causality violations are suppressed and which are enhanced.  The discussion in the next section will also make clear that the choice of sign can be understood from choosing the direction of the Wick rotation from Euclidean to Lorentzian signature: the $\psi_{L+}$ choice arises from a Wick rotation through the upper complex half-plane and the choice $\psi_{L-}$ from the lower complex half-plane for the distance square in time direction. This is the origin of our notation $\psi_{L\pm}$.

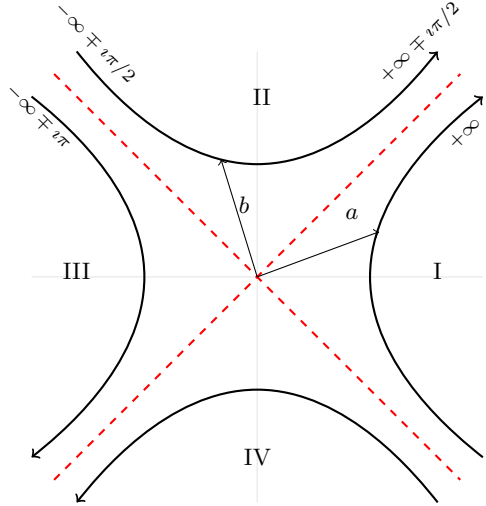
\begin{figure}[ht!]
\begin{tikzpicture}[scale=0.6]
\draw[gray!20] (-5,0)--(5,0) (0,-5)--(0,5);
\draw[dashed, red,thick] (-4.5,-4.5)--(4.5,4.5) (-4.5,4.5)--(4.5,-4.5);

\draw[thick,<-] (-4,-5) parabola bend (0,-2.5) (4,-5);
\draw[thick,<-]  (4,5) parabola bend (0,2.5)(-4,5);
\draw[thick,rotate=90 ,->] (-4,-5) parabola bend (0,-2.5) (4,-5) ;
\draw[thick,rotate=90 ,->] (4,5) parabola bend (0,2.5)(-4,5);

\draw [->] (0,0)--(2.7,1.0) ;
\draw [->] (0,0) -- (-0.8,2.6) ;

\node[rotate = 45] at (4.6,3.2) {\scriptsize $+\infty $};
\node[rotate = 45] at (3.6,5.2) {\scriptsize $+\infty \mp \imath \pi/2$};
\node[rotate = -45] at (-3.6,5.2) {\scriptsize $-\infty \mp \imath \pi/2$};
\node[below, rotate = -45] at (-4.6,3.8) {\scriptsize $-\infty \mp \imath \pi$};

\node[below] at (4,0.5) {I};
\node[left] at (0.5,4) {II};
\node[below] at (-4,0.5) {III};
\node[left] at (0.5,-4) {IV};

\node[below] at (2.1,1.7) {$ a$};
\node[right] at (-0.6,1.6) {$ b$};

\end{tikzpicture}
\caption{\label{Fig1} Lorentzian angles in the Minkowski plane.  The value of the Lorentzian angle $\psi_{L\pm}$ between $a$ and $b$, for $a$ as depicted and the end point of $b$ moving along the hyperbola in quadrant I, goes to $+\infty$. If $b$ crosses the light ray, and its end point is on the hyperbola in quadrant II, the angle $\psi_{L\pm}$ picks up a contribution $\mp \imath \pi/2$.  Moving the end point of $b$ along the hyperbola of quadrant II to the left the real part of the angle changes from positive values to negative values, whereas the imaginary part remains constant. Crossing the second light ray into quadrant III, the angle will have an imaginary part $\mp \imath \pi$.}
\end{figure}

 One way to specify the Lorentzian angles $\psi_{L\pm}$ is to define them case by case (as done in \cite{Sorkin2019}), depending on which quadrant the two vectors $a$ and $b$ are positioned in Figure \ref{Fig1}.\footnote{
 Here we will use a notion of angles without orientation, that is the angle for a wedge spanned by two vectors $a$ and $b$ is the same for clock-wise or anti-clockwise ordering of the vectors. We will furthermore assume that the wedge is convex, the angle can in this case  be uniquely specified by the vectors $a$ and $b$. The definition of angles can be extended to non-convex wedges, by demanding that the angles are additive \cite{Sorkin2019}.  We will exclude  the case that one or both of the vectors is null, see \cite{Sorkin2019} for an extensive discussion of this case.
 }
 
 Let us denote a normalized vector by $\hat a=a/|a|$ where $|a|=\sqrt{|a\cdot a|}$. The Lorentzian angles can then be defined as
\ba\label{ad1}
\psi_{L\pm}  &=& \cosh^{-1} ({\hat a \cdot \hat b} )\q \q \q\q\q \text{if}  \q\q a \, \, \text{in quadrant I} \; \;\&\;\; b\,\,  \text{in quadrant I}  \nn\\
\psi_{L\pm}  &=& \sinh^{-1} ({\hat a \cdot \hat b} ) \mp \tfrac{ \pi}{2} \imath   \q\q  \q \! \text{if} \q\q a \, \, \text{in quadrant I} \; \;\&\;\; b\,\,  \text{in quadrant II}  \nn\\
\psi_{L\pm}   &=& -\cosh^{-1} (-{\hat a \cdot \hat b} ) \mp {\pi \imath }  \,\q \text{if} \q\q a \, \, \text{in quadrant I}  \;\;\&\;\; b\,\,  \text{in quadrant III}\nn\\
\psi_{L\pm}  &=& -\cosh^{-1} (-{\hat a\cdot \hat b} ) \;  \q\q\q \text{if}  \q\q a \, \, \text{in quadrant II} \; \;\&\;\; b\,\,  \text{in quadrant II}\nn\\
\psi_{L\pm}  &=& \cosh^{-1} ({\hat a \cdot \hat b} ) \mp \pi \imath \; \, \,\; \q\q \text{if}\q\q a \, \, \text{in quadrant II} \; \;\&\;\; b\,\,  \text{in quadrant IV} \,\,  . 
\ea
Here we define $\cosh^{-1}(x) \in \mathbb{R}_+$ with $x \geq 1$.

We can also express the angles as functions of the signed\footnote{The signed squared edge lengths is positive for a space-like edge, negative for a time-like edge and null for a null edge.} squared edge lengths of the triangle spanned by the two vectors by using the relations
\ba\label{EdgeToProd}
s_a=a\cdot a  \, , \q s_b=b\cdot b  \, ,  \q s_c=a\cdot a +b\cdot b-2\,a\cdot b
\ea
where $s_a,s_b,s_c$ denote the signed squared edge lengths of the triangle, with $a$ and $b$ spanning the wedge under consideration.

The expressions (\ref{ad1}) can be in principle used to define angles for complexified variables via an analytical continuation. The case by case definition does make it however cumbersome to compute the Regge action and to analyze its analytical properties. 
It will be more convenient to start with the expressions\footnote{
One can also use a slightly different definition $\theta'$, where one replaces in (\ref{AngleLog1}) 
$+\sqrt{ (a\cdot b)^2-(a\cdot a)(b\cdot b)}$ with 
$-\sqrt{ (a\cdot b)^2-(a\cdot a)(b\cdot b)}$. 
But one can show that, with matching branches for the square roots, and an appropriate choice of the $\log$-branches, this gives just $\theta'=-\theta$. This is based  on the identity 
$(a\cdot b +\sqrt{ (a\cdot b)^2-(a\cdot a)(b\cdot b) }) 
( a\cdot b -\sqrt{ (a\cdot b)^2-(a\cdot a)(b\cdot b) }) \,=\, (a\cdot a)(b\cdot b)$ and the property $\log(z_1z_2)=\log(z_1)+\log(z_2) +2\pi \imath k$ for some $k\in \mathbb{Z}$ of the logarithm.}
\ba\label{AngleLog1}
\theta(a,b) &=&-\imath \log \frac{  a\cdot b +\sqrt{ (a\cdot b)^2-(a\cdot a)(b\cdot b)}} { \sqrt{a\cdot a} \sqrt{b \cdot b}} \q , \nn\\
\theta(s_a,s_b,s_c)&=&-\imath \log \frac{\tfrac{1}{2}(s_a+s_b-s_c)+\sqrt{\tfrac{1}{4}(s_a^2+s_b^2+s_c^2)-\tfrac{1}{2}(s_as_b+s_bs_c+s_cs_a)}}{\sqrt{s_a}\sqrt{s_b}}        \q ,
\ea
which have been also used in \cite{Sorkin2019,Ding2021}.
In fact we will show that all cases for the Lorentzian angles $\psi_{L\pm}$, as well as the Euclidean angles $\psi_E$, arise from (\ref{AngleLog1}). This has been previously shown for $\psi_{L-}$ in \cite{Ding2021}.

Adopting the principal branches for the square roots and the logarithm, the expression (\ref{AngleLog1}) has a branch cut for Euclidean signature, and for a number of the Lorentzian cases (e.g. when one of the vectors $a,b$ is time-like).   The branch cut for the Euclidean case stems from the square root in the numerator and the fact that $(a\cdot b)^2-(a\cdot a)(b\cdot b)$ is negative in Euclidean signature. Indeed the argument is given by $(-4)$ times the area square of the triangle spanned by $a$ and $b$, which we express as $-4\mathbb{A}(a,b)$ or $-4\mathbb{A}(s_a,s_b,s_c)$. This also shows that this argument of the square root is positive in Lorentzian signature.

To deal with the issue of branch cuts we can complexify either the components of the vectors $a,b$ or the squared edge lengths $s_a,s_b,s_c$. Adopting the principal branches for the square roots and logarithms (\ref{AngleLog1}) is then well defined for generic points $(a,b)\in \mathbb{C}^4$, respectively $(s_a,s_b,s_c)\in \mathbb{C}^3$.  But we will still have branch cuts, which in particular include data describing Euclidean signature configurations and part of the Lorentzian signature configurations. Approaching these branch cuts from different directions, that is different regions of analyticity, we will obtain different limit values for the angle $\theta$. To obtain a well-defined formula that encompasses also these limits, we have to specify which region of analyticity we consider.  For each region of analyticity we can furthermore define an analytical extension. Crossing a branch cut,  this analytical extension will differ in some of the choices for the branches  from the principal branches, we started with.

To analyze the possibilities of  determining regions of analyticity for (\ref{AngleLog1}), and thus provide a complete definition for the complex angle,  we will first consider a particular way to complexify the vectors $(a,b)$  and correspondingly, the edge lengths $(s_a,s_b,s_c)$. This will amount to a Wick rotation in $\mathbb{R}^2$. This will reduce\footnote{This is not a true dimensional reduction, as  in terms of edge lengths the angle is locally a function of $(s_a+s_b-s_c)/(\sqrt{s_a}\sqrt{s_b})$ and in terms of the vector components it is locally a function of $a\cdot b/(\sqrt{a\cdot a}\sqrt{b \cdot b})$, that is, of only one complex parameter.}
 our analysis from $\mathbb{C}^4$ or  $\mathbb{C}^3$ to essentially one complex dimension, which can be parametrized by the Wick rotation angle $\phi$, and \textit{e.g.} $\theta$ at $\phi=\pi$. We will then apply the insights from this analysis to the general case, and in Section \ref{SecNull} consider an example of a loop in complexified  $(s_a,s_b,s_c)$ space, which goes around a configuration with a null edge.

\subsection{The complexified dihedral angles}\label{CDA}

To define the Wick rotation we introduce the following generalization of the Euclidean and Minkowskian inner product between two vectors $a=(a_0,a_1)$ and $b=(b_0,b_1)$:
 \ba\label{Wick1}
 a\star b =e^{\imath \phi} a_0 b_0 +a_1 b_1       \q ,
 \ea
where $\phi \in (-\pi,\pi]$ can be understood as a  Wick rotation angle.  We recover the Euclidean inner product for $\phi=0$ and the Minkowskian inner product for $\phi=\pi$. Note that this just amounts to the usual, coordinate-based Wick rotation of the $x_0$--coordinate $x_0\rightarrow \exp(i\phi/2)x_0$.   We use rather $\phi$ instead of $\phi/2$ as Wick rotation angle, because we will  base the complexification on the squared edge lengths, or equivalently, on the metric components. But we should not be surprised to find that the analytical continuations lead to an extension of the $\phi$-domain to $(-2\pi,2\pi]$.

For an edge $a$ along the $x_0$-axis, we have  $s_a=e^{\imath \phi} |s_a|$. Thus, in this case, the Wick rotation angle can be identified with the polar angle of the complexification of $s_a$. In general, we can use the formulas $(\ref{EdgeToProd})$ relating the squared edge lengths to the vector inner products. The Wick rotation (\ref{Wick1}) then induces $\phi-$parametrized cycles in the space of complexified edge lengths $(s_a,s_b,s_c)$.

In the following we will therefore investigate the  definition
\ba\label{thetaW1}
\theta&=&-\imath \log \frac{  a\star b +\sqrt{ (a\star b)^2-(a\star a)(b\star b)}} { \sqrt{a\star a} \sqrt{b \star b}} 
\ea
of the dihedral angle, as a function of the Wick rotation angle $\phi$.  As mentioned above, we consider the dihedral angle of the convex wedge formed by the two vectors $a$ and $b$. The case $a=-b$ will require special attention, and has to be also defined as limit. We will exclude this case for now. Later we will differentiate between cases where $a,b$ are lying in different quadrants of $\mathbb{R}^2$. The definition of these quadrants is the same as in Fig. \ref{Fig1}.

Adopting the principal branches for the square roots and the logarithm, we see that $\theta$ as defined in (\ref{thetaW1}) has possibly branch cuts for $\phi=0$ and $\phi= \pi$. This does coincide exactly with the values describing the Euclidean or Lorentzian signatures. We therefore have to specify whether to (analytically) continue from $\phi\in (0,\pi)$ or from $\phi\in (-\pi,0)$. We will refer to $\theta$ defined for $\phi\in (0,\pi)$ as $\theta^+$ and to $\theta$ defined for $\phi\in (-\pi,0)$ as $\theta^-$.  The limit values for these angles are then described by the following

\begin{lemma}\label{lemma1} For the analytical continuation from $\phi\in (0,\pi)$ we find
\ba\label{theta+}
\lim_{\phi\rightarrow 0\downarrow} \theta^+ \,=\,-\psi_E \q \q\text{and} \q\q\q
\lim_{\phi\rightarrow \pi \uparrow} \theta^+ \,=\,-\imath \psi_{L+}  \q .
\ea
For the analytical continuation from $\phi\in (-\pi,0)$ we find
\ba\label{theta-}
\lim_{\phi\rightarrow 0\uparrow} \theta^- \,=\,+\psi_E \q\q \text{and} \q\q\q
\lim_{\phi\rightarrow -\pi\downarrow} \theta^- \,=\,-\imath \psi_{L-}  \q .
\ea
\end{lemma}

We will provide a proof together with the proof for Theorem \ref{theorem1} below. The version $\theta^+$ reproduces Sorkin's definition \cite{Sorkin2019} of the Lorentzian angle $-\imath \psi_{L+}$, and gives for Euclidean data $-\psi_E$. The version $\theta^-$ reproduces Jia's definition \cite{Ding2021} of the complex angle, it gives for Lorentzian data $-\imath \psi_{L-}$, and for Euclidean data $+\psi_E$. 
 We will see that this pattern is important in order to be able to construct a manifestly analytical expression  for the Regge action, which connects smoothly the Lorentzian to the Euclidean signature case. Although it seems we will obtain two different versions for the complex Regge action, we will see that the two versions are equivalent after they have been analytically extended.

In constructing the limit values for $\theta^+$ or $\theta^-$  we also determined which sides of the branch cuts for the square roots and the logarithms we should adopt. We use the principal branch for arguments  $z \neq \mathbb{R}_-$. To clarify the choice for the negative real axis we define
\ba
\log_+(re^{\imath \phi}) &=\log(r) + \imath \phi        \,    \quad \text{for} \quad \phi\in (-\pi,\pi]\,, \q\q
\log_-(re^{\imath \phi}) &=\log(r) + \imath \phi            \quad \text{for} \quad \phi\in [-\pi,\pi)\, , \nn\\
\sqrt{\!\!{}_{{}_+}(re^{\imath \phi})} &= \sqrt{r} e^{\imath \frac{\phi}{2} } \quad \quad\q \text{for} \q  \phi\in (-\pi,\pi] \,,  \q\q
\sqrt{\!\!{}_{{}_-} (re^{\imath \phi})} &= \sqrt{r} e^{\imath \frac{\phi}{2} } \quad \q\q \text{for} \quad \phi\in [-\pi,\pi)  \, .
\ea
Investigating the various cases where discontinuities appear, we learn that we can extend $\theta^+$ and $\theta^-$ to their limit values as follows:

\begin{lemma}\label{lemma2}
\ba
\theta^+  &=&   -\imath \log_- \frac{  a\star b +\sqrt{\!\!{}_{{}_-} \left( (a\star b)^2-(a\star a)(b\star b)\right) }} { \sqrt{\!\!{}_{{}_+}(a\star a)} \sqrt{\!\!{}_{{}_+}(b \star b)} }             \q\quad \text{for} \quad \phi\in [0,\pi] \, ,\nn\\
\theta^-&=&  -\imath \log_+ \frac{  a\star b +\sqrt{\!\!{}_{{}_+} \left( (a\star b)^2-(a\star a)(b\star b)\right) } } { \sqrt{\!\!{}_{{}_-}(a\star a)} \sqrt{\!\!{}_{{}_-}(b \star b)} } \q\quad \text{for} \quad \phi\in [-\pi,0] \, .
\ea
These choices for the log branch cut values give also the correct (according to Lemma (\ref{lemma1})) dihedral angles $\theta^\pm=\mp \pi$ for the case that $a=-b$.
\end{lemma}
We will provide a proof together with the proof for Theorem \ref{theorem1} below. Note that, considering the limit values approached from \textit{e.g.} $\phi \in (0,\pi)$, in the cases that one does encounter branch cuts, one always obtains the same sides of the branch cuts, independently of which quadrant the vectors $a$ and $b$ are situated in.

~\\
This does now allow us to define the complex angle as a function of complexified squared edge lengths, going beyond the more special case of the complexification induced by the Wick rotation (\ref{Wick1}).  To this end we specify  the regions of analyticity leading to $\theta^+$ and $\theta^-$ in terms of the squared edge lengths.
\begin{definition} \label{Def1} The complex dihedral angles as functions of the complexified edge lengths are defined by
\ba\label{thetaTria}
\theta^\pm(s_a,s_b,s_c)&=&-\imath \log_\mp \frac{\tfrac{1}{2}(s_a+s_b-s_c)+2\sqrt{\!\!{}_{{}_\mp}-\mathbb{A}(s_a,s_b,s_c)}}{\sqrt{\!\!{}_{{}_\pm}s_a}\sqrt{\!\!{}_{{}_\pm}s_b}}
\ea
with the understanding that $\theta^+$ is defined as analytical continuation from the region where $\text{Im}(s_a)>0$ if $\text{Re}(s_a)<0$ and  $\text{Im}(s_b)>0$ if $\text{Re}(s_b)>0$. We require also that $\text{Im}(\mathbb{A}(s_a,s_b,s_c))>0$ if $\text{Re}(\mathbb{A}(s_a,s_b,s_c))>0$ and that  if the argument of the logarithm has negative real part, it should have negative imaginary part. Likewise, we define $\theta^-$ from a region where $\text{Im}(s_a)<0$ if $\text{Re}(s_a)<0$ and  $\text{Im}(s_b)<0$ if $\text{Re}(s_b)<0$. We require also that $\text{Im}(\mathbb{A}(s_a,s_b,s_c))<0$ if $\text{Re}(\mathbb{A}(s_a,s_b,s_c))>0$ and that  if the argument of the logarithm has negative real part it should have positive imaginary part.
\end{definition}

\begin{figure}[ht!]
\begin{picture}(500,120)
\put(30,7){ \includegraphics[scale=0.26]{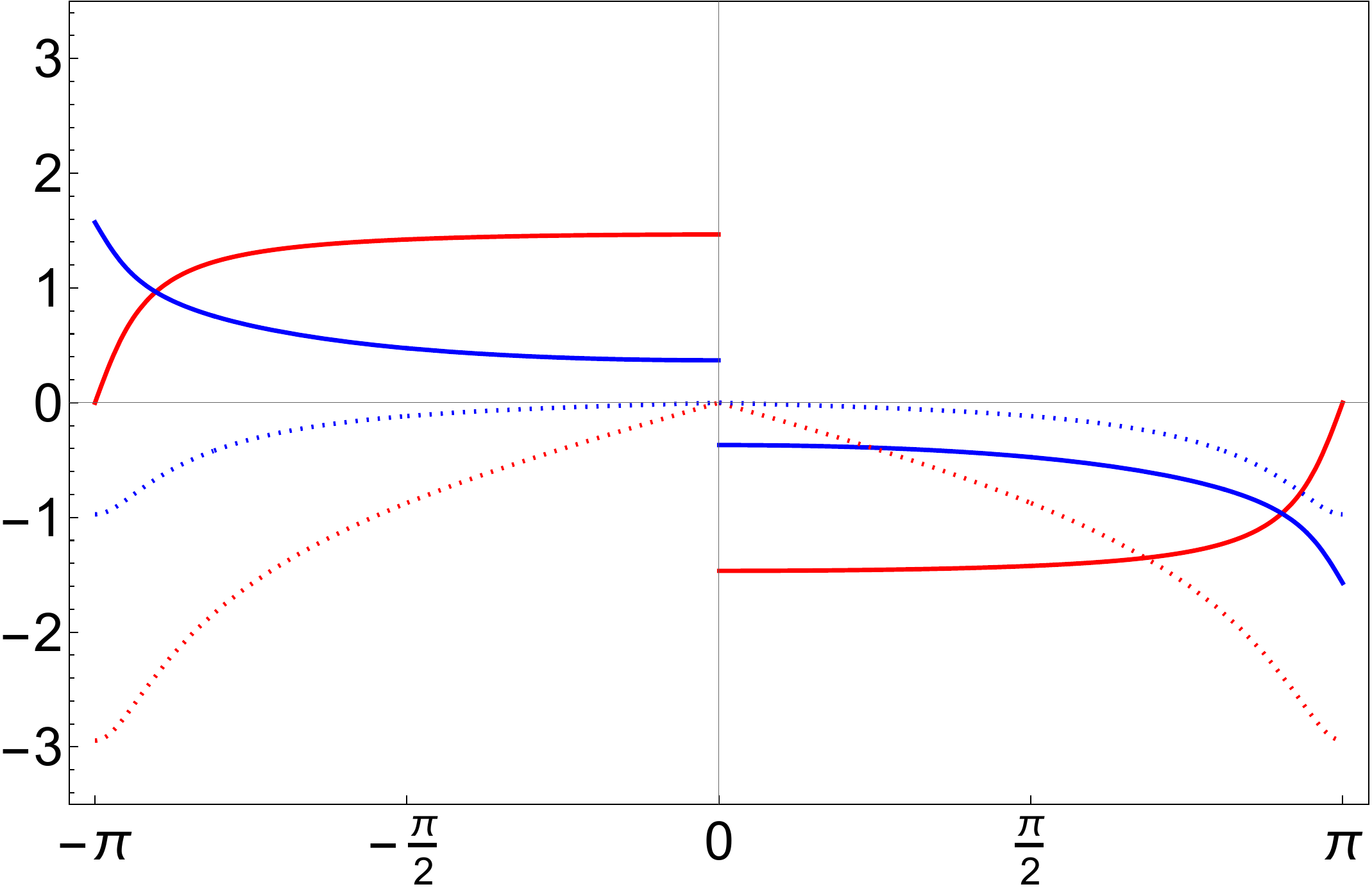} }
\put(280,7){ \includegraphics[scale=0.26]{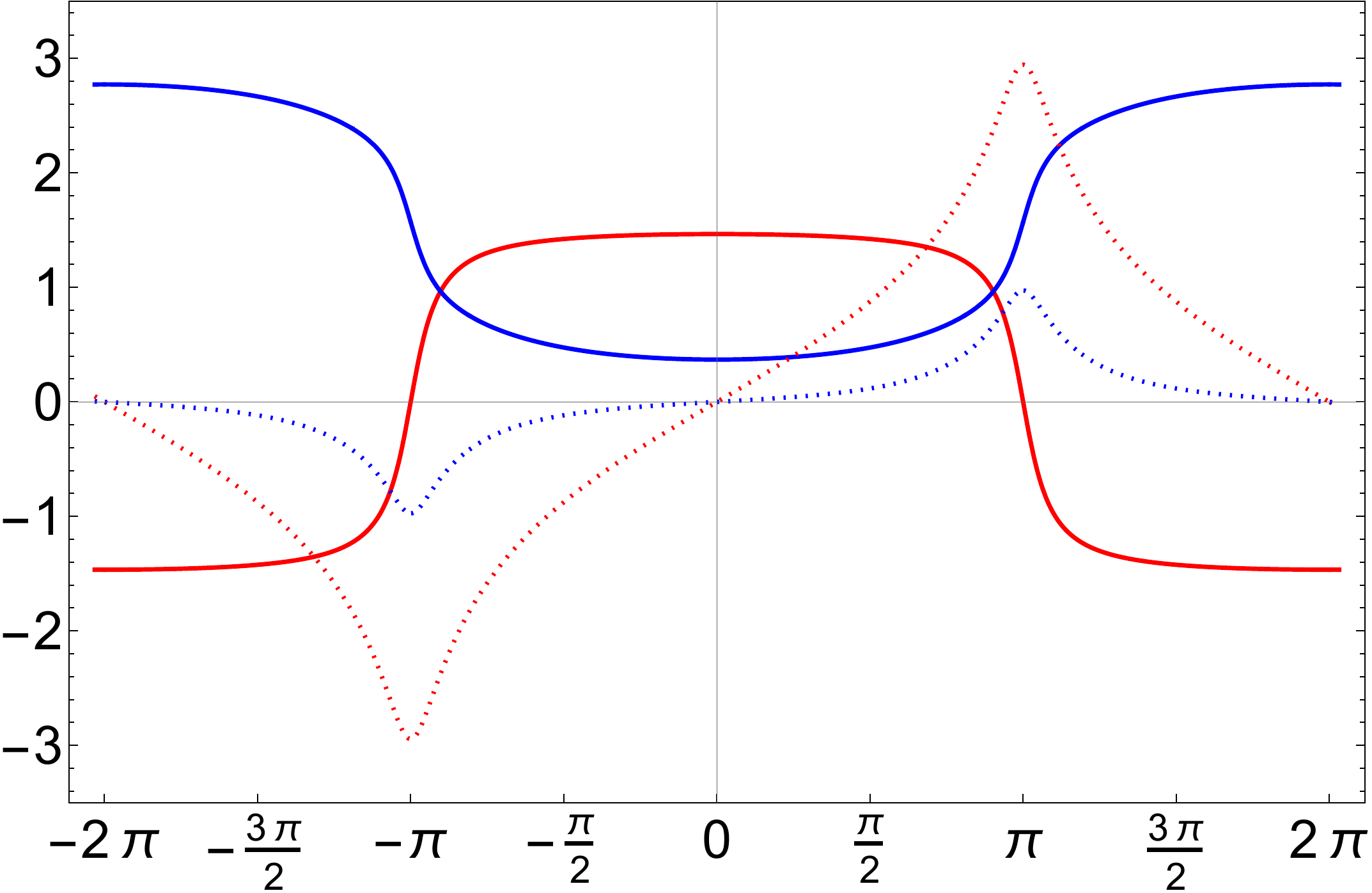} }

\put(130,0){$\phi$}
\put(370,0){$\phi$}

\put(25,110){$ \theta $}
\put(270,110){$ \theta^- $}

\end{picture}
\caption{The left panel shows the real part (solid curve) and imaginary part (dotted curve) of the dihedral angle $\theta$ defined in (\ref{thetaW1}), as a function of the Wick rotation angle $\phi$ for  two examples. These are $(a_0,a_1,b_0,b_1)=(0.9, 1, -0.9,1)$ (in red), where both vectors are in quadrant I and $(a_0,a_1,b_0,b_1)=(1.1, 1, 0.5,1)$ (in blue) with $a$ in quadrant II and $b$ in quadrant I.  The right panel shows the analytical extension to $\phi\in (-2\pi,2\pi]$, starting from the region $-\pi<\phi<0$.} \label{AnCont}
\end{figure}
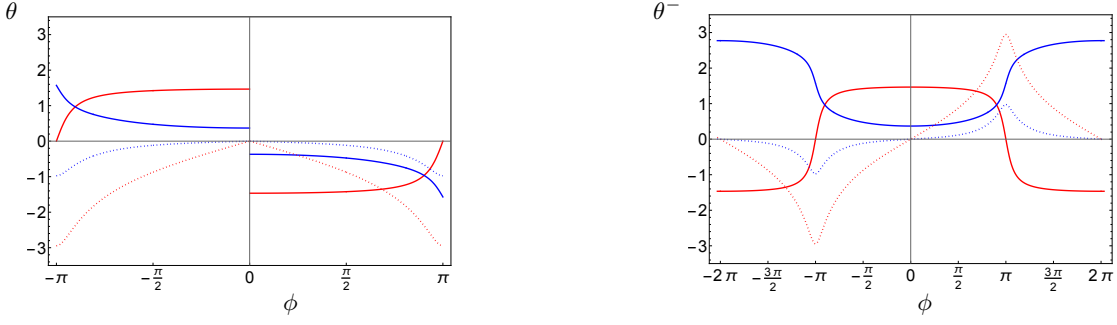

Coming back to $\theta^+$ and $\theta^-$ defined as a function of the Wick rotation angle $\phi$, we can consider an analytical continuation of $\theta^+$ to a domain which extends $[0,\pi]$, and for $\theta^-$ to a domain that extends $[-\pi,0]$, see Figure \ref{AnCont} for an illustration. This will lead to functions which are periodic and analytic on $(-2\pi,2\pi]$ and one has moreover $\theta^+=-\theta^-$.

\begin{theorem} \label{theorem1}
The analytical continuations of $\theta^+$ and $\theta^-$ in $\phi$ lead to periodic functions on $(-2\pi,2\pi]$, given by
\ba\label{eqtheo1}
\theta^\pm& =&
\begin{cases}
 -\imath \log_\mp \left(\frac{  a\star b \mp 2\imath   e^{\imath\phi/2}  |A(a,b)|  } { \sqrt{a\star a} \sqrt{  b \star b}}  \right)\, \,\,\q\q \q\q\text{for} \,\, \phi \in (-\pi,\pi] \\
  -\imath \log_\pm \left( \frac{  a\star b \mp 2\imath   e^{\imath\phi/2}  |A(a,b)|  } { \sqrt{a\star a} \sqrt{b \star b}} \right)\,\mp N_c \pi \q \,\,\,\text{for} \,\, \phi \in (-2\pi,-\pi] \cup (\pi,2\pi]       \q .
 \end{cases}
\ea
where $N_c$ denotes the number of light rays in the convex wedge spanned by $a$ and $b$. 
For these analytical continuations we have $\theta^+=-\theta^-$. The real parts of $\theta^\pm$ are  even functions in $\phi$ and the imaginary parts are  odd functions in $\phi$. Hence the imaginary parts vanish for $\phi=0$ and $\phi=2\pi$. The real parts at $\phi=\pm\pi$   are given by $\text{Re}(\theta^+(\pm\pi))= -\pi N_c/2 $ and $\text{Re}(\theta^{-}(\pm\pi))=+\pi N_c/2$.

For the $\phi$-values representing Euclidean data we have 
\ba
 \theta^\pm(0)=\mp \psi_E \q \q,\q\q     \theta^\pm(2\pi)=\pm \psi_E \mp \pi N_c \q .
 \ea  

For Lorentzian data we have 
\ba
 \theta^\pm(+\pi)=\mp \imath \psi_{L+}\q ,\q\q  \theta^\pm(-\pi)=\pm \imath \psi_{L-} \q .
 \ea
\end{theorem}

~\\

One might be surprised to find a dependence of $\theta^\pm(\phi=2\pi)$, that is for a $\phi$-value representing Euclidean signature, on the number of light ray crossings  $N_c$ in Lorentzian signature. In particular, $\theta^\pm(2\pi)$ is not invariant under simultaneous (Euclidean) rotations of the vectors $a$ and $b$. But we will see that this apparent tension will get resolved when we consider the analytical extension of the deficit angle.

~\\
{\bf Proof:} 
We first note that 
\ba\label{b2.10}
\left( (a\star b)^2-(a\star a)(b\star b)\right)&=& -e^{\imath\phi}(a_1 b_2 -a_2 b_1)^2 \,=\, -e^{\imath\phi} 4 |A(a,b)|^2
\ea
where $|A(a,b)|$ is the positive area (which is equal in Euclidean and Minkowskian signature) of the triangle spanned by $a$ and $b$. We can thus resolve the branch cut of the numerator at $\phi=0$ in the definition of $\theta$, by using the alternative representation
\ba\label{b2.11}
\theta^\pm& =& -\imath \log_\mp \left(\frac{  a\star b \mp 2\imath   e^{\imath\phi/2}  |A(a,b)|  } { \sqrt{a\star a} \sqrt{b \star b}}  \right)\q ,
\ea
which is analytic for $\phi \in (-\pi,\pi)$. Here we specified the branch cut values for the logarithm, so that we get the correct result $\theta^\pm(\phi)=\mp \pi $ for the case $a=-b$. We also used the analytical continuation for the square root in the numerator, determined from the region $\phi\in (0,\phi)$ for $\theta^+$, and from the region $\phi\in (-\phi,0)$ for $\theta^-$. This determines the use of the branch cut value $\sqrt{\!\!{}_{{}_\mp} -1}=\mp\imath$ at $\phi=0$.

Now let us consider the expression (\ref{b2.10}) for the different cases of the vectors $a$ and $b$ being space-like or time-like. (As a reminder, we exclude the case that one of the vectors is null.)

\begin{itemize}
\item
In the case that both vectors are space-like in Lorentzian signature, the arguments in the square roots in the numerator in (\ref{b2.11}) will have for all $\phi \in (-2\pi,2\pi]$ a positive real part. These are therefore analytic functions in $\phi \in (-2\pi,2\pi]$ and $2\pi$-periodic and therefore also $4\pi$-periodic. 

We have now two cases to consider: in the first case $a$ and $b$ are from the same quadrant (either quadrant I or quadrant III).  The explicit expression for the angle in this case can be found in Appendix \ref{AppA}. It shows that the argument 
\ba
Z_\pm&=&\frac{  a\star b \mp 2\imath   e^{\imath\phi/2}  |A(a,b)|  } { \sqrt{a\star a} \sqrt{b \star b}}
\ea
 of the logarithm $\ln Z_\pm$  in $\theta^\pm$ has positive real part for $\phi \in (-2\pi,2\pi]$. Hence (\ref{b2.11}) is analytic for $\phi \in (-2\pi,2\pi]$ and $4\pi$-periodic. We have also that  $\text{Arg}(Z)=\text{Re}(\theta^\pm)$.  As we avoid the branch cut of the logarithm we have  $|\text{Re}(\theta^\pm)|<\pi$.

In the second case $a$ and $b$ are from opposite quadrants. The argument of the logarithm  $Z_\pm$ has then negative real part for $\phi \in (-2\pi,2\pi]$.  The imaginary part of this argument of the logarithm is negative for $\theta^+$ (and positive for $\theta^-$) at $\phi=0$.  The imaginary part   crosses zero for $\phi=\pm \pi$, and is positive (negative) at $\phi=2\pi$.  We thus encounter the branch cut for the logarithm. The analytical continuation is therefore defined as
\ba\label{b2.12}
\theta^\pm& =&
\begin{cases}
 -\imath \log_\mp \left(\frac{  a\star b \mp 2\imath   e^{\imath\phi/2}  |A(a,b)|  } { \sqrt{a\star a} \sqrt{b \star b}}  \right)\, \,\q \q\q\text{for} \,\, \phi \in (-\pi,\pi] \\
  -\imath \log_\pm \left( \frac{  a\star b \mp 2\imath   e^{\imath\phi/2}  |A(a,b)|  } { \sqrt{a\star a} \sqrt{b \star b}} \right)\,\mp 2\pi \q \text{for} \,\, \phi \in (-2\pi,-\pi] \cup (\pi,2\pi]       \q .
 \end{cases}
\ea

\item
In the case that both vectors are time-like in Lorentzian signature, the product of  the square roots in the numerator in (\ref{b2.11}) is analytic for all $\phi \in (-2\pi,2\pi]$  and $2\pi$- and therefore also $4\pi$-periodic. 

Again we have  two cases to consider: in the first case $a$ and $b$ are from the same quadrant (either quadrant II or quadrant IV).  The argument of the logarithm in $\theta^\pm$ has then positive real part for $\phi \in (-2\pi,2\pi]$,  and (\ref{b2.11}) is analytic for $\phi \in (-2\pi,2\pi]$ and $4\pi$-periodic.

In the second case $a$ and $b$ are again from opposite quadrants. Here the same properties hold as discussed above for the space-like case. Therefore we have also the same analytical extension (\ref{b2.12}).

\item The last case to consider is that one of the vectors, say $a$ is space-like and the other, $b$, is time-like in Lorentzian signature. Here we encounter a branch cut for the square root $\sqrt{b\star b}$. This can be analytically extended from $(-\pi,\pi)$  to $\phi \in (-2\pi,2\pi]$ by
\ba\label{b2.13}
\sqrt{b\star b}\rightarrow
\begin{cases}
+\sqrt{ b\star b} \q \text{for} \,\, \phi \in (-\pi,\pi] \\
-\sqrt{ b\star b}  \q \text{for} \,\, \phi \in (-2\pi,-\pi] \cup (\pi,2\pi] \; . \\
\end{cases}
\ea
We avoid now the branch cut of the logarithm because the imaginary part is either always negative (for $\theta^+$) or always positive (for $\theta^-$). In particular, different from the cases above, we have a non-vanishing imaginary part at $\phi=\pm \pi$, originating from $\pm \sqrt{ b\star b}$.

We can however evaluate the logarithm for the sign factor $(-1)$ in (\ref{b2.13}), which we used to define the analytical extension of the square root. This leads to 
\ba\label{b2.14}
\theta^\pm& =&
\begin{cases}
 -\imath \log_\mp \left(\frac{  a\star b \mp 2\imath   e^{\imath\phi/2}  |A(a,b)|  } { \sqrt{a\star a} \sqrt{  b \star b}}  \right)\, \,\q \q\q\text{for} \,\, \phi \in (-\pi,\pi] \\
  -\imath \log_\pm \left( \frac{  a\star b \mp 2\imath   e^{\imath\phi/2}  |A(a,b)|  } { \sqrt{a\star a} \sqrt{b \star b}} \right)\,\mp \pi \q \,\,\,\text{for} \,\, \phi \in (-2\pi,-\pi] \cup (\pi,2\pi]       \q .
 \end{cases}
\ea

\end{itemize}

~\\
In summary, we can express the analytical continuation to $\phi \in (-2\pi,2\pi]$ for all cases as 
\ba\label{b2.15}
\theta^\pm& =&
\begin{cases}
 -\imath \log_\mp \left(\frac{  a\star b \mp 2\imath   e^{\imath\phi/2}  |A(a,b)|  } { \sqrt{a\star a} \sqrt{  b \star b}}  \right)\, \,\,\q\q \q\q\text{for} \,\, \phi \in (-\pi,\pi] \\
  -\imath \log_\pm \left( \frac{  a\star b \mp 2\imath   e^{\imath\phi/2}  |A(a,b)|  } { \sqrt{a\star a} \sqrt{b \star b}} \right)\,\mp N_c \pi \q \,\,\,\text{for} \,\, \phi \in (-2\pi,-\pi] \cup (\pi,2\pi]       \q .
 \end{cases}
\ea
where $N_c$ does denote the number of light rays in the convex wedge spanned by $a$ and $b$. Figure \ref{AnCont} shows examples of the analytical extension of $\theta$ from the region $\phi \in(-\pi,0]$ to $\phi \in(-2\pi,2\pi]$. 

~\\

To show $\theta^+=-\theta^-$ we realise that for $\phi \in (-\pi,\pi)$ we can write
\ba\label{b3.15}
\frac{a\star b -\imath \sqrt{ (a\star a)(b\star b)-(a\star b)^2}}{\sqrt{ a\star a}\sqrt{b\star b}} \, \cdot\,
\frac{a\star b +\imath \sqrt{ (a\star a)(b\star b)-(a\star b)^2}}{\sqrt{ a\star a}\sqrt{b\star b}} 
&=&\frac{ (a\star a)(b\star b) }{(a\star a)(b\star b)} =1 \q .
\ea
As $0=\log(zz^{-1})=\log_-(z)+\log_+(z^{-1})$, we have that $\theta^+=-\theta^-$ for $\phi \in (-\pi,\pi)$, and by analytical extension for $\phi \in (-\pi,2\pi]$.

That the real part of $\theta^\pm$ is an even function of $\phi$ and the imaginary part an odd function of $\phi$, follows from
\ba
(Z_\pm(\phi))^{-1} = \overline{Z_\pm(-\phi)}\q ,
\ea
which holds for the arguments $Z_\pm(\phi)$ of the logarithms appearing in the definition of $\theta^\pm$. Indeed, note that \\ $(x\star y)(-\phi)=\overline{ (x\star y)(-\phi)}$. Therefore, making use of (\ref{b3.15}),
\ba
\overline{Z_\pm(-\phi)} &= &\left( \frac{  a\star b \pm 2\imath   e^{\imath\phi/2}  |A(a,b)|  } { \sqrt{a\star a} \sqrt{  b \star b}}  \right)\,=\,
\left( \frac{  a\star b \mp 2\imath   e^{\imath\phi/2}  |A(a,b)|  } { \sqrt{a\star a} \sqrt{  b \star b}}  \right)^{-1}\,=\, (Z_\pm(\phi))^{-1} \q .
\ea

The analytical continuation (\ref{b2.15}) gives for Euclidean data at $\phi=0$
\ba
\theta^\pm(0)&=&
-\imath \log_\mp \left(\frac{  a\cdot_E b \mp 2\imath     |A(a,b)|  } { \sqrt{a\cdot_E a} \sqrt{  b \cdot_E b}}  \right)\,=\, \mp \psi_E \q 
\ea
where $a\cdot_E b$ denotes the Euclidean inner product.
For $\phi=2\pi$ we obtain
\ba
\theta^\pm(2\pi)&=&
-\imath \log_\mp \left(\frac{  a\cdot_E b \pm 2\imath     |A(a,b)|  } { \sqrt{a\cdot_E a} \sqrt{  b \cdot_E b}}  \right)\mp N_c\pi\,=\, \pm \psi_E \mp N_c\pi \q .
\ea
The evaluation for Lorentzian data at $\phi=\pm \pi$ can be verified case by case.
See also Appendix \ref{AppA} for a listing of the explicit analytical continuations for the various cases.

\subsection{Circling around a configuration with a null edge}\label{SecNull}

In the previous subsection we excluded the case that the vectors $a$ or $b$ are null in Lorentzian signature. The complex angle is indeed singular for such configurations. But we can consider a path in the space of complex edge length squared around such a configuration. 

Consider for instance the following path in the space of squared edge lengths
\ba\label{sa1}
s_a(\psi)=r_a \exp(\imath \psi) \, ,\q s_b=1 \,\q s_c=4 \, ,
\ea
parametrized by $\psi\in (-\pi,\pi]$. If  $r_a<1$, we have for both $\psi=0$ and $\psi=\pi$ a Minkowskian triangle, that is the Lorentzian triangle inequalities are satisfied. We will therefore restrict to $r_a<1$.

The dihedral angle (\ref{AngleLog1}) for the configurations (\ref{sa1}) is given by
\ba
\theta(s_a)=-\imath \log \frac{ s_a-3+\sqrt{s_a^2-10s_a+9}}{2\sqrt{s_a}}  \q .
\ea
Let us consider $\theta(s_a)=\theta(r_a \exp(\imath \psi))$ for $\psi\in(0,\pi)$. For the square root in the numerator we do not encounter a branch cut. We approach the branch cut of the square root $\sqrt{s_a}$ in the denominator for $\psi\rightarrow \pi$ from the upper complex half-plane. We approach also the branch cut of the logarithm when $\psi\rightarrow 0$, with the argument $Z$ of the logarithm having negative imaginary part. This is consistent with our Definition \ref{Def1} for the complex dihedral angles. For the analytical continuation from  $\psi\in(0,\pi)$ to $\psi\in (-2\pi,2\pi)$, we obtain
\ba
\theta^+(\psi)&=&
\begin{cases} 
-\imath \log \left(\frac{ r_a  \exp(\imath \psi)-3+\sqrt{( r_a\exp(\imath \psi))^2-10 r_a \exp(\imath \psi)+9}}{2\sqrt{r_a}  \exp(\imath \psi/2)}\right) -2\pi \q \text{for}\, \,-2\pi<\psi\leq 0\\
-\imath \log \left(\frac{ r_a  \exp(\imath \psi)-3+\sqrt{( r_a\exp(\imath \psi))^2-10 r_a \exp(\imath \psi)+9}}{2\sqrt{r_a}  \exp(\imath \psi/2)} \right)\,\,\q \q\q\text{for}\,\, \,\, 0<\psi<2\pi   \q .
\end{cases}
\ea
The real part of this function is however {\it not} periodic in $\psi$. We rather have a difference of $2\pi$ between the limits of $\psi\rightarrow -2\pi$ and $\psi\rightarrow +2\pi$ (that is we encircle $s_a=0$ twice), and a difference of $\pi$ between the values at $\psi=-\pi$ and $\psi=\pi$ (that is we encircle $s_a=0$ once).  The imaginary part of the angle is  $2\pi$-periodic and therefore also $4\pi$-periodic. 
Thus $s_a=0$, which describes an angle with a null edge can be seen as a branch point of type $\log\sqrt{z}$.

This discussion fore-shadows the appearance of branch points and branch cuts in the Regge action. Indeed in our examples in Section \ref{SecEx} such branch points are connected to the change of signature of building blocks, which lead to a change of signature for the edges defining the dihedral angles.

\subsection{The complexified deficit angles}\label{CEps}

 The deficit angle arises when one glues several wedges to a cone. It is a measure for the curvature concentrated on the tip of the cone. It is defined as the difference of the value for the ``flat cone" and the sum of the angles of the wedges. The flat cone angle   can be defined as 4 times the angle between $a=(0,1)$ and $b=(1,0)$.  Now we have  for {\it both}  signatures, and actually for all values of the Wick rotation angle $\phi$, that   
\ba
\theta^\pm(\text{flat cone})&=&\mp2\pi    \q , \nn\\
\ea
This allows as to define  complex deficit angles $\delta^\pm$ (as a function of $\phi \in (-2\pi,+2\pi]$) by  
\ba
\delta^\pm(\phi)&=&(2\pi\pm\sum_{w_\sigma} \theta^\pm_{w_\sigma}(\phi)) \,\, ,
\ea
where $w_\sigma$ runs over the wedges associated to a given cone. As we previously found that $\theta^+(\phi)=-\theta^-(\phi)$, we have $\delta^+(\phi)=\delta^-(\phi)$. But we will keep the super-index $\pm$ to remember how these deficit angles are constructed.

To compare to the deficit angles defined for Euclidean and Lorentzian triangulations, we will shortly review their definitions.
For Euclidean signature the  deficit angle is defined as
\ba\label{DefE}
\epsilon_E=2\pi-\sum_{w_\sigma} (\psi_E)_\sigma       \q .
\ea
For Lorentzian signature, remember that there are two different choices $\psi_{L\pm}$ for the angles, which only differ in their value $\mp \imath \pi/2$ for a light ray crossings. Accordingly, we can define two versions for the deficit angle:
\ba\label{DefL}
\epsilon_{L\pm}=\mp 2\pi \imath-\sum_{w_\sigma} (\psi_L\pm)_{w_\sigma}          \q .
\ea
These Lorentzian deficit angles agree in their real part. The imaginary part is vanishing, if and only if the wedges do include exactly four light ray crossings, that is $N_c=4$. In this case we refer to the deficit angle as causally regular. If this is not the case, we do have a non-vanishing imaginary part, given by ${\rm Im}(\epsilon_{L\pm})= \pm N_c \pi/2$.

We can now compare the evaluation of the complex deficit angles $\delta^\pm$ at $\phi=0$ and $\phi=2\pi$, as well as at $\phi=\pm \pi$ to the Euclidean and Lorentzian deficit angles, respectively.  

For  $\delta^\pm$ at $\phi=0$ we have
\ba
\delta^\pm(0)= \epsilon_E \q .
\ea
For $\phi=2\pi$ we use that $\theta^\pm(2\pi)=-\theta^\pm(0)\mp N_c\pi$ and obtain
\ba\label{2.25}
\delta^\pm(2\pi)=-\epsilon_E +(4-N_c)\pi       \q .
\ea
The deficit angle for Euclidean data at $\phi=2\pi$  does therefore depend on a Lorentzian property, namely whether the Lorentzian deficit angle is causally regular or not. We will however see that causally irregular data lead to branch cuts and thus an associated analytical extension, that can be attached to these branch cuts for the causally irregular data in the Lorentzian domain.

For $\phi=\pi$ we have that $\theta^\pm(\pi)=\mp\imath \psi_{L+}$, thus
\ba
\delta^\pm(+\pi)=+ i\epsilon_{L+} \q .
\ea
Similarly, as $\theta^\pm(-\pi)=\pm\imath \psi_{L-}$, we obtain
\ba
\delta^\pm(-\pi)=- i\epsilon_{L-} \q .
\ea

Consider a cone configuration whose edge lengths squared are parametrized by a complex variable $z$, such that (a) the polar angle of $z$ can be identified with the Wick rotation angle $\phi$ and (b) configurations with $|z|>c$ at $\phi=\pm \pi$ are causally regular and configurations with $|z|<c$ at $\phi=\pm \pi$ are causally irregular, where $c\in \mathbb{R}_+$ is some positive number. 

We have seen that we can extend $\delta^+(z)$ from $\phi\in (0,\pi)$ to $\phi \in (-2\pi,2\pi]$ and $\delta^-(z)$ from $\phi\in (-\pi,0)$ to $\phi \in (-2\pi,2\pi]$, at least if we do not encounter any wedges whose boundary edges are null (which might however apply to the configuration with $|z|=c$). But if we move from $|z|>c$ along $\phi=\pm \pi$ to $|z|<c$ we see that the real part of $\delta^\pm$ changes from $0$ to $- N_c\pi/2$.  The points with $|z|=c$ and $\phi=\pm \pi$ are therefore singular, and we can expect a branch cut for $|z|\leq c$ if we consider the analytical extension defined from the region $|z|>c$, and a branch cut for $|z|\geq c$ if we consider the analytical extension defined from the region $|z|<c$. We will encounter such a behaviour in the examples in Section \ref{SecEx}. We will see there, that the analytical extensions of $\delta^\pm$ defined from the regular region have indeed branch cuts, where the values on the two sides of the branch cut are given by $+\imath\epsilon_{L\pm}$ for $\phi=+\pi$ and  by $-\imath\epsilon_{L\mp}$ for $\phi=-\pi$.

\subsection{The complex Regge action}\label{CRA}

Using the definition of the complex dihedral angle as function of the complexified edge lengths (\ref{thetaTria}) we are now going to define the complex Regge action. We will then specify to the complexification induced by a global Wick rotation and determine the complex Regge action for the various values of the Wick rotation angle describing Euclidean or Lorentzian configurations.

To start we will shortly review the construction of the Regge action for four-dimensional triangulations.  Above we defined deficit angles for two--dimensional conical spaces, which arise from gluing two-dimensional wedges around a shared vertex.  This allows to define the deficit angle at vertices in two--dimensional triangulations. 

For higher--dimensional triangulations, the deficit angles are attached to the so-called hinges of the triangulation. In a $d$--dimensional triangulation the hinges are given by the $(d-2)$--simplices. Thus, in a four-dimensional triangulation the triangles are the hinges. Given a triangle $t$, we consider all $4$--simplices $\sigma$ sharing this triangle. Each $\sigma$ can be embedded isometrically into (flat) Euclidean or Minkowskian 4-space. Using the induced flat metric, we can project a given 4-simplex $\sigma$ onto the plane orthogonal to the triangle $t$. This defines a wedge $w_\sigma$, with the triangle $t$ being projected onto the tip of the wedge. We can now define a dihedral angle for each wedge $w_\sigma$ and define the Euclidean or Lorentzian deficit angle according to (\ref{DefE}) or (\ref{DefL}), respectively. 

In terms of the length square variables associated to a 4-simplex $\sigma$, the projection leads to the following (Euclidean, Lorentzian or star) inner products between the vectors $a$ and $b$ forming the triangle onto which the 4-simplex is projected (see Appendix \ref{AppProj}):
\ba\label{3.1}
a\cdot a \,=\, 3^2 \frac{\mathbb{V}_{\tau_a}}{\mathbb{V}_t} \, ,\q \q  b\cdot b \,=\, 3^2 \frac{\mathbb{V}_{\tau_b}}{\mathbb{V}_t} \, ,\q\q  a\cdot b\,=\, \frac{3^2\cdot 4^2\,}{\mathbb{V}_t}\frac{\partial \mathbb{V}_\sigma}{\partial s_{\bar{t}} } \q .
\ea
Here $\tau_a,\tau_b$ are the two tetrahedra in $\sigma$, which share the triangle $t$, and $s_{\bar{t}}$ is the squared length of the edge opposite the triangle $t$ in $\sigma$.  
The volume squares $\mathbb{V}_{X}$ of a $d$-simplex $X$ can be defined as polynomials in the lengths squares $s_e$, see Appendix \ref{AppProj}. This allows us to define the angle $\theta$ as a function of the complexified variables $s_e$.

Here we have of course, for triangulations with Euclidean signature, only Euclidean angles. For a Lorentzian signature triangulation we can obtain either Euclidean or Lorentzian angles:  if $t$ is space-like, the projection onto the space orthogonal to $t$ will result in a Lorentzian signature geometry, and if $t$ is time-like we will obtain a Euclidean signature geometry.  Null-triangles do not contribute to the Regge action.

The four-dimensional\footnote{This can be easily generalized to other dimensions: \textit{E.g.} for three-dimensional Regge calculus we replace the 4-volumes (4-simplices) with 3-volumes (tetrahedra) and the areas (triangles) with lengths (edges). For two-dimensional Regge calculus all vertices are taken as space-like with volume square equal to one. See also Appendix \ref{AppProj}  for the generalisation of (\ref{3.1}) to other dimensions.} Regge action (for a triangulation without boundary) for Euclidean and Lorentzian signature can then be defined as \cite{Hartle1985,Sorkin2019}
\ba
8\pi G \,\,S^{E}_{R} = -\sum_t |A_t| (\epsilon_E)_t      \,+\, \Lambda \sum_\sigma |V_\sigma| \,\, ,\q\q 
8\pi G S^{L\pm}_{R} = \sum_{t:t-like}  |A_t| (\epsilon_E)_t + \sum_{t:s-like}  |A_t| (\epsilon_{L\pm})_t \,-\, \Lambda \sum_\sigma |V_\sigma|
\ea

where $|A_t|$ denotes the (positive) area of a space-like or time-like triangle $t$ and $|V_\sigma|$ the (positive) volume of a four-simplex $\sigma$.  We defined the Euclidean Regge action to correspond to the one of Euclidean quantum gravity with amplitudes $\exp(-S^E_R)$. Note that the Lorentzian action does depend on the choice of either the $\epsilon_{L+}$  or the $\epsilon_{L-}$ version of the deficit angle.  These differ only for causally irregular configurations:   The Lorentzian amplitudes $\exp(iS^{L+}_R)$ for a positively oriented triangulation will lead to a suppression factor for trouser-like causally irregular configurations (where $N_c>4$) and an enhancement factor for yarmulke-like causally irregular configurations (where $N_c<4$).  We obtain the opposite behaviour, if we  choose the amplitudes 
 $\exp(iS^{L-}_R)$ instead.

We aim to extend the definition of the Regge action to complex squared lengths variables, such that it reproduces the Euclidean or Lorentzian Regge action, if the variables are real. But we have also seen, that the Euclidean and Lorentzian data lead exactly onto the branch cuts for the square roots and the logarithm, which appear in the definition of the angles and therefore the Regge action.  In Lemma \ref{lemma2}  we identified two consistent choices for these branch cut values. These arose by considering a complexification of the angle induced by a Wick rotation angle $\phi$: the $\theta^+$ choice for the branch cut values arose from reaching the branch cuts from the region $\phi \in (0,+\pi)$, whereas the $\theta^-$ choice arose from reaching the branch cuts from the region $\phi \in (-\pi,0)$. We generalized this construction to dihedral angles $\theta^\pm(s_a,s_b,s_c)$  as functins of complexified squared edge lengths in Definition \ref{Def1} and Equation \ref{thetaTria}.

Similarly, we can define two versions for the complex Regge exponent $W:=\tfrac{\imath}{\hbar} S$,
\ba\label{3.3}
\ell_{\rm P}^2\, W^\pm&=&      \sum_t \sqrt{\!\!{}_{{}_\pm}\mathbb{V}_t} \,(2\pi \pm\sum_{\sigma \supset t} \theta^\pm_{\sigma,t})    - \Lambda\sum_\sigma  \sqrt{\!\!{}_{{}_\pm}\mathbb{V}_\sigma}  \q ,
\ea
where $\ell_{\rm P}=\sqrt{8\pi \hbar G}$ is the Planck length.
The dihedral angles are defined in terms of the squared edge lengths of the projected triangles, which leads to 
\ba\label{3.4}
\theta^\pm_{\sigma,t} &=&
-\imath \log_\mp  \frac{
 \frac{4^2}{  \mathbb{V}_{t}}  \frac{\partial \mathbb{V}_\sigma}{\partial s_{\bar{t}}}    \mp\imath
\sqrt{\!\!{}_{{}_{\pm}} \,\,- \left( \frac{4^2}{  \mathbb{V}_{t}}  \frac{\partial \mathbb{V}_\sigma}{\partial s_{\bar{t}}} \right)^2+ \frac{\mathbb{V}_{\tau_a} }{   \mathbb{V}_{t}  }\frac{\mathbb{V}_{\tau_b}}{    \mathbb{V}_{t} }
} 
 } 
{ \sqrt{\!\!{}_{{}_\pm} \,\,\frac{\mathbb{V}_{\tau_a} }{   \mathbb{V}_{t}  } }   \sqrt{\!\!{}_{{}_\pm}\,\,\frac{\mathbb{V}_{\tau_b}}{    \mathbb{V}_{t} }} }     
\ea
where $\tau_a,\tau_b$ indicate the two tetrahedra sharing the triangle $t$ and $\bar{t}$ denotes the edge opposite $t$ in $\sigma$. 
For the sake of argument, which we will make below, we have used a representation of the complex angle, which is analytical around generic Euclidean data.  That is we just shifted the branch cut for the numerator of (\ref{thetaTria}) But one can also use a representation as in Equation (\ref{thetaTria}), as both representations agree in the region where 
\ba
4\mathbb{A}_{\rm proj}=  - \left( \frac{4^2}{  \mathbb{V}_{t}}  \frac{\partial \mathbb{V}_\sigma}{\partial s_{\bar{t}}} \right)^2+ \frac{\mathbb{V}_{\tau_a} }{   \mathbb{V}_{t}  }\frac{\mathbb{V}_{\tau_b}}{    \mathbb{V}_{t}}
\ea
has positive imaginary part (for $\theta^+$), respectively negative imaginary part (for $\theta^-$).

$W^+$, respectively $W^-$,  are defined in the regions for the complexified edge lengths where the corresponding branch cut values for the square roots and the logarithm arise, if one considers the limit where these variables do define  Lorentzian data. \textit{E.g.} for $\theta^+$ we demand that $\mathbb{A}_{\rm proj}$ has positive imaginary part, if the real part is negative. The same has to hold for $\mathbb{V}_{\tau_a}/ \mathbb{V}_{t}$ and $\mathbb{V}_{\tau_b}/\mathbb{V}_{t}$, whereas we require for the argument of the logarithm that the imaginary part is negative, if the real part if negative.

Note that this requirement informed our choice for $ \sqrt{\!\!{}_{{}_\pm}\mathbb{V}_t}$ and  $\sqrt{\!\!{}_{{}_\pm}\mathbb{V}_\sigma}$ for $W^\pm$. The reason is that $W^\pm$ does involve $\theta^\pm_{\sigma',t'}$, and thus terms like $ \sqrt{\!\!{}_{{}_\pm} \mathbb{V}_{\tau'_a}/\mathbb{V}_{t'}}$. Considering Lorentzian data satisfying the generalized triangle inequalities, we hit the branch cut of the latter square root in the case that $t'$ is space-like and $\tau'_a$ is time-like. (If $t'$ is time-like, then $\tau'_a$ has also to be time-like or null.) Now, if we consider a complexification induced by a global Wick rotation, we will have that the squared volumes of time-like simplices are approaching the negative real axis all from the same side in the complex plane. This suggest to use only $ \sqrt{\!\!{}_{{}_+}\mathbb{V}_X}$ for any simplices $X$ or to use only  $\sqrt{\!\!{}_{{}_-}\mathbb{V}_X}$. Otherwise one might not be able to find a region from which the required branch cut sides can be reached.

For the construction of the complexified Regge exponent $W^\pm$ we thus need to identify a region where $W^\pm$ is analytic and where the branch cut values as specified in (\ref{3.3}) and (\ref{3.4}) are approached from.
 We will do so explicitly for the example considered in section \ref{SecEx}.  

The two versions $W^\pm$ for the Regge exponent are defined from different regions of analyticity, but both regions include (generic) Euclidean data. In this case $\theta^+$ reproduces $-\psi_E$ and $\theta^-$ reproduces $+\psi_E$. From the definition (\ref{3.3}) of $W^\pm$ we can see that $W^+=W^-=-S^E_R$ for Euclidean data. Thus $W^\pm$ coincide on the intersection of their regions of analyticity, and thus lead to the same analytical extension. 

One also finds that $W^+$, if we approach Lorentzian data from its region of analyticity described above, reproduces $+\imath S^{L+}_R$ and $W^-$, if we approach Lorentzian data from its region of analyticity described above, reproduces $-\imath S^{L-}_R$. These are generally different, as the actions $W^\pm$ as defined in (\ref{3.3}) and (\ref{3.4}), and before any further analytical extension, have a branch cut for squared edge lengths describing generic Lorentzian data. But after analytical extension of $W^+$ from its region of analyticity defined above, and of $W^-$ from its region of analyticity, we do obtain equivalent Riemann surfaces and analytical extensions. One has on the other hand intersecting regions of analyticity for $W^+$ and $+\imath S^{L+}_R$,  as well as for $W^-$ and $-\imath S^{L-}_R$, showing that these also define equivalent analytical extensions, if defined from their respective intersections of regions of analyticity. 

Thus, although it seems we have two different Lorentzian actions $S^{L+}_R$ and $S^{L-}_R$, they do lead to equivalent analytical extensions, that also agree with the analytical extensions constructed from $W^\pm$.

We will illustrate this for the case that the complexification can be described by a global Wick rotation. 
As before  we assume a Wick rotation angle $\phi\in (-\pi,\pi]$ where $\phi=0$ describes an Euclidean signature triangulation and $\phi=\pm \pi$ describe a Lorentzian signature triangulation. We also assume that we can construct an analytical extension that enlarges the range of $\phi$ to $\phi\in (-2\pi,2\pi]$, with $\phi=2\pi$ describing an Euclidean signature triangulation.

We furthermore assume that the Wick rotation does affect the angles $\theta^{\pm}_{\sigma,t}$, for a triangle $t$ that is space-like in the Lorentzian signature triangulation, and therefore comes with Lorentzian dihedral angles, in the same way as the angles $\theta^\pm$ defined in section \ref{CDA}.  

If the triangle $t$ is time-like in the Lorentzian signature triangulation, and thus the associated dihedral angles Euclidean, we assume that $\theta^\pm_{\sigma,t}=\mp\psi_E$ for $\phi=0,\phi=2\pi$ and for $\phi=\pm\pi$.

 For the analytical  continuations of the square roots $\sqrt{\!\!{}_{{}_{\pm}} \mathbb V_X}$, for simplices $X$ that are time-like in Lorentzian signature we assume
 \ba
 \sqrt{\!\!{}_{{}_{\pm}} \mathbb V_X}\,=\, 
 \begin{cases} 
   -\imath |V_X|  \q \,\phi = -\pi  \, ,\\
 \,\,\,+ |V_X|   \q \phi= 0 \, ,\\
 +\imath |V_X|  \q \,\phi=\pi \, , \\
  \,\,\,- |V_X| \q \phi= 2\pi \, .\\
  \end{cases}
 \ea
Here $|V_X|$ denotes the (positive) Euclidean area or volume for $\phi=0,\phi=2\pi$ and the (positive) Lorentzian area or volume for $\phi=\pm\pi$.     
 
 For the  volume squares of simplices that are space-like in Lorentzian signature we assume $\sqrt{\!\!{}_{{}_{\pm}}   \mathbb V_X}=|V_X|$ for $\phi=0,\phi=2\pi$ and for $\phi=\pm \pi$.

From the results in section \ref{CEps} 
 we can then conclude, that for causally regular configurations we have
\ba\label{Wpm}
\hbar \,W^+\,=\,\hbar \, W^-=
\begin{cases}
-iS^{L-}_R \q \phi= -\pi  \,, \\
-S^E_R  \q \q\phi= 0 \,, \\
+iS^{L+}_R \q \phi= \pi  \,,\\
+S^E_R  \q\q \phi= 2\pi \,. \\
\end{cases} \q\q\q
\ea
Note that for causally regular data $S^{L-}=S^{L+}$. We thus find that $\phi=-\pi$ leads to minus the Regge action. This can be also identified as the action for triangulations with negative orientation.\footnote{For causally irregular configurations, using $-S^{L-}_R$ as action for negative orientation and $-S^{L+}_R$ for positive orientation, we keep trousers suppressed and yarmulkes enhanced for both orientations. We also keep the property that the sum over orientations yields a real partition function.}

We have argued in section \ref{CEps}, 
that we should expect a branch cut along the Lorentzian causally irregular data for the analytical extension defined from the causally regular data. We will confirm this for the examples in Section \ref{SecEx}.  These examples will also show that $W^+$ and $W^-$ do define equivalent analytical extensions, even throughout the branch cuts  caused by the irregular Lorentzian data.

\section{Local causality conditions}\label{CCond}

So far we referred to causal irregular triangulations as triangulations where the number of light cones at one or more space-like hinges is different from two. 

But there are additional local causality conditions that one can impose, and have been introduced  to define  three-dimensional Causal Dynamical Triangulations with a generalized slicing structure \cite{LollJordan}. Here we propose a generalization of these conditions \cite{LollJordan} to four dimensions. We exclude the case that we have null triangles or null edges.  In four dimensions we have three types of conditions:
\begin{itemize}
\item {\bf Triangle or hinge causality} demands that for each space-like triangle the sum of the wedges resulting from projecting out this triangle from the attached 4-simplices, includes exactly two light cones. This condition is equivalent to our notion of light cone or causal regularity. We refer to hinge causality violations, where the sum of wedges include more than two light cones as trouser-like hinge violations. Hinge causality violations, where the sum of wedges include less than two light cones are called yarmulke-like hinge violations. Note however, that hinge causality violations in dimensions larger than two, are not necessarily connected to spatial topology change, see the examples in Section \ref{SecEx}.
\item {\bf Edge causality} applies to space-like edges. For a given space-like edge $e$ we consider for each 4-simplex $\sigma$, which includes $e$, a cut through this simplex orthogonal to $e$, and attached to the midpoint of $e$.  This leads to a three-dimensional Lorentzian piecewise flat manifold $\Sigma$ with spherical boundary and only one bulk vertex. We then demand that there are exactly two light cones\footnote{See \cite{DittrichLoll} on how to construct null geodesics and thus light cones for Lorentzian piecewise flat manifolds.} attached to this bulk vertex. That is the intersection of the light cones with the spherical boundary of $\Sigma$ leads to two disconnected disks. 
\item {\bf Vertex causality} applies to all vertices in the four-dimensional triangulation. Here we consider the star of a given vertex $v$, that is the union of all the simplices containing the vertex $v$. We then demand that the intersection between the light cones attached to $v$ and the boundary of the star of $v$ leads to two disconnected three-dimensional balls. 
\end{itemize}

The interdependence of these conditions is to a large extent an open issue. We however note that one of the examples considered in Section \ref{SecEx} has a parameter range, where it satisfies hinge causality, but violates vertex causality at its one and only bulk vertex. The reason is that the intersection of the light cones emanating from this bulk vertex with the boundary of the star of this vertex include the full boundary, in the parameter range where the bulk edges are time-like.

This Lorentzian configuration, in the parameter range where hinge causality is satisfied, will have a real Regge action, that is, there are \textit{a priori} no imaginary terms. The deformation of the integration contour into the complex plane will however lead to a suppressing choice, in the sense that the saddle point, which is visited by the deformed contour, contributes an amplitude $\exp(+S^E_R)$ with $S^E_R$ being negative, and given by the Euclidean Regge action. 

For the parameter range where triangle  causality is violated, we will have an imaginary part for the Lorentzian Regge action. Here we will show that  the triangle causality violations lead to a branch cut, and the imaginary part on the two different sides of the branch cut differs in sign. The deformation of the integration contour will force a certain choice for the side of the branch cut, leading to  suppressed contributions for these configurations violating triangle causality.

This example, in which vertex causality is violated, will also present a case where we have a topology change for the space-like hypersurfaces during time evolution. We go from ``nothing", that is a spatial hypersurface with vanishing volume, to a 3-spherical hypersurface with finite volume. 

We will also consider an example where we evolve from a 3-spherical hypersurface of finite volume to a 3-spherical hypersurface of finite volume. Here we will have a parameter range where hinge causality is violated, but the evolution does not describe a topology change.

The conditions we discussed are local, and as discussed in \cite{LollJordan} might not exclude \textit{e.g.} the existence of closed time-like curves.

\section{Examples: A simplicial approximation to homogeneous and isotropic cosmology  }\label{SecEx}

\subsection{A simplicial model for the  de Sitter universe }\label{SSecA}

Here we consider two simple models, which describe one evolution step for a homogeneous and isotropic cosmology with a positive cosmological constant. We refer to \cite{DGS} for a more detailed description of these models and  its  dynamics.  

The first model approximates the early universe via a regular 4-polytope. The boundary represents a spatial spherical slice of the de Sitter universe, with a finite scale factor, whereas the centre of the 4-polytope represents the beginning of the universe, where the scale factor vanishes. The 4-polytope describes thus one discrete time step starting from the `big bang'. This model, with the 4-simplex chosen as 4-polytope, was also used in the context of Euclidean quantum gravity, for the construction of a simplicial no-boundary wave function \cite{Hartle1985}. 

Here we will be using the 600-cell, which gives the best approximation to the continuum description \cite{DGS}. The boundary of the 600-cell consists of 600 tetrahedra.  It furthermore contains 1200 triangles, 720 edges and 120 vertices. This 600-cell is subdivided into 4-simplices by placing a vertex in the centre and by connecting this vertex to all 120 vertices on the boundary of the 600-cell, thus obtaining 120 bulk edges inside a triangulation consisting of 600 4-simplices.

To approximate a homogeneous and isotropic universe we will set all length square variables for the edges in the boundary to be equal and given by $s_l$. We also set the length square variables for the bulk edges to be equal and given by $s_m$.  It will be more convenient to replace the variable $s_m$ with the square of the height $s_h$ of the 4-simplices, which is given by
\ba
s_h=s_m-\tfrac{3}{8}s_l        \q .
\ea
For positive $s_l$ the Lorentzian generalized triangle inequalities  then impose $s_h\leq 0$, whereas the Euclidean generalized triangle inequalities impose $s_h\geq 0$. 

The second model describes a discrete time evolution step from a spatial hypersurface, given by the boundary of the 600-cell, and finite volume, to the same kind of spatial hypersurface, with a finite (but generically different) volume. Here it is more convenient to use instead of 4-simplices four-dimensional frusta, whose basis is tetrahedral, see \cite{Collins1973,DGS}. The Regge action does easily generalize to this case.\footnote{An alternative construction consists of subdividing the frusta into simplices, and then to set all bulk deficit angles, which do not appear for the frusta to zero, see \cite{DGS}.} We set again all length squares $s_{l_1}$ in the first hypersurface equal, as well as all length squares $s_{l_2}$ in the second hypersurface. Likewise we set the length squares $s_m$ of all bulk edges equal.  We replace again the bulk lengths square with the square of the height $s_h$ in the frusta, which are given by
\ba
s_h=s_m-\tfrac{3}{8}(\sqrt{s_{l_2}}-\sqrt{s_{l_1}})^2 \q .
\ea
The Euclidean triangle inequalities are satisfied if $s_h$ is positive, and the Lorentzian triangle inequalities  \cite{Visser,EffSF3} are satisfied if $s_h$ is negative.

We will refer to the two models as ball model and shell model respectively. For both models, the Lorentzian data include 
 three different regimes, characterized by the (Euclidean or Lorentzian) signature of the bulk sub-simplices \cite{DGS}. We will state these regimes for the shell model; to obtain the cases for the ball model, one just needs to set $s_{l_1}=0,s_{l_1}=s_l$. Note that for the shell model the two-dimensional building blocks in the bulk are trapeziums, and the three-dimensional building blocks are 3-frusta, with a triangular base.
\begin{itemize}
\item[$(a)$]  The bulk trapeziums are time-like (and the bulk edges can be either time-like or space-like). This is the case for $s_h<-\tfrac{1}{8}(\sqrt{s_{l_2}}-\sqrt{s_{l_1}})^2$ and gives causally regular configurations. As the bulk trapeziums are time-like the deficit angles attached to these trapeziums are Euclidean. The extrinsic curvature angles attached to the boundary triangles are Lorentzian and include one light cone.
\item[$(b)$]  The bulk trapeziums are space-like, but the bulk 3-frusta are time-like. This is the case for $-\tfrac{1}{8}(\sqrt{s_{l_2}}-\sqrt{s_{l_1}})^2<s_h<-\tfrac{1}{24}(\sqrt{s_{l_2}}-\sqrt{s_{l_1}})^2$. This leads to configurations with irregular light cone structure. As the bulk trapeziums are now space-like the deficit angles attached to the bulk trapeziums are now Lorentzian. But they do not include any light cones. The extrinsic curvature angles attached to the boundary triangles are also Lorentzian and include one light cone.
\item[$(c)$]  Here the bulk frusta are space-like. This is the case for $-\tfrac{1}{24}(\sqrt{s_{l_2}}-\sqrt{s_{l_1}})^2<s_h$. This leads to configurations with irregular light cone structure. More specifically the deficit angles attached to the bulk trapeziums are Lorentzian and do include three light cones (instead of two light cones), and the extrinsic curvature angles attached to the boundary triangles include no light cone (instead of one light cone). { We remark that using the Sorkin prescription ($\theta^+$) these configurations would be enhanced overall in the path integral \cite{DGS}.}
\end{itemize}

For the shell model, neither of the three cases does describe a topology change of the spatial hypersurfaces: for all three cases there does exist a slicing with a continuous global time function, such that the constant time hypersurfaces are space-like and piecewise flat, and topologically equivalent to 3-spheres.

The ball model does represent an example of topology change, as one evolves from a hypersurface with vanishing volume to a hypersurface with finite volume. For the ball model all three cases violate vertex causality: \textit{e.g.} in the case that the bulk edges are time-like we do {\it not} have two light cones spreading out from the bulk vertex but only one light cone: the entire outer boundary is in the future of the inner bulk vertex.

We wish to compute the path integral over Lorentzian  data, for fixed boundary lengths $s_l>0$, respectively $s_{l_2},s_{l_1}$. As we have only $s_h$ as bulk variable, this integral is given by 
\ba
Z=\int_{s_h<Y}\mu(s_h)ds_h \exp( W(s_h))          \q ,
\ea
where $\mu(s_h)$ is a measure factor.  Here we choose a  measure suggested from the continuum,\footnote{See eg. \cite{TurokEtAl}, which uses $\mu({\cal N}) d{\cal N}\sim d{\cal N}/\sqrt{\cal N}$ where ${\cal N}=Na$ with $N$ the lapse in proper time gauge and $a$ the scale factor. We then identify $N\sim \sqrt{-s_h}$ and $a\sim \sqrt{s_l}$.} given by $\mu(s_h)={\frac12\sqrt\frac{3\pi\imath}2}s_l^{1/4} s_h^{-3/4}$ for the ball model and $\mu(s_h)=\frac12\sqrt\frac{3\pi\imath}2 s_{\bar{l}}^{1/4} s_h^{-3/4}$, where $s_{\bar{l}} =(\tfrac{1}{2}(\sqrt{s_{l_1}}+\sqrt{s_{l_2}}))^2$ for the shell model.
 
 We will consider two different values for $Y$:  if we include causally irregular data, that is integrate over the entire Lorentzian domain, we choose $Y=0$. If we exclude causally irregular configurations, we choose $Y=-\tfrac{1}{8}s_l$ or $Y=-\tfrac{1}{8}(\sqrt{s_{l_2}}-\sqrt{s_{l_1}})^2$ respectively.

 In order to possibly deform the integration contour into the complex plane, we consider $s_h$ as complex variable. Using a polar representation for $s_h$, that is $s_h=r_h \exp(\imath \phi)$ with $r_h>0$, we do obtain also a notion of Wick rotation, which acts, with the symmetry reduction introduced above, on the entire triangulation. Working with the height squared (as opposed to the bulk length square $s_m$), we have that $\phi=0$ leads to a viable Euclidean configuration for all  $r_h>0$, and that $\phi=\pm \pi$ leads to a viable Lorentzian triangulation.

For the ball model one finds that the action $S(s_h)=-\imath \hbar W(s_h)$ has, for sufficiently small $s_l<s_l^{\rm crit}$, no saddle points for Lorentzian data. It has however a saddle point  for Euclidean data, that is for $s_h>0$.  We will later see that the analytical continuation of $W$ leads to a double cover of the complex plane for $s_h$ and further analytical extensions related to  branch cuts for $W$ that appear along Lorentzian data describing causally irregular configurations.  This double cover includes  a further saddle point along the additional copy of the Euclidean data (which can be interpreted to correspond to negative orientation). The actions evaluated at these two saddle points differ by a global minus sign.

Here we have an important difference to the continuum case, \textit{e.g.} \cite{TurokEtAl,DGS}, where one finds a pair  of saddle points (instead of a single saddle point) along Euclidean data with plus the Euclidean action and another pair of saddle points along Euclidean data with minus the Euclidean action. These pairs arise because there are generically two trajectories on the 4-sphere (or Euclidean deSitter space), which connect two 3-spheres of given radii: one trajectory remains inside a half-sphere, the other crosses the equator. One discrete evolution step can model only the first type of trajectory, but not the second.

The critical value for $s_l^{\rm crit}$ below which we have saddle points in the Euclidean is given by \cite{DGS}
\ba\label{spc}
s_l^{\rm crit}=24 \sqrt{2}\frac{720\pi-3\cdot 600 \cos^{-1}(\tfrac{1}{3})}{600 \Lambda} \approx \frac{2.61}{\Lambda} \q .
\ea
But the action evaluated on this solution gives only a reasonable approximation to the continuum for $s_l<1.74/\Lambda$ \cite{DGS}. 
We will therefore restrict to this regime for $s_l$.

For $s_l>s_l^{\rm crit}$ one does have a saddle point for negative $s_h$ (and another one after the analytical extension, representing configurations with negative orientation). This is different from the continuum, where the saddle points for boundary conditions which include a vanishing scale factor, are either purely Euclidean or complex. We therefore see this behaviour  as a discretization artefact.

For the shell model we will consider sufficiently large boundary values $s_{l_1}, s_{l_2}$, with a small difference $s_{l_1} \approx s_{l_2}$, such that we have  saddle points for Lorentzian data and approximate the continuum evolution for a small finite time step. We will again have two saddle points, whose Regge exponent will differ by a global minus sign.

It will be convenient to introduce the following dimensionless versions of the cosmological constant and the length square 
\ba
\tilde \Lambda = \Lambda \ell_P^2 \q ,\q\q \tilde s \,=\, s\Lambda \,=\,  s \frac{\tilde \Lambda}{\ell_{\rm P}^2}        \q .
\ea
We then have \textit{e.g.} for the Euclidean Regge action
\ba
\frac{1}{\hbar} \,S^{E}_{R} \,=\,  \frac{1}{8\pi G \hbar} \left( -\sum_t  A_t (\epsilon_E)_t      \,+\, \Lambda \sum_\sigma  V_\sigma \right)\,=\,  \frac{1}{\tilde \Lambda} \left( -\sum_t \tilde A_t (\epsilon_E)_t      \,+\,  \sum_\sigma \tilde V_\sigma \right) \q ,
\ea
where $\tilde A_t$ and $\tilde V_\sigma$ denote the area and 4-volume expressed in units of $\Lambda^{-1}$ and $\Lambda^{-2}$ respectively. The deficit angles are dimensionless and thus invariant under rescaling of the lengths, and we defined $\ell_{\rm P}^2=8\pi G \hbar$.

Using these dimensionless variables, a change in the cosmological constant amounts to a rescaling of the action. In the following, we will use these variables  but drop the tilde for the geometric quantities.


\subsection{Analytical extension of the Regge exponent for the ball model}\label{ssec:ball_extension}

Here we will discuss the analytical extension of the Regge exponent for the ball model, and establish that branch cuts appear for configurations violating hinge causality. A very similar discussion applies to the shell model.

The Regge exponent $W^\pm(s_h)$ for the ball model is given by
\ba
 \tilde \Lambda W^\pm(s_h)&=& 720 \sqrt{\!\!{}_{{}_\pm} \mathbb{A}_{\rm blk}} \,\, \delta^\pm_{\rm blk} +1200 \sqrt{\!\!{}_{{}_\pm} \mathbb{A}_{\rm bdry}} \, \,\delta^\pm_{\rm bdry} -600 \sqrt{\!\!{}_{{}_\pm}\mathbb{V}_\sigma}    \q .
\ea
Here we denote by $\mathbb{A}_{\rm blk}$ and $\delta_{\rm blk}$ the area square and the complex deficit angle attached to the bulk triangles, and with  $\mathbb{A}_{\rm bdry}$ and $\delta_{\rm bdry}$  the area square and the complex extrinsic curvature angle\footnote{The extrinsic curvature angle is similarly defined to the deficit angle, just that the additive constant $2\pi$ is replaced by $\pi$.} attached to the boundary triangles. $\mathbb{V}_\sigma$ denotes the squared 4-volume of the simplices.  We specify all these functions in Appendix \ref{AppC}, and detail there the analytical extensions explicitly.

But to illustrate the main mechanism, consider the term
\ba
\sqrt{\!\!{}_{{}_\pm} \, \mathbb{A}_{\rm blk}} &=&\frac{\sqrt{s_l}}{4\sqrt{2}}\, \sqrt{\!\!{}_{{}_\pm} \,s_l+8 s_h} \q .
\ea
The square root $\sqrt{s_l+8 s_h}$ has branch cuts for $r_h:=|s_h| \geq \tfrac{ 1}{8}s_l$ and $\text{Arg}(s_h)=\phi= \pi$.  The analytical continuation from the region $r_h>\tfrac{ 1}{8}s_l$ and $\phi\in (-\pi,\pi)$ is given by
\ba\label{4.5A}
\sqrt{\!\!{}_{{}_\pm} \,s_l+8 s_h} &\rightarrow& R_a(r_h,\phi):=e^{\imath \phi/2} \sqrt{ s_l  e^{-\imath \phi}  + 8 r_h}  \q .
\ea
and extends $\phi$ to $\phi \in (-2\pi,2\pi)$. We take the coordinates $r_h>0$ and  $\phi \in (-2\pi,2\pi)$ as parameters of a Riemann sheet ${\cal S}_a$. The square root on the right hand side of (\ref{4.5A}) defines an analytic function  $R_a(r_h,\phi)$ in $\phi$ for $r_h>\tfrac{1}{8}s_l$, but has a branch cut for $r_h\leq \tfrac{1}{8}s_l$ and $\phi=\pm \pi$.

We can extend   $e^{\imath \phi/2} \sqrt{ s_l  e^{-\imath \phi}  + 8 r_h}$ through these branch cuts. Here we have \textit{a priori} four further possibilities to extend: for both branch cuts from each of their two sides. But we can form two pairs, and glue the members of each pair to one analytical patch. That is, the extension from the region $r_h<\tfrac{1}{8}s_l$ and $\phi\in (-\pi,\pi)$ is given by
\ba\label{4.5B}
\sqrt{\!\!{}_{{}_\pm} \,s_l+8 s_h} &\rightarrow&  R_b(r_h,\phi_b):= \sqrt{ s_l   + 8 r_h e^{\imath \phi_b} }  \q .
\ea
Here, we introduced a copy $\phi_b\in (-2\pi,2\pi]$ of the polar coordinate for $s_h$. (We extend to a range $\phi_b\in (-2\pi,2\pi]$ because that is the minimal range for which \textit{e.g.} the analytically extended deficit angles are periodic.) This leads to a further Riemann sheet ${\cal S}_b$ parametrized by $r_h<0$ and $\phi_b\in (-2\pi,2\pi]$. $R_b(r_h,\phi_b)$ is analytic on $S_b$ in the region $r_h<\tfrac{1}{8}s_l$, but has branch cuts for $r_h\geq \tfrac{1}{8}s_l$ at $\phi=\pm \pi$.  We have furthermore that $R_b(r_h,\phi_b)$ does coincide with $R(r_h,\phi)$, if we identify $\phi$ with $\phi_b$ and consider $\phi=\phi_b\in (-\pi,\pi)$. We can thus identify the Riemann sheet ${\cal S}_a$ with the Riemann sheet ${\cal S}_b$ for $\phi=\phi_b\in (-\pi,\pi)$. The analytical extensions are however different for $\phi,\phi_b \in (-2\pi,-\pi) \cup (\pi,2\pi]$ as $R_a$ has a branch cut for $r_h\leq \tfrac{1}{8}s_l$ and $\phi=\pm\pi$ and $R_b$ has a branch cut for $r_h\geq \tfrac{1}{8}s_l$ and $\phi=\pm\pi$.

$R_b$ defines an analytical extension of $R_a$ through the branch cuts of $R_a$ at $\phi=\pm \pi$, from the region $\phi \in (-\pi,\pi)$. We can also define an analytical extension $R_{b'}$ of $R_a$,  through the same branch cuts, but from the region $\phi \in (-2\pi,-\pi) \cup (\pi,2\pi]$. This is given by
\ba
R_{b'}(r_h,\phi_{b'}):= -\sqrt{ s_l   + 8 r_h e^{-\imath \phi_{b'} }}  \q .
\ea
Here we defined a third Riemann sheet ${\cal S}_{b'}$, parametrized by $r_h>0$ and $\phi_{b'}\in (-2\pi,2\pi]$. Now $R_{b'}(r_h,\phi_{b'})$ does coincide with $R(r_h,\phi)$, if we identify $\phi$ with $\phi_{b'}$ and consider $\phi=\phi_{b'}\in (-2\pi,-\pi) \cup (\pi,2\pi]$. We can therefore identify ${\cal S}_a$ with the Riemann sheet ${\cal S}_{b'}$ for $\phi=\phi_{b'}\in (-2\pi,-\pi) \cup (\pi,2\pi]$.

Note that the condition $r_h>\tfrac{1}{8} s_l$ defines causally regular configurations for Lorentzian signature, described by case $(a)$ in Section \ref{SSecA}. In case $(a)$ the bulk triangles are time-like.  We had furthermore case $(b)$, $\tfrac{1}{24} s_ l  > r_h>\tfrac{1}{8} s_l$, where the bulk triangles are space-like, but the bulk tetrahedra are time-like.  In case $(c)$, $r_h< \tfrac{1}{24}$ the bulk tetrahedra are also space-like. We have seen that the case $(b)$ leads to branch cuts for the analytical extension of the bulk triangle area (and also $W^\pm$) defined from case $(a)$. This holds also for case $(c)$ --- it leads to branch cuts for the analytical extensions of $W^\pm$ defined from case $(a)$ as well as for those defined from case $(b)$. 

The analytical extension of $W^\pm$ can therefore be defined on the following Riemann surface: We will have the main sheet ${\cal S}_a$ parametrized by $r_h>0$ and $\phi \in (-2\pi,2\pi]$.  On this main sheet we can define the analytical extension $W^\pm_a$, defined from the region $r_h>\tfrac{1}{8}s_l$. This analytical extension  $W^\pm_a$ has branch cuts for $r_h\leq\tfrac{1}{8}s_l$ at $\phi=\pm \pi$. 

$W^\pm_a$ can be further extended through these branch cuts. This will lead to a pair of Riemann sheets ${\cal S}_b$ and ${\cal S}_{b'}$, on which we define the analytical extensions $W^\pm_b$ and $W^\pm_{b'}$, respectively. Both these extensions are defined from regions where $\tfrac{1}{24}s_l <r_h<\tfrac{1}{8}s_l$. But $W^\pm_b$ is defined from $\phi=\phi_{b} \in (-\pi,\pi)$, whereas $W^\pm_{b'}$ is defined from $\phi=\phi_{b'}\in (-2\pi,-\pi) \cup (\pi,2\pi]$.  One also finds that $W^\pm_{b'}(\phi')=-W^\pm_{b}(\phi'-2\pi)$ for  all $\phi' \in (-2\pi,2\pi]$.  

\begin{figure}[ht!]
\begin{picture}(500,280)
\put(30,7){ \includegraphics[scale=0.26]{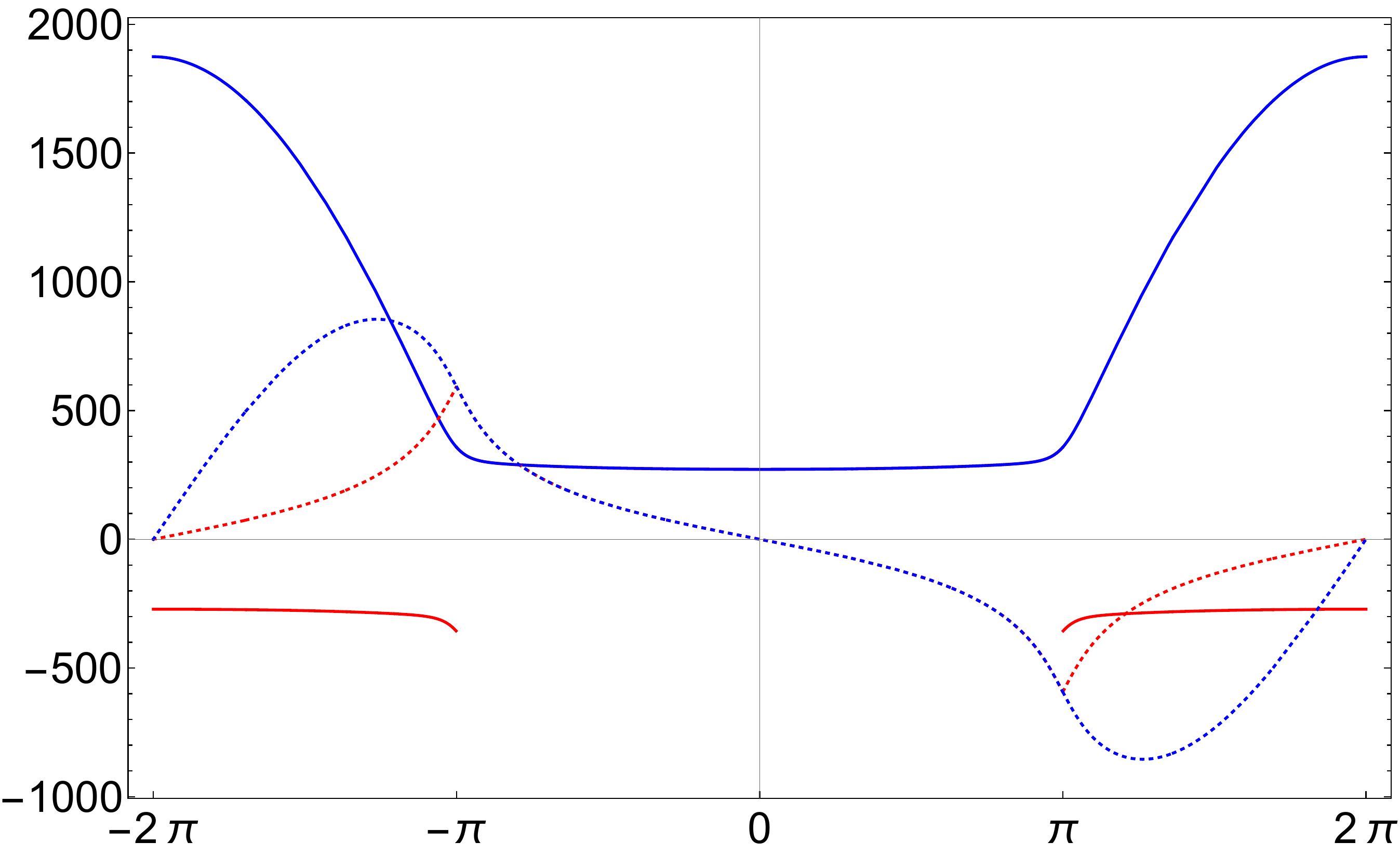} }
\put(280,7){ \includegraphics[scale=0.26]{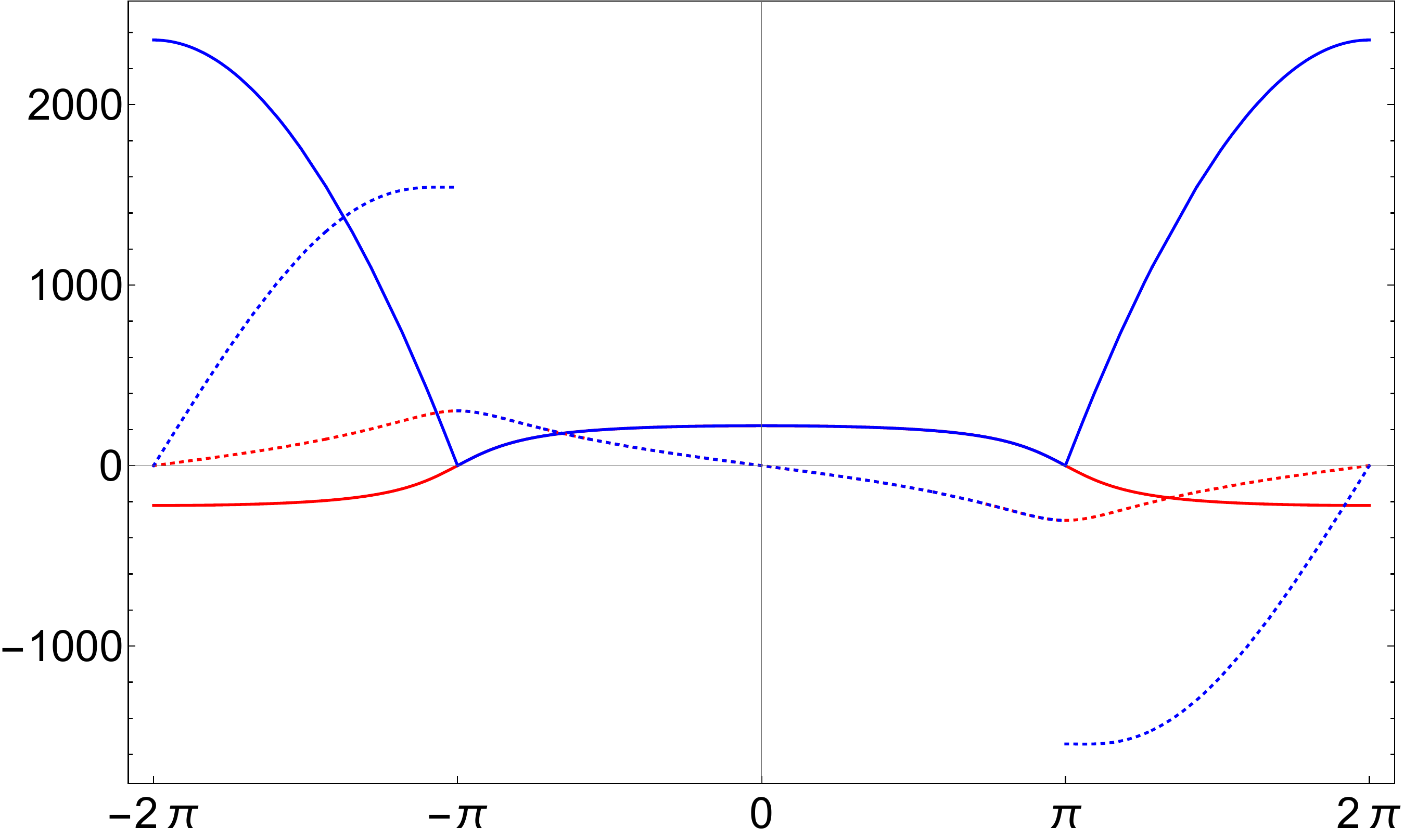} }

\put(30,145){ \includegraphics[scale=0.26]{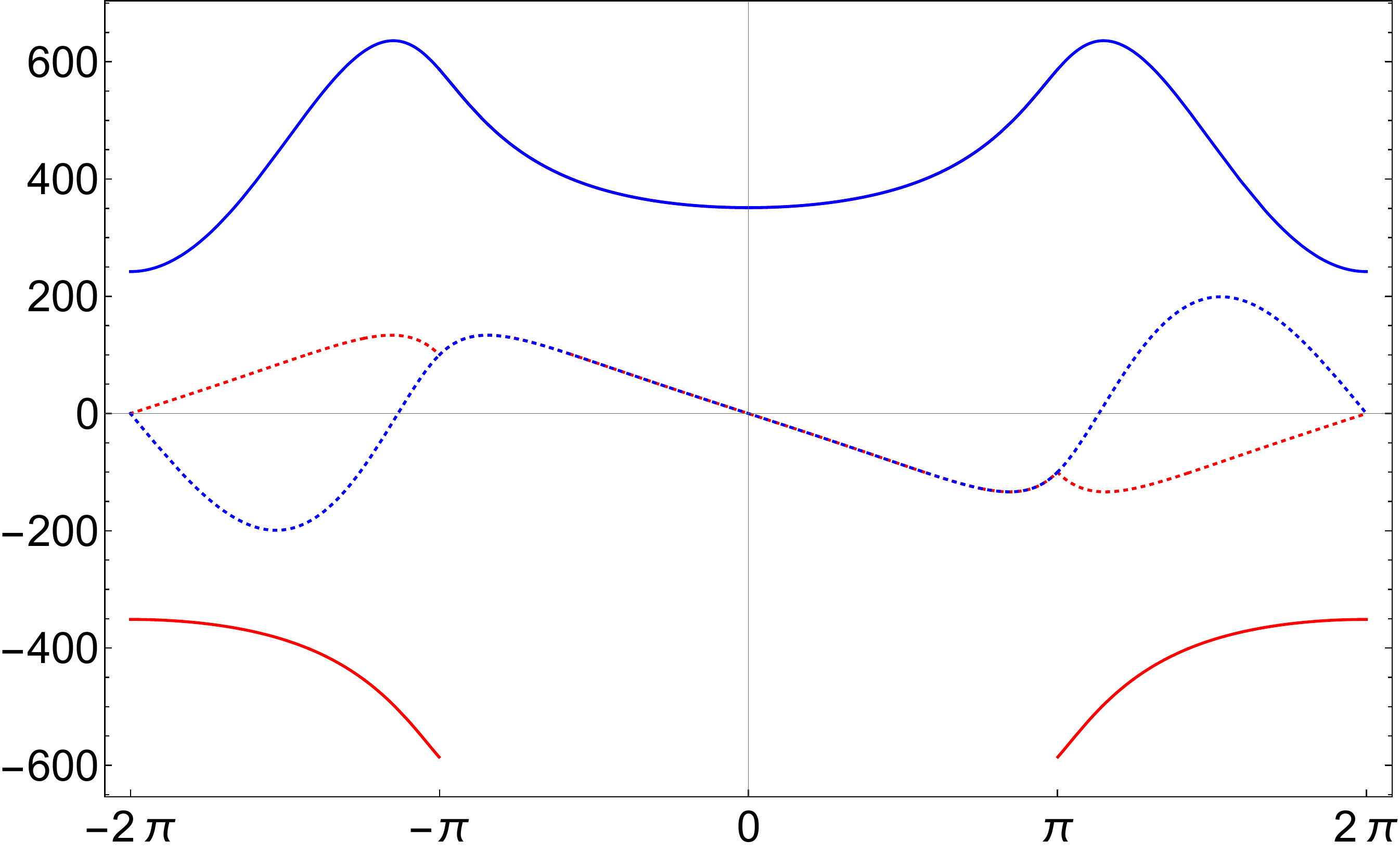} }
\put(280,145){ \includegraphics[scale=0.26]{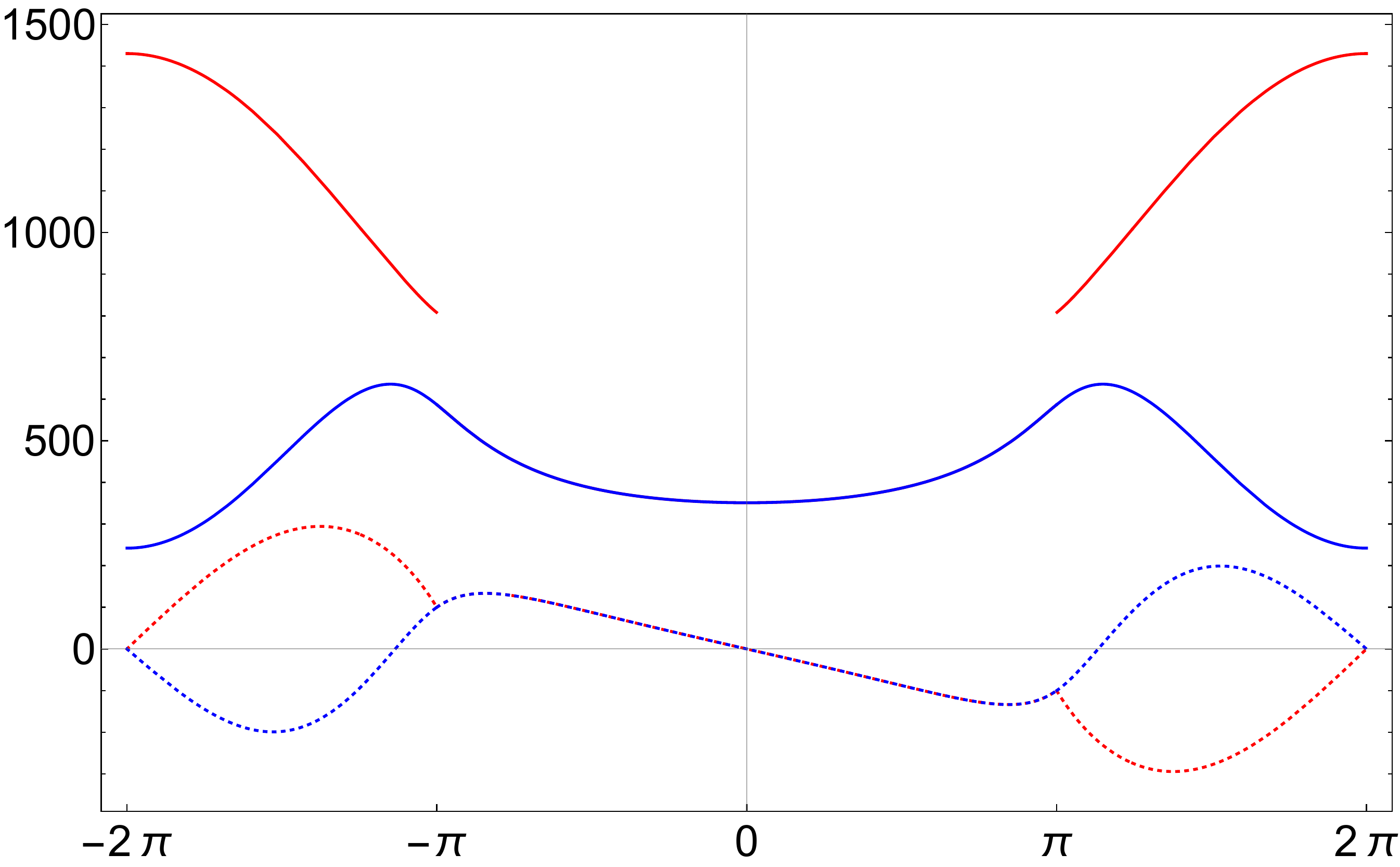} }

\put(130,0){$\phi$}
\put(400,0){$\phi$}

\put(20,115){\color{red} $ W_a $}
\put(20,100){\color{blue} $ W_b $}

\put(270,115){\color{red} $ W_a $}
\put(270,100){\color{blue} $ W_b $}

\put(20,245){\color{red} $ W_a $}
\put(20,230){\color{blue} $ W_c $}

\put(270,245){\color{red} $ W_b $}
\put(270,230){\color{blue} $ W_c $}

\end{picture}
\caption{ \label{WExtns}The various panels show  plots of the real part (solid curve) and imaginary part (dotted curve) of $W_a,W_b$ and $W_c$ as functions of $\phi$ for $\tilde\Lambda=1,s_l=1$ and for various values of $s_h$. The functions $W_a,W_b,W_c$ agree for $-\pi<\phi<\pi$.  Left and right top panel: For $s_h=0.03<s_l/24$ the function $W_c$ is continuous but $W_a$ and $W_b$ have branch cuts. Left bottom panel:  For $s_h=0.1<s_l/8$ the function $W_b$ is continuous but $W_a$ has branch cuts. Right bottom panel: For $s_h=0.2>s_l/8$ the function $W_a$ is continuous but $W_b$ has branch cuts. }
\end{figure}
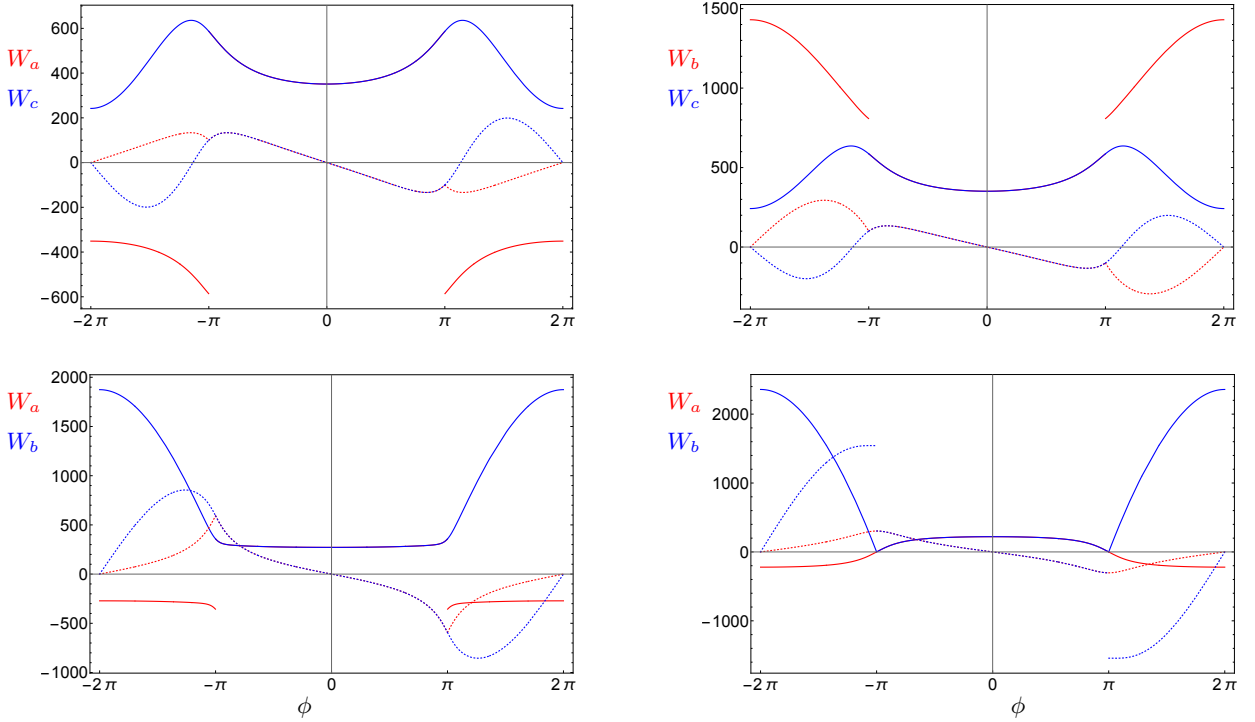

In the same way we can construct another pair of Riemann sheets ${\cal S}_c$ and ${\cal S}_{c'}$, with analytical extensions $W^\pm_c$ and $W^\pm_{c'}$, respectively. These are defined from the region $r_h<\tfrac{1}{24} s_l$, and from $\phi=\phi_{c} \in (-\pi,\pi)$, respectively 
$\phi=\phi_{c'}\in (-2\pi,-\pi) \cup (\pi,2\pi]$. Here we have also $W^\pm_{c'}(\phi')=-W^\pm_{c}(\phi'-2\pi)$ for  $\phi' \in (-2\pi,2\pi]$. See Figure \ref{WExtns} for plots of the various extensions.

As expected we find that the analytical extensions defined by $W^+$ and $W^-$ are equivalent, that is $W^+_x=W^-_x$ in all three cases $x=a,b,c$. 

The explicit construction of the analytic extensions in Appendix \ref{AppC}, shows that $W^\pm_a$ does evaluate for Lorentzian and Euclidean data as follows
\ba\label{Wa1}
\hbar W^+_a(\phi)=\hbar W^-_a(\phi)
 \longrightarrow
\begin{cases}
-iS^{L+}_R  \q \text{for}\,\,\,\,  {\phi\rightarrow -\pi\uparrow} \\
-iS^{L-}_R  \q \text{for}\,\,\,\,  {\phi\rightarrow -\pi\downarrow}\\
-S^E_R \q\,\,\,\,\, \text{for} \,\,\,  {\phi\rightarrow 0}\\
+iS^{L+}_R  \q \text{for}\,\,\,\,  {\phi\rightarrow +\pi\uparrow}\\
+iS^{L-}_R   \q \text{for}\,\,\,\,  {\phi\rightarrow +\pi\downarrow} \\
+S^E_R \q\,\,\,\,\, \text{for} \,\,\,  {\phi\rightarrow 2\pi}      \q .
\end{cases}            
\ea
This confirms the general discussion in section \ref{CRA} for the causally regular case. Formula  (\ref{Wa1}) makes explicit the branch cuts for irregular data where $S^{L+}_R \neq  S^{L-}_R$, and illustrates that the choice between $S^{L+}_R$ and $S^{L-}_R$ amounts to the choice for one of the sides of the branch cut.

The analytical extensions $W^\pm_{b}$  and $W^\pm_{c}$ coincide with $W^\pm_a$ for $\phi \in (-\pi,\pi)$ and therefore have the same limit values. But for $\phi=2\pi$ we have
\ba
\hbar W^\pm_b(2\pi)&=& +S^E_R\,+\, 4\pi \cdot 720|A_{\rm blk}|   \q\q\q\q\q \q\q\q\;\;\, \text{for}\, \, \tfrac{1}{24}<r_h< \tfrac{1}{8}\, ,\nn\\
\hbar W^\pm_{c}(2\pi)&=&+S^E_R-2\pi\cdot 720|A_{\rm blk}|+2\pi \cdot 1200 |A_{\rm bdry}|    \q \text{for}\,\,  r_h<\tfrac{1}{24} \, .
\ea
Correspondingly, we have for $W^\pm_{b'}$  and $W^\pm_{c'}$
\ba\label{dec4.14}
\hbar W^\pm_{b'}(0)&=& -S^E_R\,-\, 4\pi \cdot 720|A_{\rm blk}|     \q\q\q\q\q \q\q\q\;\;\, \text{for}\, \, \tfrac{1}{24}<r_h< \tfrac{1}{8}\, ,\nn\\
\hbar W^\pm_{c'}(0)&=&-S^E_R+2\pi\cdot 720|A_{\rm blk}|-2\pi \cdot 1200 |A_{\rm bdry}|   \q \text{for}\, \, r_h<\tfrac{1}{24} \, .
\ea

 This explains the dependence on the number of light ray crossings for the analytical continuation of the deficit angle to Euclidean data, which we found in Equation (\ref{2.25}). The analytical continuation goes across branch cuts associated to causally irregular data in the Lorentzian domain, which introduces a dependence on the number of light ray crossings appearing for the Lorentzian data.

Now we do have for $W^\pm_{b},W^\pm_{b'}$ branch cuts for $r_h\leq \tfrac{1}{24}s_l$ and $\phi=\pm \pi$ and for $r_h\geq \tfrac{1}{8}s_l$ and $\phi=\pm \pi$. For  $W^\pm_{c},W^\pm_{c'}$ we have branch cuts for $r_h\geq \tfrac{1}{24}s_l$ and $\phi=\pm \pi$.  

These lead to further analytical continuations. 
\textit{E.g.} starting from the region $\phi\in(-\pi,\pi)$ and circling either the branch point at $(r_h=s_l/8,\phi=\pi)$ or at $(r_h=s_l/24,\phi=\pi)$ or both, one finds expressions that differ from $W^\pm_{a}$  only by  adding terms of the form $2\pi n_{\text{blk}}$ and $2\pi n_{\text{bdry}}$  to the deficit angles. The values of $n_{\text{blk}},n_{\text{bdry}}\in (\mathbb{Z},\mathbb{Z})$ depend on the path along which one analytically continues and on whether one considers the bulk or boundary deficit angle, see Appendix \ref{AppC} for details.

We conjecture that this is also generally the case:  circling branch points leads to a change, that can be described by adding $2\pi n$, for some triangle dependent $n\in \mathbb{Z}$, to the deficit or extrinsic curvature angles.

Finally, let us comment on the existence of saddle points for the ball model. For $W^\pm_a$ and $s_l$ satisfying the condition $s_l<s_l^{\rm crit}$, see (\ref{spc}),  one finds two saddle points, one along $\phi=0$ and another one at $\phi=2\pi$. These saddle points describe therefore Euclidean data, and the Regge exponents $\hbar W^\pm$ evaluates to $-S^E_R$ for $\phi=0$ and $+S^E_R$ for $\phi=2\pi$ on these saddle points.

A numerical search did not identify any additional saddle points for $W^\pm_b$ and $W^\pm_c$ (and thus no saddle points for $W^\pm_{b'}$ and $W^\pm_{c'}$), apart from those in the region where these functions coincide with $W_a^\pm$.  This does not exclude the possibility that there are saddle points for the analytical continuations obtained by crossing the branch cuts of $W^\pm_b$ and $W^\pm_c$.

We furthermore note that the real parts of $W^\pm_a,W^\pm_b,W^\pm_c$ are  even functions of $\phi \in (-2\pi,2\pi)$, whereas the imaginary parts are  odd functions. The same holds for $W^\pm_x(\phi-2\pi),x=a,b,c$.  As we will discuss in the next section, the Lefschetz thimbles can be determined by using the vector field defined by (minus) the gradient of the real part of the Regge exponents. The flow of this vector field minimizes the real part of $W^\pm_a$ and leaves the imaginary part invariant. This flow will therefore have a reflection symmetry with respect to the $\phi=0$ axis and a reflection symmetry with respect to the $\phi=2\pi$ axis. Therefore the flow does not cross these  Euclidean axes.

~\\

\subsection{Deforming the integration contour to approximate the Lefschetz thimble}\label{Sec:PLIntegral}

{Here we will compute the integral, using a deformation of the contour towards a Lefschetz thimble \cite{PLTheory}.}

{Lefschetz thimbles are solutions of the initial value problem\footnote{ More generally one can consider the equation $-g^{ij}\partial_j\text{Re}[W]=\frac{\mathrm dz^i}{\mathrm dt}$, for some metric $g$ \cite{Witten1}, but we have set it to be given by the identity matrix in $(\text{Re}(z),\text{Im}(z))$ coordinates.}
\begin{equation}
    -\nabla\text{Re}\left[W(z=:x+\imath y)\right]=\frac{\mathrm dz}{\mathrm dt},\quad \lim_{t\to-\infty}=z_*,
    \label{eq:Lefschetz_thimble}
\end{equation}
associated to (complex) saddle points $z_*$ of the flow.\footnote{Since we are dealing with the real part of an analytic function, critical points of the action and of the flow are in one-to-one correspondence. Further, they are all saddles. }

The Lefschetz prescription is then {to complexify the integration variables and to} deform the original contour to one consisting of, or approximating,  a combination of Lefschetz thimbles. In particular, it can be shown that only Lefschetz thimbles for critical points whose \textit{anti-thimbles} (defined by \eqref{eq:Lefschetz_thimble} but with $+\infty$) cut the original integration domain, {contribute}. Since by construction Lefschetz thimbles have the property that the real part of the exponent decays as quickly as possible along them, while the imaginary part remains constant, integrals of {the form} $\sim e^W$ along them can become absolutely convergent, ease oscillatory behaviour for numerical simulations and be subject to Monte-Carlo simulations \cite{QCDReview,HanLefschetz,Ding2021}.

We now turn to applying said prescription to the ball and shell models.

\subsubsection{The ball model}

The flow \eqref{eq:Lefschetz_thimble} for the ball model's {Regge exponent} $W_a$, {defined in detail in Appendix \ref{AppC}},  is shown in {Figure} \ref{fig:ball_flow} in a $(r_h,\phi)$ coordinatization of the Riemann surface sheet. From it, we note the two Euclidean saddles forecasted in {Section} \ref{ssec:ball_extension}, one for $\phi=0\equiv4\pi$ and one for $\phi=2\pi$. The Lefschetz anti-thimble for the former is manifestly the parametrized line at $\phi=0\equiv4\pi$, whereas for the latter it is shown in black. Clearly, only this last anti-thimble cuts the original Lorentzian integration domain at $\phi=\pi$, independently of whether we include {or} exclude the irregular regions(s). Thus, we will (partially) deform the original integration contour into the Euclidean line at $\phi=2\pi$,  corresponding to the Vilenkin sign for the exponent (in opposition to the Hartle-Hawking sign) and in agreement with the continuum findings of \cite{TurokEtAl}. Along this line, the integrand is bell-shaped. Note that since the Riemann surface of $W$ covers multiple copies of the complex plane, {the final contour could in principle not be mappable onto the complex plane} and in any case, the Riemann surface portion in which the integrand is to be evaluated must be specified.
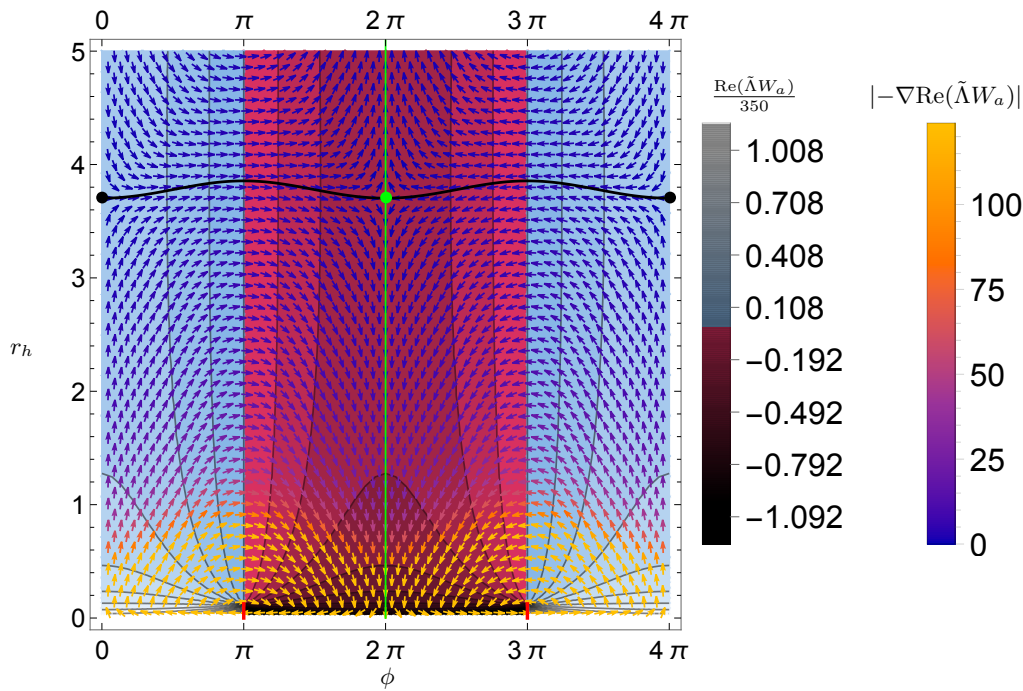
\begin{figure}[ht!]
\begin{picture}(500,270)
\put(80,7){ \includegraphics[height=250pt]{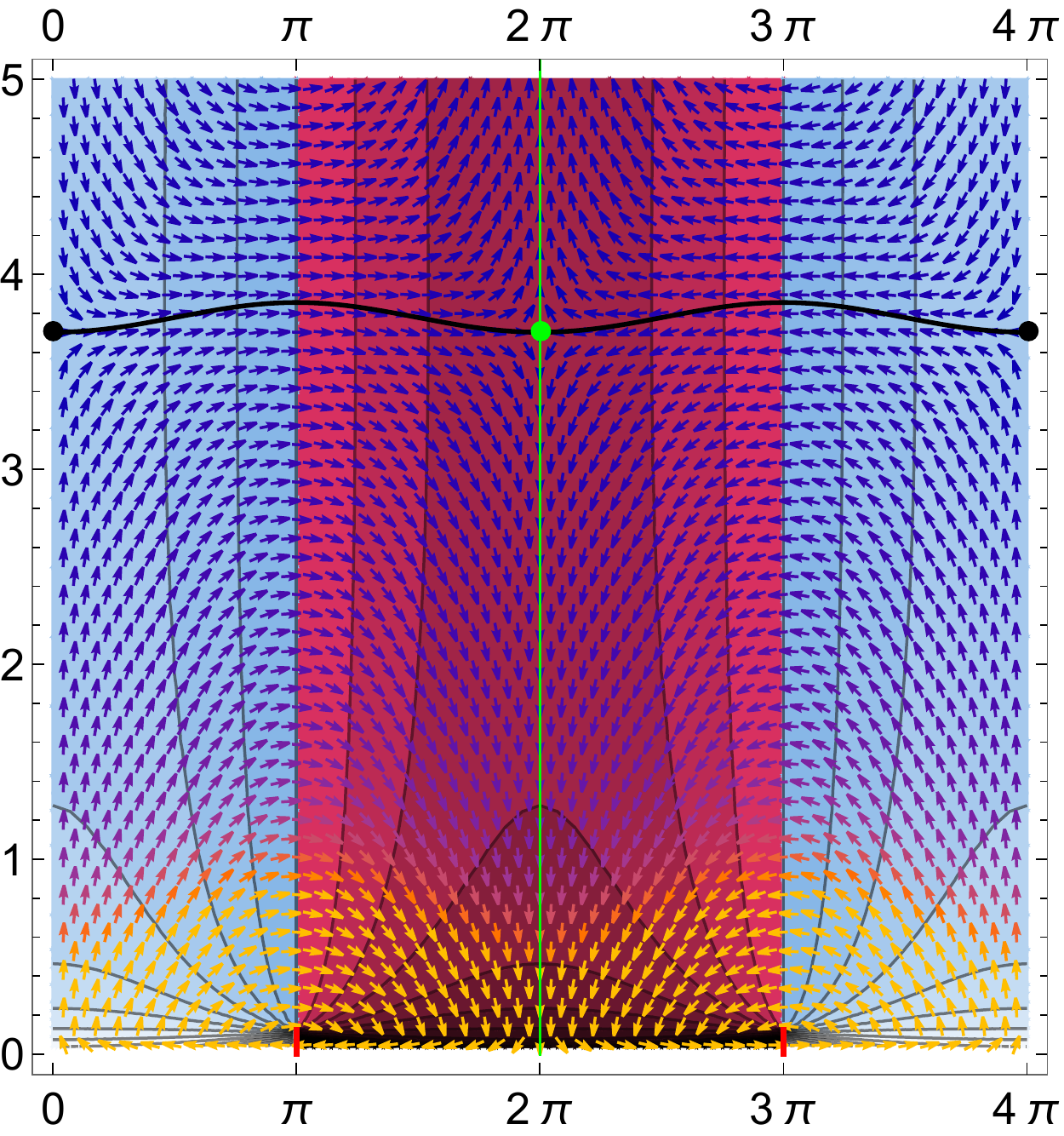} }

\put(315,36){ \includegraphics[height=180pt]{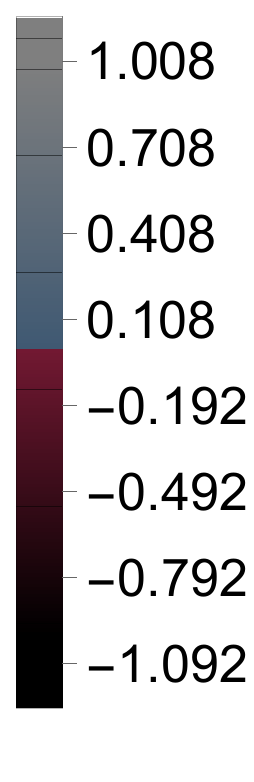} }
\put(400,36){ \includegraphics[height=180pt]{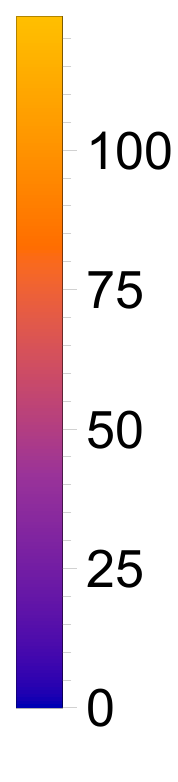} }

\put(200,0){$\phi$}

\put(60,125){$r_h$}

\put(325,220){${\frac{\text{Re}(\tilde\Lambda W_a)}{350}}$}

\put(385,220){$|{-}\nabla \text{Re}({\tilde\Lambda}W_a)|$}

\end{picture}
\caption{ Contour plot of the {(re-scaled)} real part of the Regge exponent (left legend) and of the {(re-scaled)} flow given by \eqref{eq:Lefschetz_thimble} in $(\phi,r_h)$ coordinates —magnitudes shown in the right legend. The exponent has two saddles, one in each Euclidean branch, the relevant for the (positive lapse) Lorentzian path integral is shown in green and its thimble and anti-thimble in green and black, respectively. For this plot $s_l$ has been set to one.
\label{fig:ball_flow}
}
\end{figure}

If we choose to exclude the irregular region from the integration, then we can take the deformed contour to be given by two parts (\textit{cf.} Fig. \ref{fig:ball_contours}):\footnote{In the following expressions we will abuse notation treating $s_h$ as a complex number, although in principle it lives on the Riemann surface of $W$. The ambiguity is fixed by the values of $\phi$ that we use to evaluate the integrand, \textit{e.g.} for the Euclidean portion of the deformed contour, we will use $\phi=2\pi$, \textit{cf.} eq. \eqref{eq:euclidean_line}.}
\begin{equation}
    s_h(\varphi)=-\frac{s_l}{16}+\left(\frac{s_l}{16}+\epsilon\right)e^{\imath\varphi},\quad\varphi\in[\pi,2\pi],\epsilon>0\quad\text{and}\quad \phi\in(x,x+2\pi)\supset[\pi,2\pi],x\in\mathbb R,
    \label{eq:medium_arc}
\end{equation}
and
\begin{equation}
    s_h(r_h)=r_h,\quad r_h\in[\epsilon,R]\quad\text{and}\quad\phi=2\pi.
    \label{eq:euclidean_line}
\end{equation}
And take the limits in which $\epsilon\to0$ and $R\to\infty$ to define the improper integral. 

The contour can then be closed with the Lorentzian domain without the irregular region by adding an arc
\begin{equation}
    s_h(\phi)=R e^{\imath\phi},\quad\phi\in[\pi,2\pi]
    \label{eq:large_arc}
\end{equation}
so that in the end, we have a closed contour that can be mapped into the complex plane, and is shown in Figure \ref{fig:ball_contours}. At each finite $R$ and $\epsilon$, the integral along said contour is subject to Cauchy's theorem and therefore, the Lorentzian integral will be given by the limiting contributions of the contours $\eqref{eq:medium_arc}$ and $\eqref{eq:euclidean_line}$ provided that the contribution of \eqref{eq:large_arc} goes to zero as $R\to\infty$, which is indeed the case, as can be shown by expanding $W_a$ for large $R$, performing the integral and then taking the limit.
\begin{figure}[ht!]
\begin{center}
\includegraphics[height=4.5cm,angle=-0]{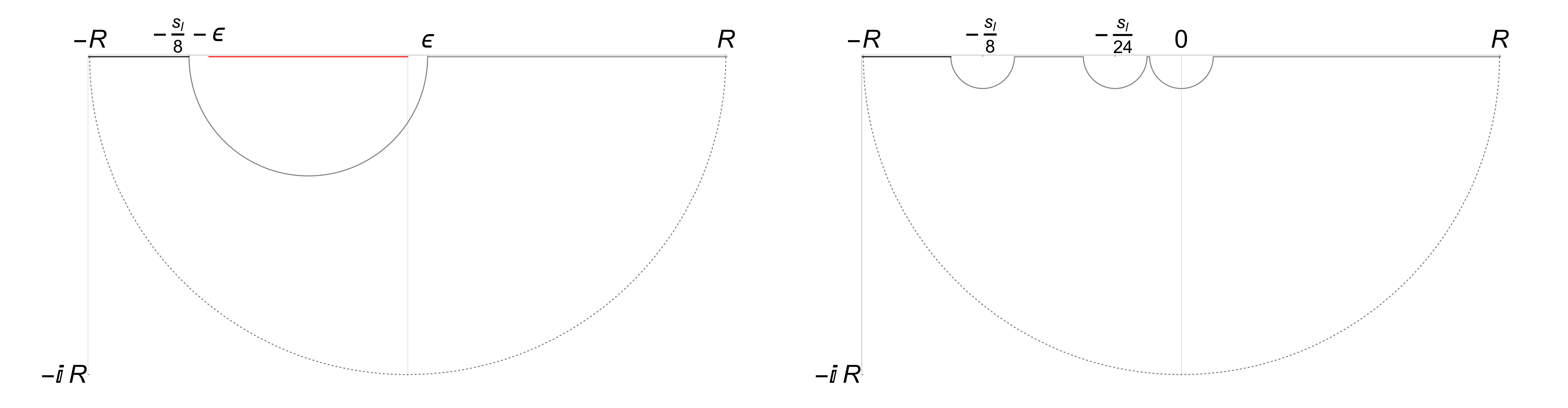}
\caption{ Contours in the complex $s_h$-plane used for the ball model path integral: in the left when excluding hinge causality violations by going around the branch cut (red) and in the right when including them, only going around the branch points. The final integral is taken as $R\to\infty$, so that the contribution of the dashed arc goes to zero.}
\label{fig:ball_contours}
\end{center}
\end{figure}

Therefore, for the ball model with the irregular region excluded, we have that
\begin{equation}
    \small
    Z_\text{Ball}=Z_1+Z_2:=\lim_{\epsilon\to0}\frac{s_l^{1/4}}2\sqrt{\frac{3\pi\imath}2}\left(\int_{r_h=\epsilon}^{r_h\to\infty}\mathrm d s_h(-s_h)^{-3/4} e^{W(s_h)}\Biggr|_{s_h=(r_h,2\pi)}+\int_{\phi=\pi}^{\phi=2\pi}\mathrm d s_h (-s_h)^{-3/4}e^{W(s_h)}\Biggr|_{s_h=-\frac{sl}{16}+\left(\frac{sl}{16}+\epsilon\right)e^{i\phi}}\right),
    \label{eq:regular_path_integral}
\end{equation}
where the square and fourth roots are to be taken so that arguments $\phi$ lay in $[\pi,2\pi]$.

The two contributions of $\eqref{eq:regular_path_integral}$ are shown in \ref{fig:ball_SOH} as a function of $s_l$ and we appreciate a sharp decay as $s_l$ increases. We also note that the thimble contribution $Z_1$ goes from dominating to sub-leading as $s_l$ increases  {(\textit{cf.} Figure \ref{fig:ball_SOH_abss})}.
\begin{figure}
    \centering
    \begin{subfigure}[t]{0.45\textwidth}
        \centering
        \includegraphics[width=\linewidth]{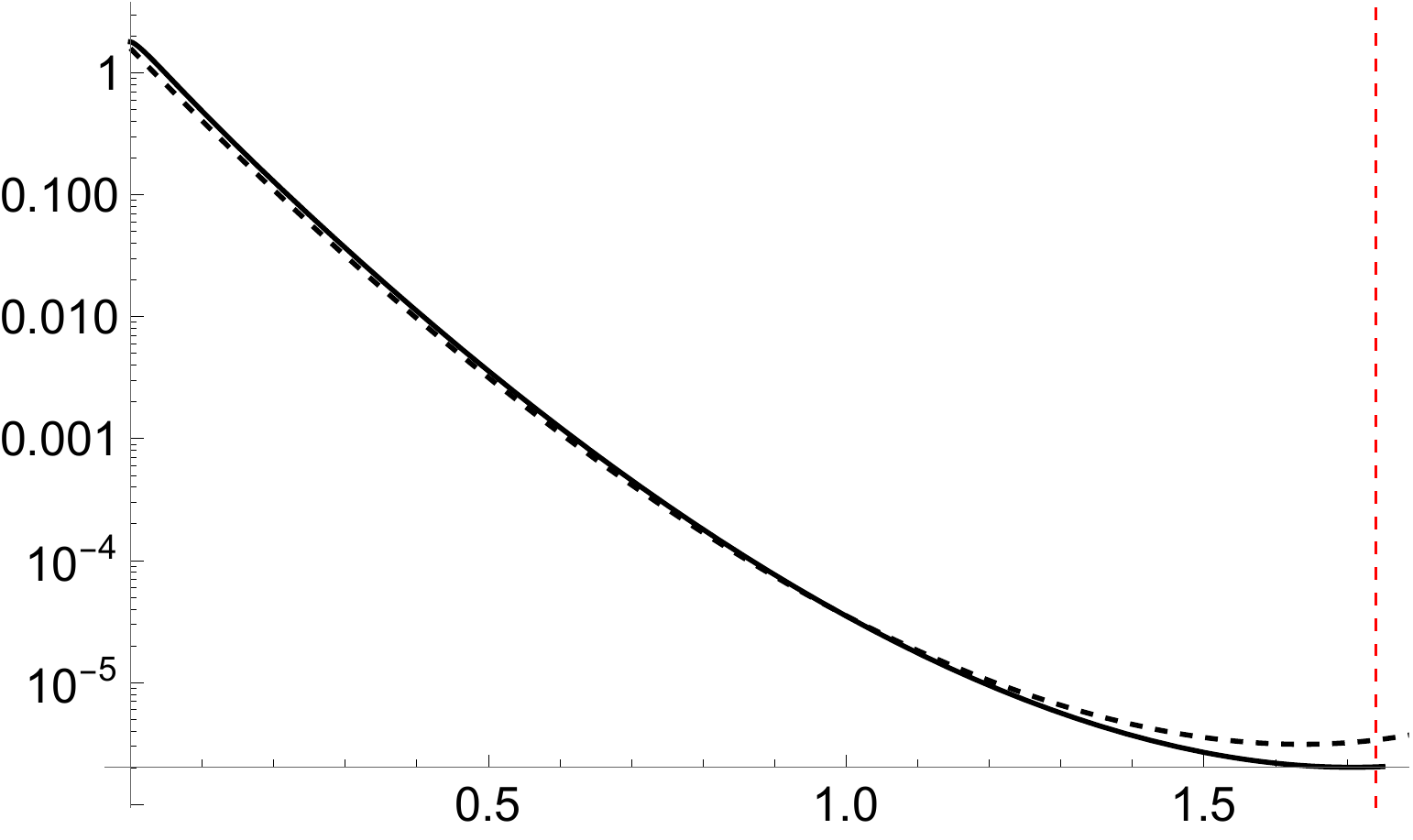} 
        \put(-220,140){$Z_1,Z_\text{Continuum}'$}
        \put(3,10){$s_l$}
        \caption{ In solid: the logarithmic graphic of the thimble contribution to $Z_\text{Ball}$, \textit{i.e.} $Z_1$. When including the causally irregular region, this is the only contribution. Note that it is is real. In gray-dashed: the logarithmic graphic of $Z_\text{Continuum}'$.}
        \label{fig:ball_euclidean_SOHs}
    \end{subfigure}
    \hfill
    \begin{subfigure}[t]{0.45\textwidth}
        \centering
        \includegraphics[width=\linewidth]{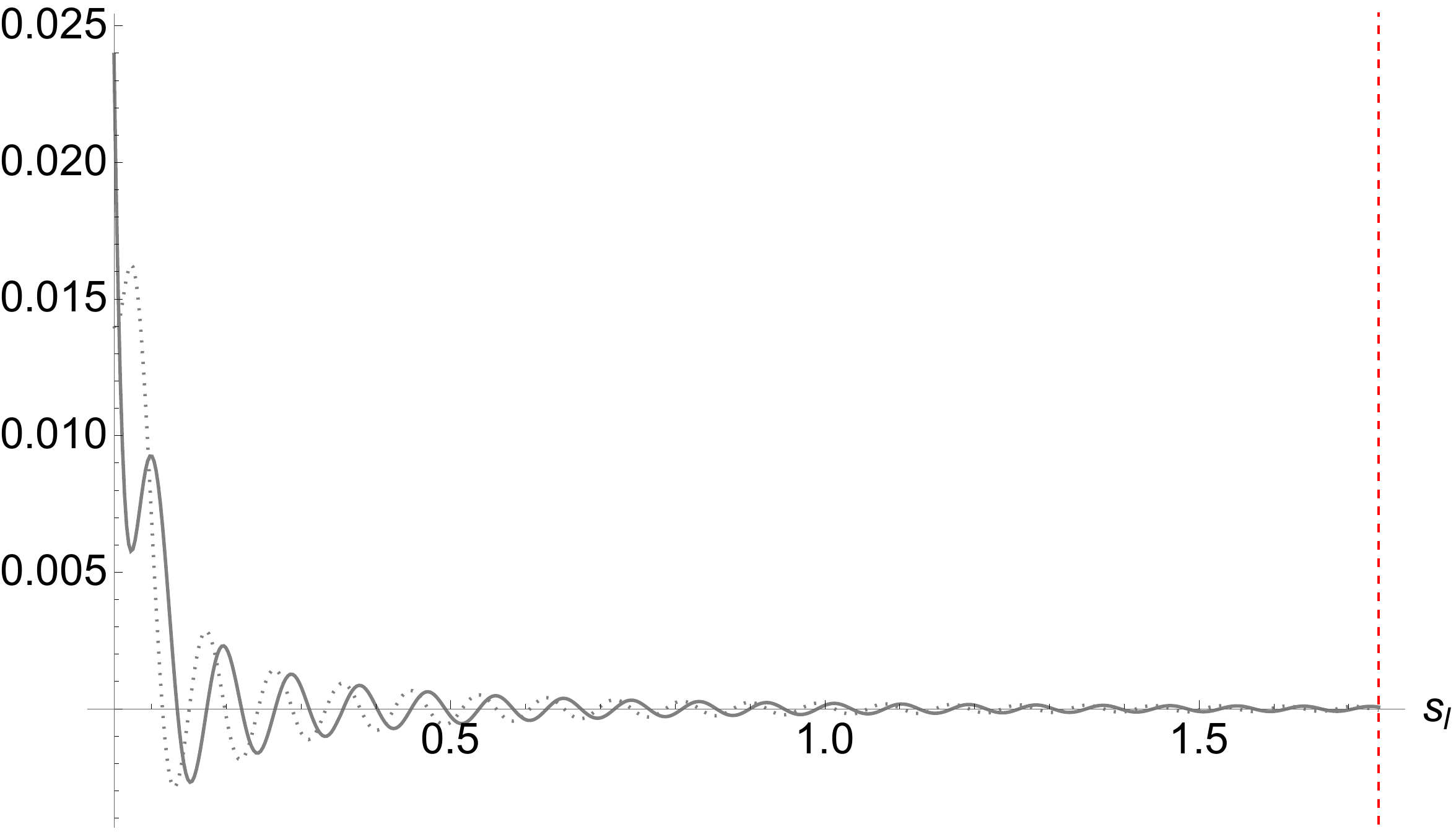} 
         \put(-220,140){$Z_2$}
        \caption{Contribution to $Z_\text{Ball}$ from the arc around the irregular region, as a function of $s_l$. The real part is solid and the imaginary part dashed.}
    \end{subfigure}

    \vspace{1cm}
    \begin{subfigure}[t]{\textwidth}
    \centering
        \includegraphics[width=0.45\linewidth]{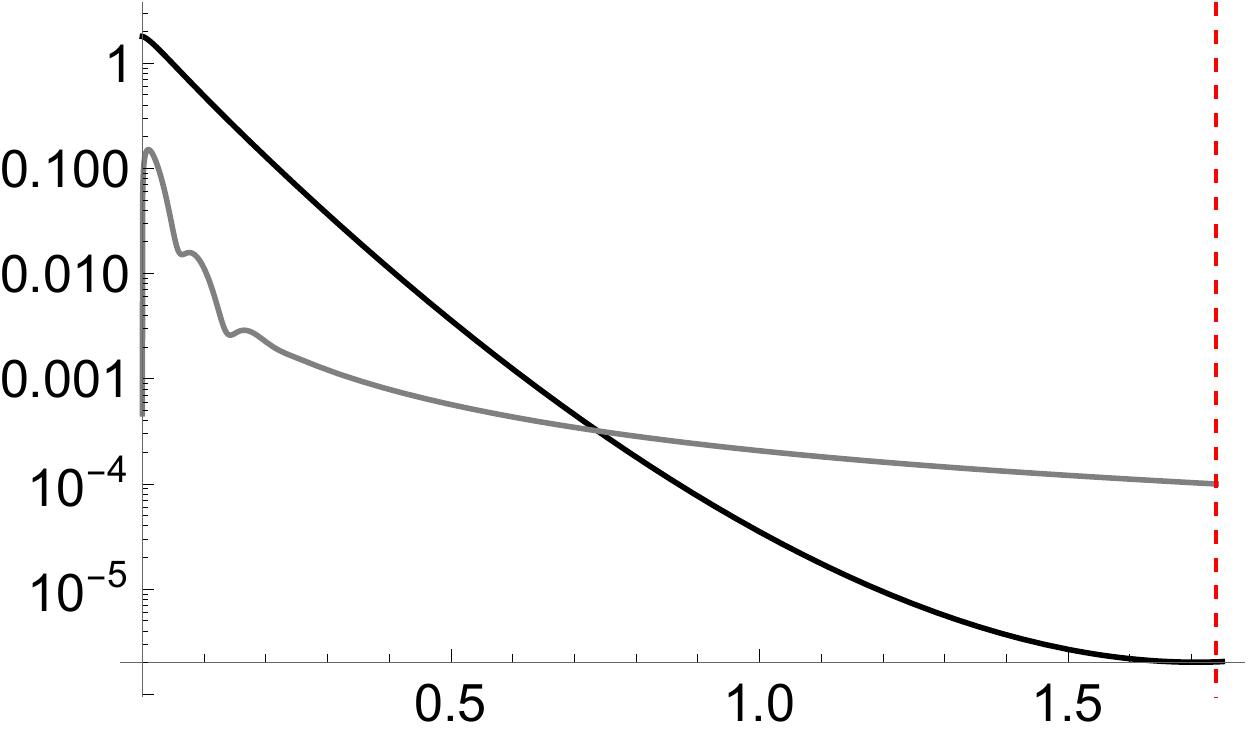} 
        \put(3,10){$s_l$}
        \put(-220,144){$|Z_1|,|Z_2|$}
        \caption{ Logarithmic plot of the absolute values of the two contributions in \eqref{eq:regular_path_integral} as functions of $s_l$. The thimble contribution is shown in black and the arc's in gray.} \label{fig:ball_SOH_abss}
    \end{subfigure}
    \caption{ Path integral numerical results for the ball model. For these plots we have set $\tilde\Lambda= 10$.  The red-dashed vertical line denotes the point from which the discretization loses validity, \textit{i.e.} $s_l=1.74$.}
    \label{fig:ball_SOH}
\end{figure}

We can now turn to the problem of computing the Lorentzian integral including causally irregular data. This presents more subtleties. Firstly, one has decide on the sheet of the Riemann surface that will be associated to the causally irregular region and secondly, go around the branch-points at $s_h=0,s_l/24,s_l/8$ with \textit{e.g.} semi-circles of radii $\epsilon$ going to zero. If one is to apply Cauchy's theorem as above, then we must close all arcs below the real line: closing some above and some below would result in a not-null-homotopic curve, while closing all above is impossible since the arc around zero cannot reach $\phi=2\pi$, as it would go into the sheet of $W_c$ as $\phi\to-\pi\equiv3\pi$ from above.

Importantly, the fact that we must close below implies that the Riemann sheet associated to the causally irregular {region} is determined by the Lefschetz prescription, which knows about the dynamics of the theory. Namely for this case it corresponds to the one gotten from approaching $\pi$ from above and thus, notably, the {(overall \cite{DGS})} suppressing sheet.

Therefore, at finite $\epsilon$ an $R$ we have a contour like the one shown in Figure \ref{fig:ball_contours}. After analysing the picture, the reader might wonder whether for $Z_\text{Ball}$ one should subtract the contribution of the arcs around the branch-points, \textit{i.e.} whether it is defined as a principal value of an integral over the Lorentzian line $\phi=\pi$, or not. It can be shown that the contributions from the arcs around the branch-points are zero, {(\black note  that these branch points are not actually poles of the integrand)}. And therefore, the Lorentzian integral is simply given by the first term of \eqref{eq:regular_path_integral}.  The result is therefore the  solid one in Figure \ref{fig:ball_euclidean_SOHs}.

Thus, we see that the Lefschetz prescription, in addition of offering a tool to compute highly oscillatory integrals (which in this case were simply computed by a Gaussian-like integral and at most a complex integral over a finite-volume domain in addition) can also instruct us on how to analytically continue the Regge action in regions that violate {hinge} causality.

In order to check the quality of these results one would like to compare with the continuum results, where it can be shown that\footnote{This action arises from integrating out the scale factor $a(t)$ from the continuum action. But as one ignores factor ordering issues in this procedure, it should be rather seen as an approximation \cite{TurokEtAl}.} \cite{TurokEtAl}
\ba
Z_\text{Continuum}=\sqrt\frac{3\pi\imath}2\int_0^\infty\frac{\mathrm d \mathcal{N}}{\sqrt\mathcal{N}}e^{2\pi^2\imath S_0},
\label{eq:Z_Continuum_Ball}
\ea
with
\ba
8\pi G S_0=\mathcal{N}^3\frac{\Lambda^2}{36}+\mathcal{N}\left(-\frac\Lambda2a_1^2+3k\right)-\frac1{\mathcal{N}}\frac34a_1^4,
\ea
where the spatial curvature $k=1$, $a_1$ is the `final' scale factor, the `no-boundary boundary condition' $a_0=0$ has already been applied {and $\mathcal{N}=N a$ is the lapse multiplied by the scale factor}.

For $a_1$ below a critical value $a_1^\text{crit}(\Lambda)$ (analogue to $s_l^\text{crit}$) $S_0$ has four critical points \cite{DGS}, two for each branch of Euclidean data, corresponding to the two pairs of saddles mentioned in \ref{SSecA}. However, as explained there, the triangulation used only captures one critical point (for a given branch of Euclidean data). Therefore, in order to benchmark we modify $S_0$ so that it also has only one critical point. The way we do so is by neglecting the $\mathcal{N}^3$ term, which also results in a behaviour better resembled by that of $W_a$ in both the Lorentzian and Euclidean branches. That is, we will consider
\ba
 \frac{\tilde\Lambda}{\hbar} S_0'=\tilde{\mathcal{N}}\left(-\frac12\tilde{a}_1^2+3\right)-\frac1{\tilde{\mathcal{N}}}\frac34\tilde{a}_1^4,
\ea
to compute $Z_\text{Continuum}'$ using \eqref{eq:Z_Continuum_Ball}. {Here we introduced the dimensionless variables $\tilde{a}_1=a_1 \sqrt{\tilde\Lambda}$ and $\tilde{\mathcal{N}}=\mathcal{N}\tilde\Lambda$.}

We can now use the Lefschetz prescription for this integral. In this continuum case, the integration contour can be deformed at once to the thimble, which corresponds to simply the Euclidean branch given by positive-imaginary $\tilde{\mathcal N}$, giving again the Vilenkin sign and a Gaussian-like integral. In this way, we can easily compute $Z_\text{continuum}'$ as a function of $a_1$. The fact that the final integration contour is simply this Euclidean branch already suggests that there could be better agreement with the continuum by including causally irregular regions in the discrete path integral, this is then supported by the facts that the continuum path integral is purely real  and non-oscillatory.

To relate the scale factor $a_1$ with the edge length square $s_l$ we use the following identification \cite{DGS}
\ba
\tilde{a}_1^2=\left(\frac12\left(\frac1{0.654}+\frac1{0.641}\right)\right)^2s_l \, ,
\ea
which is obtained as an average of equating the three-volume and the three-curvature of the 600-cell's boundary, with edge length square $s_l$, and a 3-sphere with radius $\tilde{a}_1$.
This allows us to obtain $Z_\text{Continuum}'$ as a function of $s_l$ and finally compare with the discrete path integral. This is also done in Figure \ref{fig:ball_euclidean_SOHs}, where we note that the discrete path integral does indeed approximate the (adapted) continuum result reasonably, specially for $ s_l\ll 1.74$. {(We remind the reader that we restrict our consideration to $ s_l<1.74$, as the discrete model approximates the continuum only up to this point, \textit{cf.} \ref{SSecA}.)} Notably, we see that including the causally irregular region in the integration seems to lead to a better approximation of the continuum  (\textit{cf.} Figures \ref{fig:ball_euclidean_SOHs} and \ref{fig:ball_SOH_abss}).

{Let us conclude by remarking that if one insists on choosing the other `side' (or sheet) of the branch-cut to be the one integrated over, then such contribution would totally dominate in the path integral, as its absolute value ranges from $\approx 3$ to $\mathcal O(10^{36})$, increasing with $s_l$ (\textit{cf.} fig. \ref{fig:ball_SOH}) —we note that $Z_1(s_l=0)\approx 2$. For example, at $s_l\approx 0.3$, the absolute value of the contribution from the irregular region is $\approx 10^6$}. This is therefore not a viable option.

\subsubsection{The shell model}
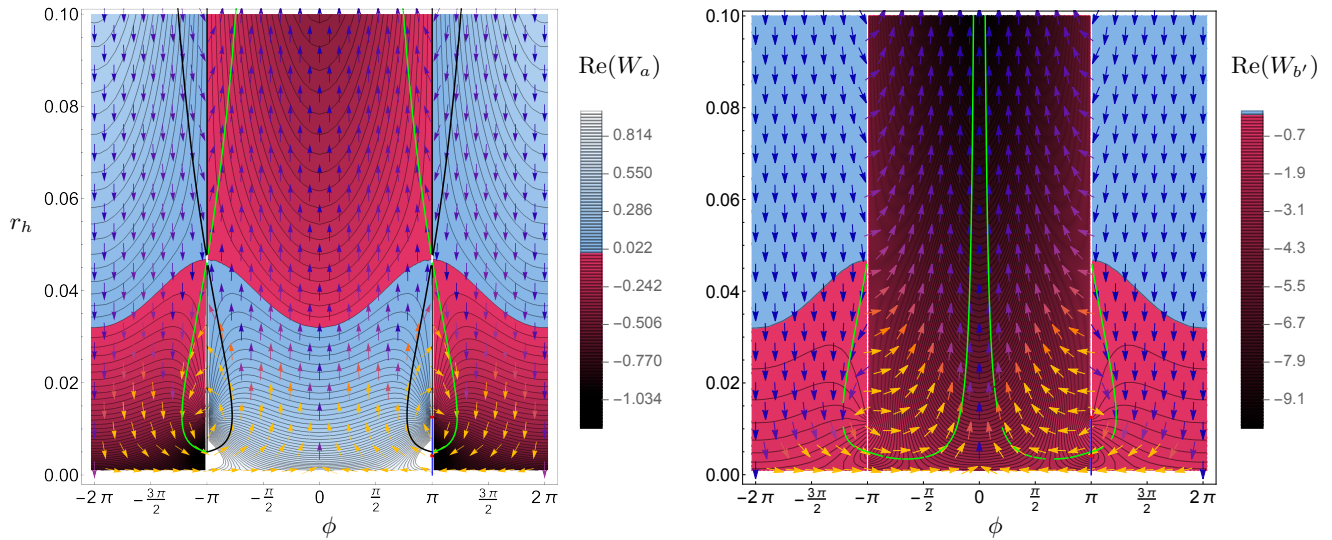
\begin{figure}[ht!]
\begin{picture}(500,190)
\put(20,7){ \includegraphics[scale=0.281]{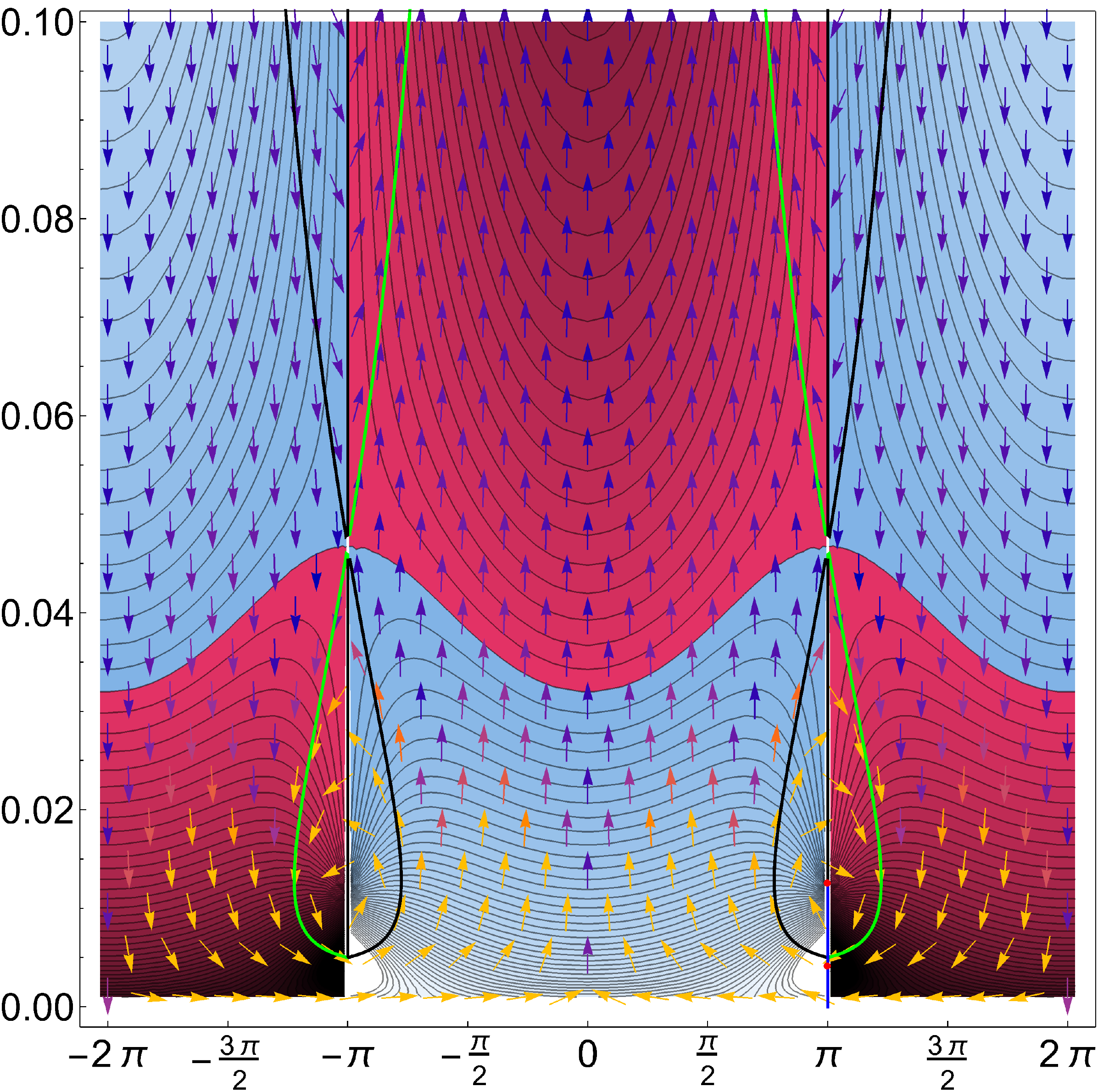} }
\put(220,27){ \includegraphics[scale=0.6]{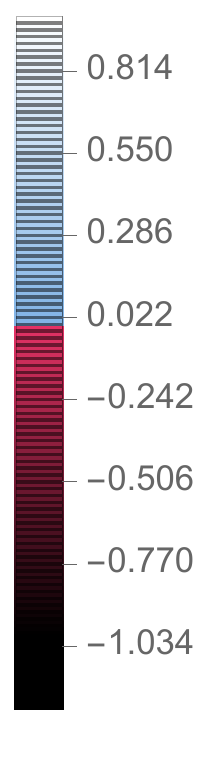} }

\put(270,7){ \includegraphics[scale=0.28]{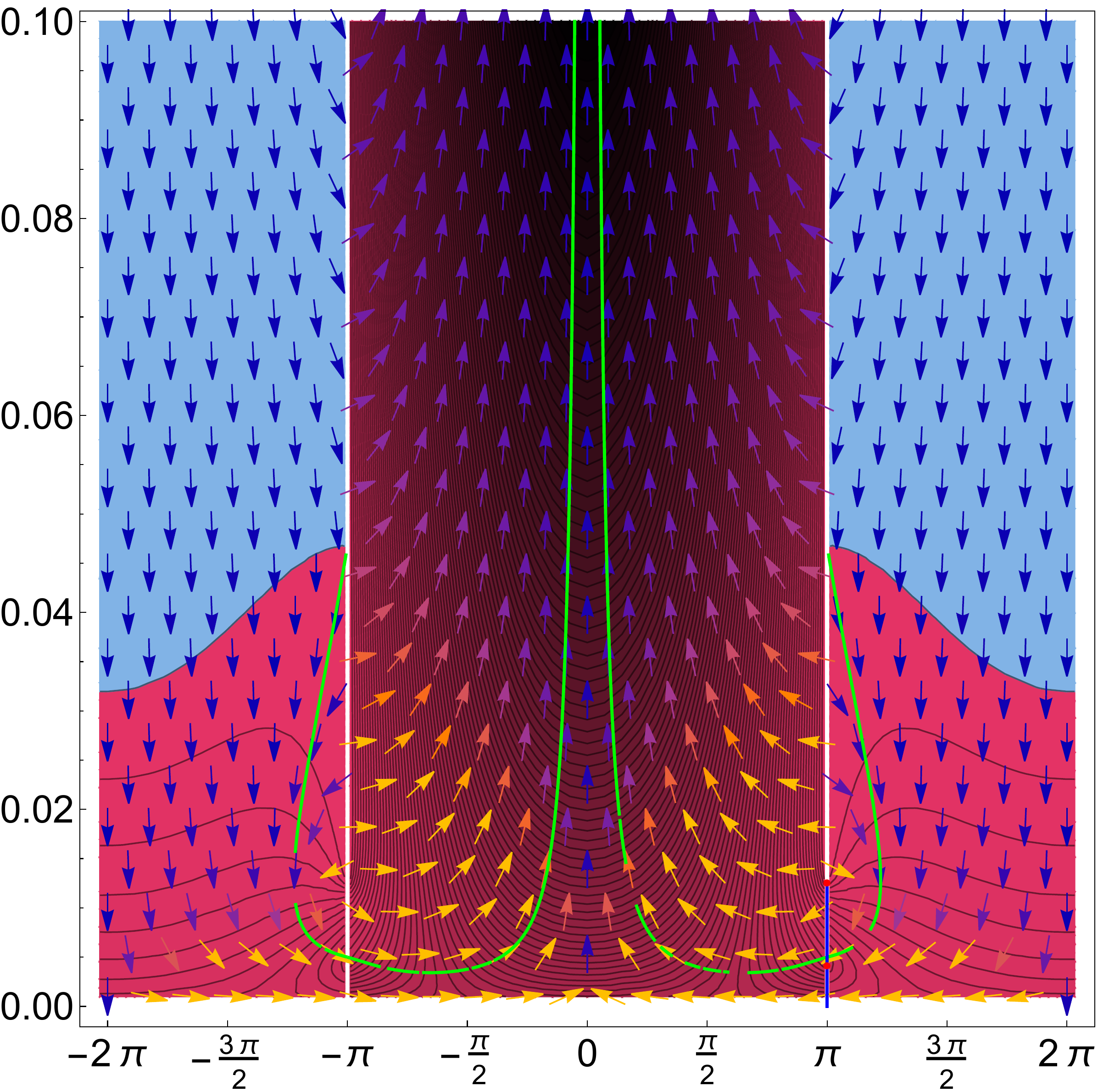} }

\put(470,27){ \includegraphics[scale=0.6]{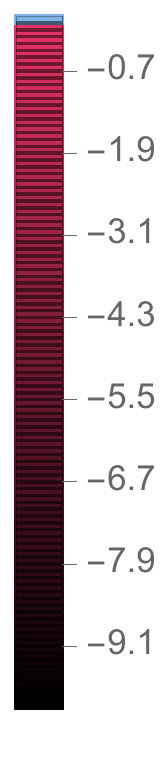} }

\put(225,174){$\text{Re}(W_a)$}
\put(472,174){$\text{Re}(W_{b'})$}
\put(128,0){$\phi$}
\put(380,0){$\phi$}

\put(10,115){$ r_h $}

\end{picture}
\caption{ Contour plots for the real parts of the Regge exponents $W_a$ (left) and $W_{b'}$ (right) for the shell model with the gradient vector field, the Lefschetz thimbles (green lines) and the anti-thimbles (black lines) laid on top. We have used as boundary values $s_{l_1} = 9, s_{l_2} = (3+\sqrt{0.1})^2$ and a cosmological constant $\tilde \Lambda = 1000$. The saddle point on the Lorentzian axis is at $\phi= \pi, r_h^* \approx 0.047$. The left panel shows the thimbles on the Riemann sheet ${\cal S}_a$. The thimbles continue from $\phi >\pi$ to the region $0\leq \phi<\pi$ on the Riemann sheet ${\cal S}_{b'}$,  as shown on the right panel. The blue line at $\phi=\pi$ indicates configurations violating hinge causality. \label{Fig:Shell1}}
\end{figure}

We will now discuss the shell model, where we will choose boundary values $s_{l_1},s_{l_2}$ so that the saddle points occur for Lorentzian data. We will see that this leads to a more involved behaviour of the Lefschetz thimbles.

The Regge exponents $W_a,W_b,W_{b'}$ etc. for the shell model are defined in Appendix \ref{AppD}. The Riemann sheet for the shell model has the same description as for the ball model, discussed in Section \ref{ssec:ball_extension}. 

As mentioned above we choose boundary values $s_{l_1},s_{l_2}>s_l^{\rm crit}$, so that we have saddle points for Lorentzian data: we will have one saddle point for $\phi=\pi$ and one for $\phi=-\pi$. There will be again a reflection symmetry for the Lefschetz thimbles and anti-thimbles, as well as for the gradient flow, with respect to the $\phi=0$ axis and with respect to the $\phi=2\pi$ axis. We can therefore restrict our discussion to the region $0\leq \phi\leq 2\pi$. 

 Figure \ref{Fig:Shell1} shows the Lefschetz thimble for an example with $s_{l_1}=9$ and $s_{l_2}=(3+\sqrt{1/10})^2\approx 11$  on the Riemann sheet ${\cal S}_a$  (where the Regge exponent is represented by $W_a$, see Section \ref{ssec:ball_extension}). We use a cosmological constant $\tilde \Lambda=1000$, but remind the reader that with our use of dimensionless lengths variables, a change of $\tilde \Lambda$ leads only to a re-scaling of the action, and therefore does not change the Lefschetz thimbles.  We see that, starting from the saddle point  at $\phi=\pi, r_h=r_h^* \approx 0.047$ the Lefschetz thimble enters the region  $0\leq \phi<\pi$  and for large $s_h$ asymptotes towards the Euclidean axis at $\phi=0$. (It cannot reach the Euclidean axis, as there the imaginary part of the Regge exponent vanishes, whereas the imaginary part of the Regge exponent along the Lefschetz thimble is finite and constant and thus equal to  the imaginary part of the Regge exponent of the saddle point.)
 
 Indeed, different from the ball model in the Euclidean regime where $s_{l}<s_l^{\rm crit}$, the Regge exponent for the shell model in the Lorentzian regime is negative for sufficiently large $r_h$, and monotonically decreasing for increasing $r_h$.
 
 Following the Lefschetz thimble from the saddle point towards smaller $r_h$ values, we now enter the region $r_h<r_h^*,\,\pi<\phi<2\pi$, where the Regge exponent has negative real part. Indeed, the real part of the Regge exponent at the saddle point is vanishing.  The Lefschetz-thimble, which is defined by following the flow of the gradient vector field along which the real part can get only smaller, can therefore only enter regions where the real part is negative. Thus, it will go along the suppressing side of the branch cut resulting from the causally irregular configurations.
 
 The Lefschetz thimble does however {\it not} end at $r_h=0$. Rather, after a turn with respect to the $\phi$ variable,  it approaches the branch cut at $\phi=\pi$ and  finite value of $r_h=r_h^{**}\approx 0.0052$ from the side with $\phi>\pi$. For the class of boundary values we will consider here, this occurs for the region $(b)$ (defined in Section \ref{SSecA}), that is for $(\sqrt{s_{l_2}}-\sqrt{s_{l_1}})^2/24<r^{**}_h<(\sqrt{s_{l_2}}-\sqrt{s_{l_1}})^2/8$.  Thus, the Lefschetz thimble continues on the Riemann sheet ${\cal S}_{b'}$, where the Regge exponent is given by $W_{b'}$, see Figure \ref{Fig:Shell1}. It continues to follow decreasing $\phi$ and $r_h$, but then turns towards increasing $r_h$, and for large $r_h$ asymptotes towards the Euclidean axis at $\phi=0$.  Note that along this Euclidean axis at $\phi=0$ on ${\cal S}_{b'}$, the Regge exponent is even decaying faster than along the Euclidean axis on ${\cal S}_{a}$, see Figure \ref{Fig:WaWb0}. This is due to additional contributions proportional to the areas, see Equation (\ref{dec4.14}).
 
 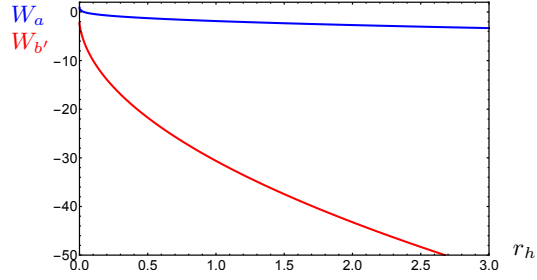
\begin{figure}[ht!]
\begin{picture}(400,105)
\put(100,0){ \includegraphics[scale=0.22]{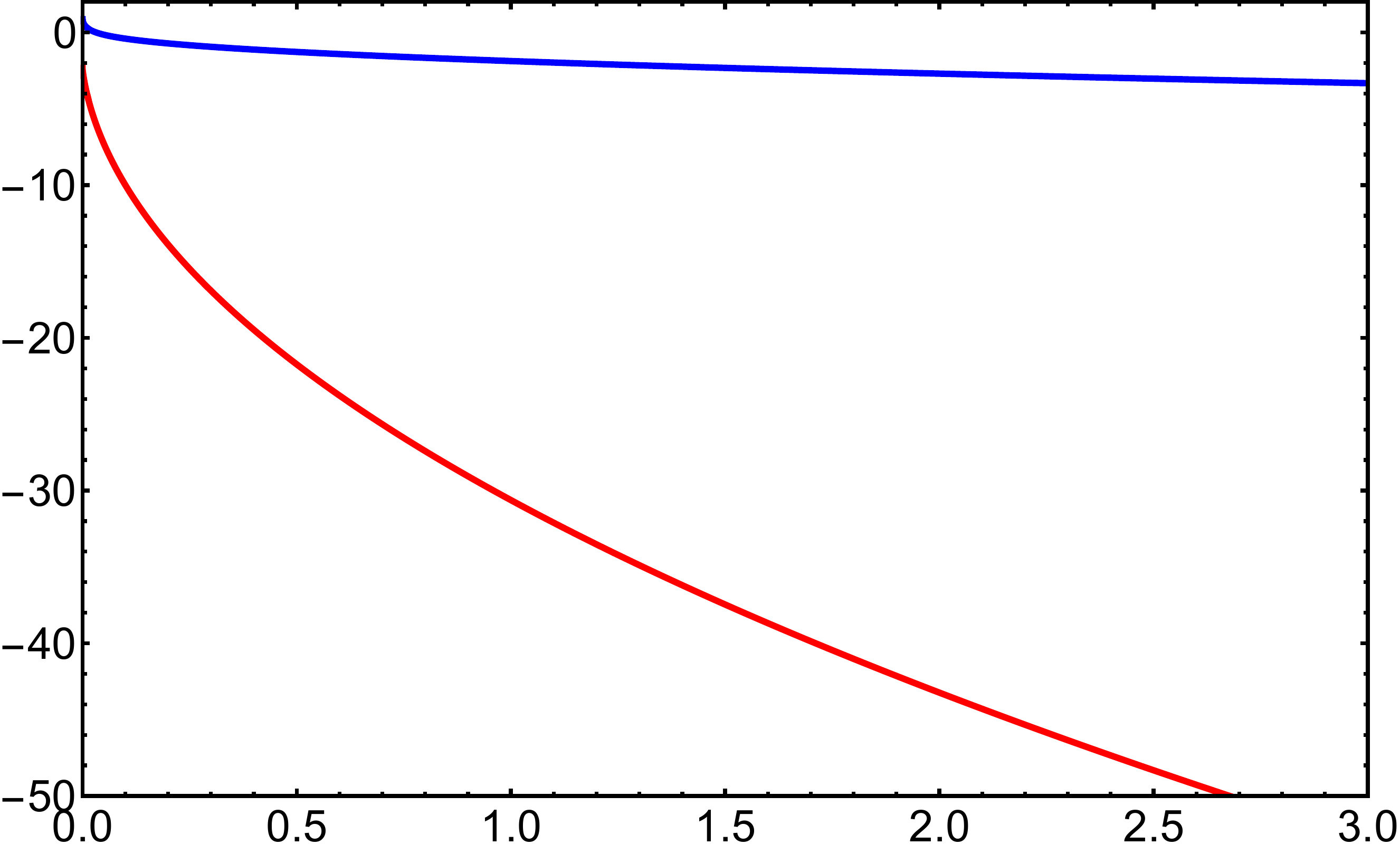} }

\put(87,95){\color{blue} $ W_{a} $}
\put(87,85){\color{red} $ W_{b'} $}
\put(277,7){$r_h$}

\end{picture}
\caption{ The Regge exponents $W_{a}$ (in blue) and $W_{b'}$ (in red) along the Euclidean axis at $\phi=0$ as a function of $r_h$ for $s_{l_1}=9,s_{l_2}=(3+\sqrt{1/10})^2, \tilde \Lambda=1000$.  \label{Fig:WaWb0}}
\end{figure}
 
 The Lefschetz thimble is therefore divided into two parts: the first part is in on the sheet ${\cal S}_a$ and goes from the saddle point at finite $r^*_h$ and $\phi=\pi$ towards  $r_h\rightarrow \infty$ and $\phi\rightarrow 0$. The second part starts from the saddle point on the sheet ${\cal S}_a$ in an arc in the region $\pi\leq \phi<2\pi$ towards a point $(r_h=r_h^{**},\phi=\pi)$, which is on the branch cut for $W_a$, in the region $(b)$ of  irregular Lorentzian data. It then enters the sheet ${\cal S}_{b'}$ of the Riemann surfaces, where it eventually goes to $r_h\rightarrow \infty$ and asymptotes to the Euclidean axis. 
 
To be able to connect a contour along the Lorentzian axis (including irregular data) to this Lefschetz thimble, we have to choose this contour along the Lorentzian axis as follows: for regular data we go along $\phi=\pi$ on the sheet ${\cal S}_a$. Going down towards smaller $r_h$ we choose to circumvent the first branch point (at $r_h=(\sqrt{s_{l_2}}-\sqrt{s_{l_1}})^2/8$)  on the side  with $\phi>\pi$, but cross $\phi=\pi$ for $r_h=(\sqrt{s_{l_2}}-\sqrt{s_{l_1}})/8-\epsilon$ and enter the sheet ${\cal S}_{b'}$. (Remember that $W_{b'}$ is analytical around $(r_h,\phi=\pi)$ with $(\sqrt{s_{l_2}}-\sqrt{s_{l_1}})^2/24<r_h<(\sqrt{s_{l_2}}-\sqrt{s_{l_1}})^2/8$.) On this sheet ${\cal S}_{b'}$ we continue towards $r_h=0+\epsilon$, going left from the branch cut  that starts for $r_h=(\sqrt{s_{l_2}}-\sqrt{s_{l_1}})^2/24$, , that is we go along a Wick rotation angle $\phi$ slightly smaller than $\pi$.

Note that with this choice of contour for the irregular region, which changes to the sheet ${\cal S}_{b'}$ (on the side $\phi<\pi$) we obtain an even more suppressing amplitude for the region $(c)$ of irregular data, as compared to staying on ${\cal S}_a$ on the side $\phi>\pi$, where the real part of the Regge exponent is negative, but increases in the region $(c)$, see Figure \ref{Fig:ShellExponents}.  

\begin{figure}[ht!]
\begin{picture}(400,105)
\put(100,0){ \includegraphics[scale=0.27]{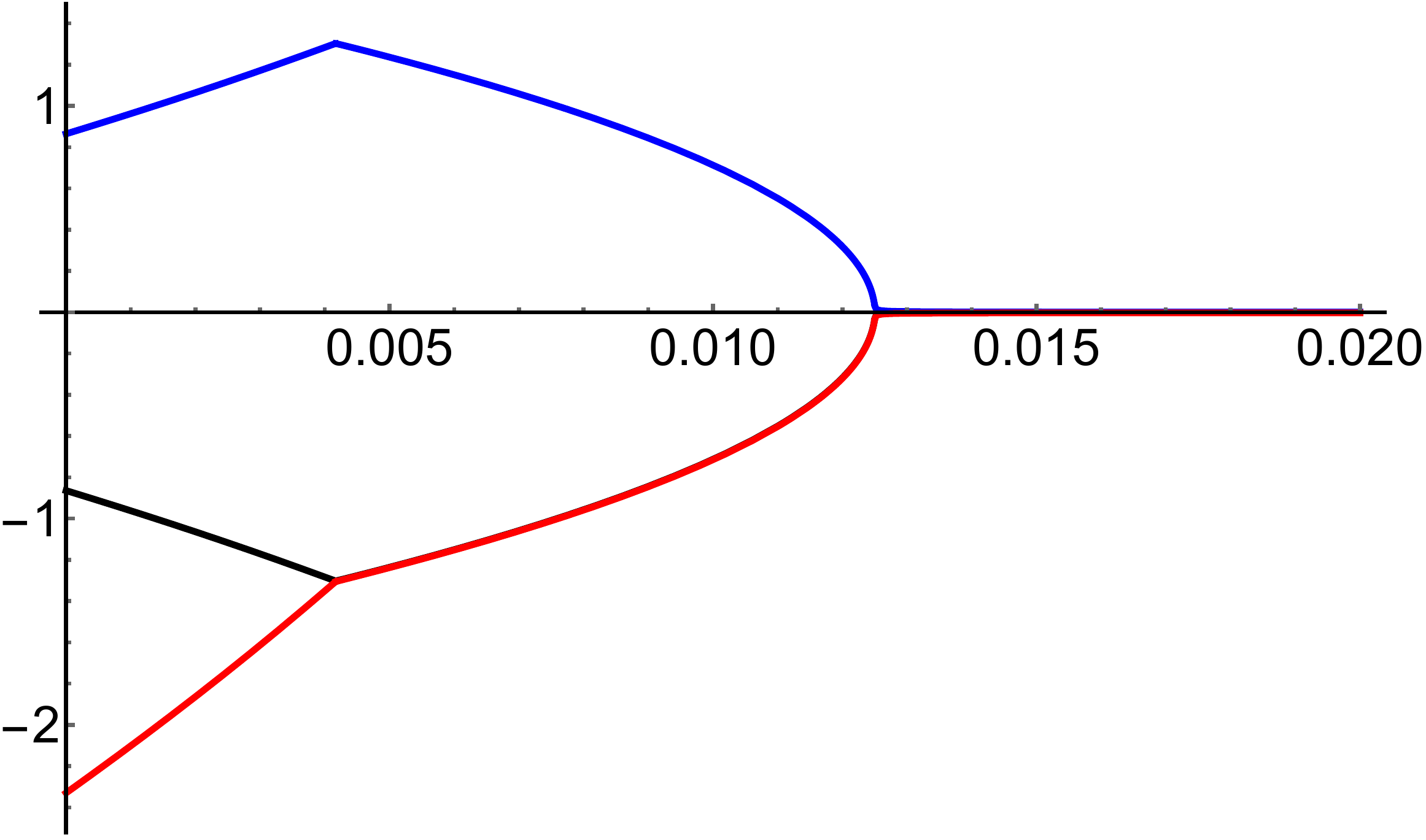} }

\put(90,95){$ W $}
\put(297,64){$r_h$}

\end{picture}
\caption{ This figure shows for $s_{l_1}=9,s_{l_2}=(3+\sqrt{1/10})^2, \tilde \Lambda=1000$ the real parts of the Regge exponent $W_a$ at $\phi=\pi-10^{-3}$ (in blue), of $W_a$ at $\phi=\pi+10^{-3}$ (in black), and of $W_{b'}$ at $\phi=\pi-10^{-3}$ (in red). The branch points are at $r_h=0.0125$ and at $r_h\approx 0.0042$.}    \label{Fig:ShellExponents}
\end{figure}

To connect the Lefschetz thimble with this contour along the Lorentzian data, we need now two additional pieces: On sheet ${\cal S}_a$ we need an arc connecting the Lefschetz thimble for large $r_h=R$ at $\phi\approx 0$ to the Lorentzian axis at $\phi=\pi$. The contribution from this arc vanishes for $R\rightarrow 0$. On the sheet ${\cal S}_{b'}$ we need a contour ${\cal C}$ that connects $r_h=0,\phi=\pi-\epsilon$ to the Lefschetz thimble, again at some large $r_h=R$ at $\phi\approx 0$. But this will give a closed contour on the sheet ${\cal S}_{b'}$, summing the various contributions we obtain zero. Thus, instead of evaluating the Lefschetz thimble on ${\cal S}_{b'}$ and on the connecting contour ${\cal C}$, we directly evaluate the Lorentzian axis contour described above, from the point where it crosses the Lefschetz thimble (in region $(b)$) to the point $r_h=0+\epsilon$.

In fact, using that we can deform the contour as long as the integrand is analytical, we do not need to follow exactly the Lefschetz thimble. Indeed, different from the ball model in the Euclidean regime,  the precise form and parametrization of the Lefschetz thimble now depends on the precise values of the boundary data $s_{l_1}, s_{l_2}$, and working out this parametrization is cumbersome. 

Instead we choose the following much simpler contour: On the sheet ${\cal S}_a$, from the Lorentzian axis at $\phi=\pi$ and $r_h=R$, with $R$ large, we go along $r_h=R$ toward $\phi=0$, that is the Euclidean axis. The integral over this arc vanishes for $R\rightarrow \infty$. We then go along the Euclidean axis from $r_h=R$ to $r_h=r_h^*$. The Regge exponent along the Euclidean axis is real, and for sufficiently large $r_h$ negative and monotonically decreasing. This integral does therefore converge for $R\rightarrow \infty$. 

We then connect the point $(r_h=r_h^*,\phi=0)$ with the point $(r_h=r_h^*,\phi=\pi)$ going along the line of constant $r_h=r_h^*$. From the latter point we follow a half-circle, in the region $\phi>\pi$, connecting to a point $(r_h= (\sqrt{s_{l_2}}-\sqrt{s_{l_1}})^2/8-\epsilon,\phi=\pi)$ on the Lorentzian axis. Here we enter the sheet ${\cal S}_{b'}$ and on this sheet use another half-circle, now in the region $\phi<\pi$ to connect the previous point to the final point $(r_h=0+\epsilon, \phi=\pi)$.

All oscillatory integrals are over finite integration domains and therefore well defined. The integration over the $r_h>r_h^*$ part of the Euclidean axis is over an unbounded domain, but the Regge exponent is real and negative and moreover decreases rapidly, so that the integral is absolutely convergent. The integral converges also rapidly for $\epsilon\rightarrow 0$. 

The above contour applies if we include irregular data into our Lorentzian path integral. If we exclude these data, we can simple leave out the integration over the last half-circle (from $(r_h= (\sqrt{s_{l_2}}-\sqrt{s_{l_1}})^2/8-\epsilon,\phi=\pi)$ to $(r_h=0+\epsilon, \phi=\pi)$). Also here, the contributions from the branch points is vanishing in the limit $\epsilon\rightarrow 0$.

\begin{figure}[ht!]
\begin{picture}(500,125)
\put(30,1){ \includegraphics[scale=0.33]{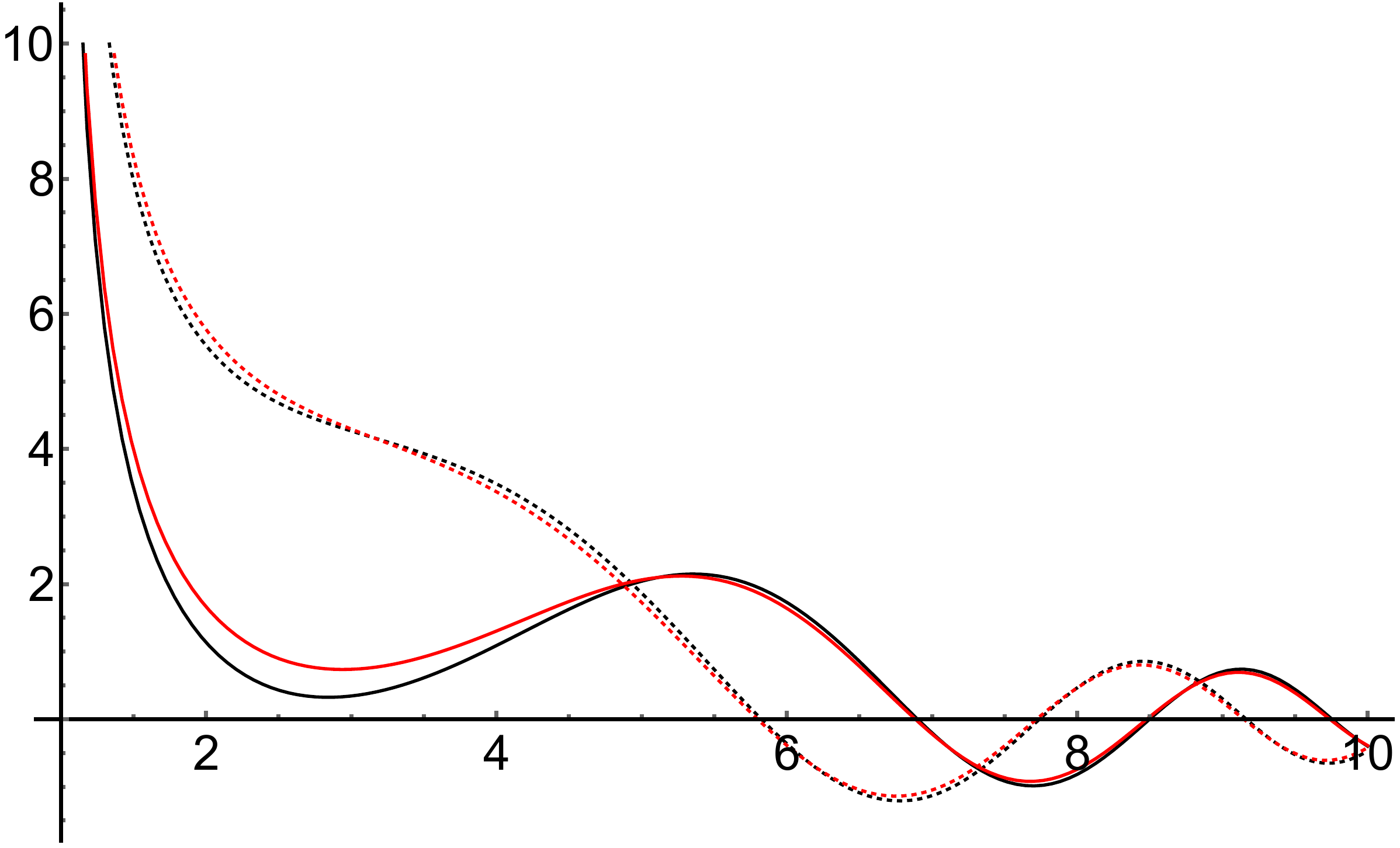} }

\put(275,7){ \includegraphics[scale=0.3]{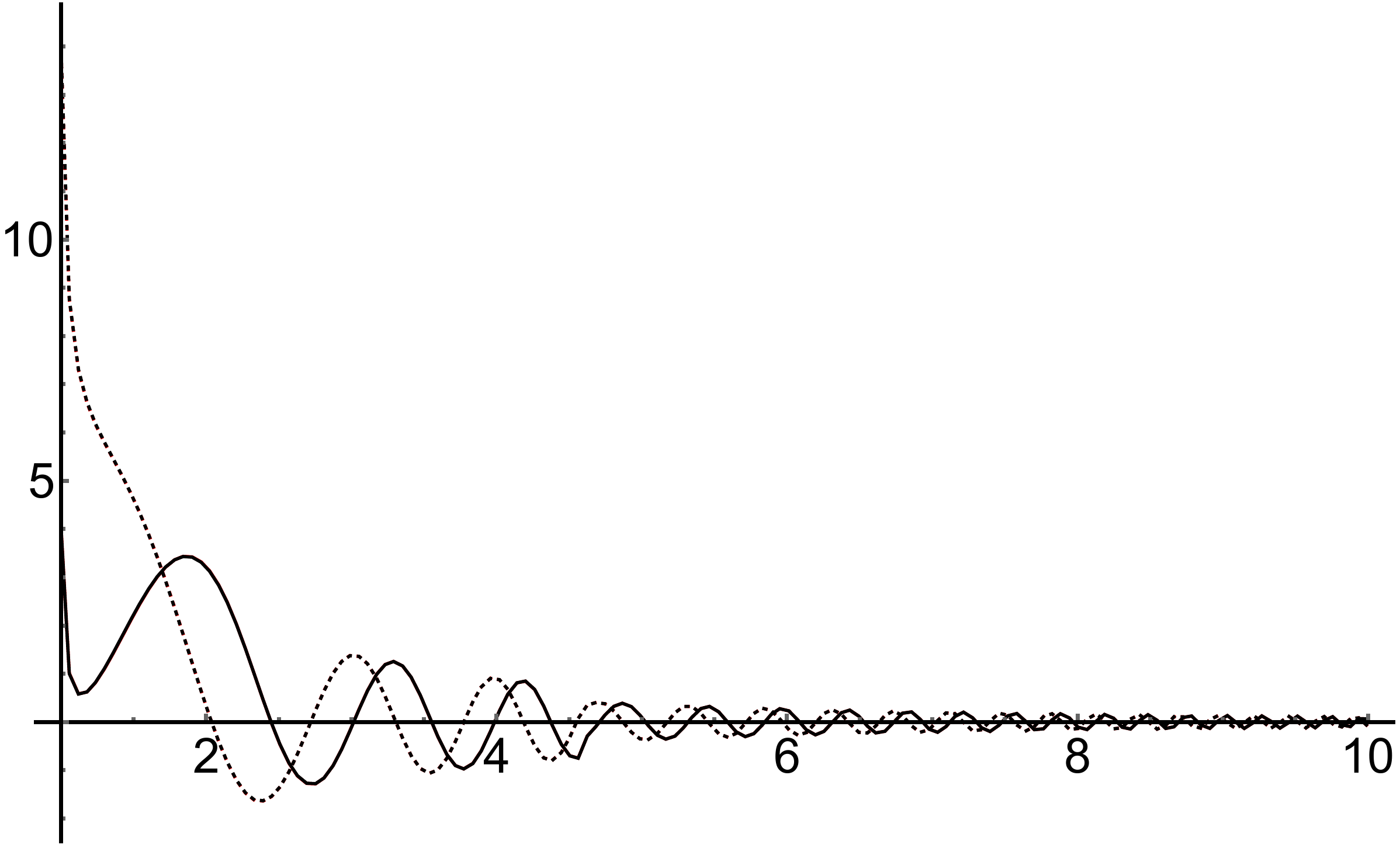} }

\put(182,115){ \line(1,0){7}}
\put(195,112){$Y = 0$}
\put(182,100){\color{red}  \line(1,0){7}}
\put(195,97){$Y = -\frac{0.1}{8}$}

\put(432,115){ \line(1,0){7}}
\put(445,112){$Y = 0$}
\put(432,100){ \color{red} \line(1,0){7}}
\put(445,97){$Y = -\frac{0.1}{8}$}

\put(128,0){$\sqrt{s_{l_1}}$}
\put(380,0){$\sqrt{s_{l_1}}$}

\put(7,115){$ Z_{\rm Shell} $}
\put(262,115){$ Z_{\rm Shell} $}

\end{picture}
\caption{
The two panels show the partition function including (in black) and excluding (in red) configurations violating hinge causality. Both panels use $s_{l_2}=(\sqrt{s_{l_1}}+\sqrt{1/10})^2$, for the left panel we used  $\tilde \Lambda=1000$, for the right panel we used  $\tilde \Lambda=100$. The real parts are shown as solid curves, the imaginary parts as dotted curves. In the right panel there is no visible difference between the black and red curves describing the cases of including and excluding irregular configurations, respectively.
\label{Fig:ShellResults}
}
\end{figure}
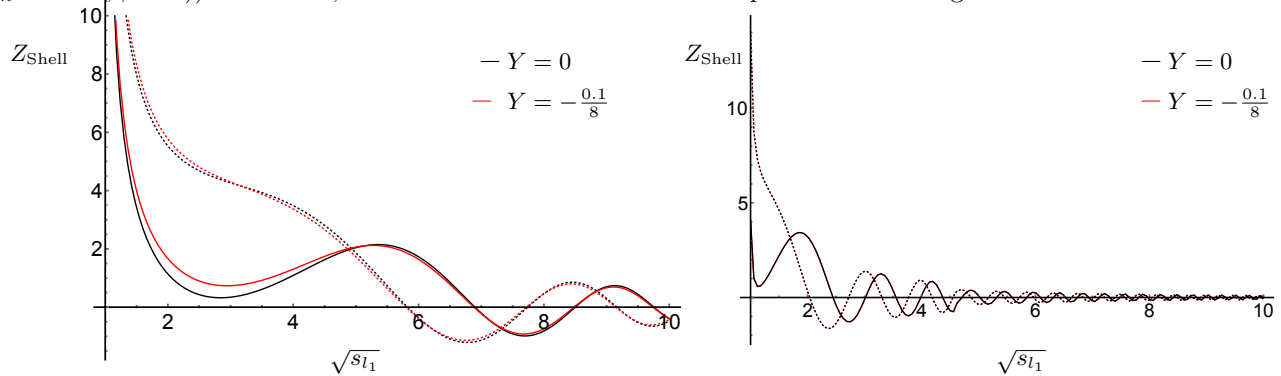

 Figure \ref{Fig:ShellResults}  shows the results for the path integral using a range of boundary data, and for the two choices of including and excluding irregular data. We see that for these examples the differences are small, more so for larger $s_{l_1},s_{l_2}$ and for smaller values of $\tilde \Lambda$.

In summary the Lorentzian regime of the shell model provided us with an example where we have saddle points on the Lorentzian axis. The Lefschetz thimble does again suggest a deformation of the contour that leads to a converging integral. It leads also to a suppressing choice for the irregular data -- for region $(c)$ of irregular data even more suppressing than in the Euclidean regime for the ball model. In general, if we have only saddle points along regular Lorentzian data, the real part of the Regge exponent evaluated on these saddle points will be vanishing. The Regge exponent along the associated Lefschetz thimbles will therefore have  a negative real part. Irregular data can therefore only enter on the suppressing sides of the branch cuts.

 ~\\
 We conclude with a comment on the sum over orientations: here we only considered positive orientation, and equivalently a (path) integral over positive lapse. That is the starting point for our contour is provided by the Lorentzian axis at $\phi=\pi$. The Lorentzian axis at $\phi=-\pi$ leads to the Regge exponent for a triangulation with negative orientation or equivalently negative lapse. 
 
 Due to the mirror symmetry of  the gradient flow at $\phi=0$, performing the integral for negative orientations, that is along $\phi=-\pi$ will lead to almost the same result. The only difference is that we will obtain the complex conjugated of the partition function for $\phi=\pi$. The reason is that the imaginary parts of the Regge exponents are odd with respect to $\phi=0$ (whereas the real parts are even).
 
 Thus as usual, a sum over orientations, or equivalently an integral over positive and negative lapse, amounts to taking the real part of the partition function obtained from taking only the positive orientation into account.

\section{Summary and Discussion}\label{discussion}

We discussed here  a Lorentzian path integral for quantum gravity, based on Regge calculus \cite{Regge}, and applied it to a triangulation modeling homogeneous and isotropic cosmology. Regge calculus underlies both Causal Dynamical Triangulations \cite{CDT} and spin foams \cite{SFReview,EPRL-FK,EffSF,EffSF3}, we therefore expect that the discussions here do have implications for a range of approaches. 

To this end we analyzed first the properties of the complex Regge action, based on a careful consideration of the complex angles.  We established that the two variants of the Lorentzian Regge action discussed in the literature do define equivalent analytical extensions.  These analytical extensions also include a further copy for the Lorentzian data, where the Regge exponent gives $(-\imath)$ times the (Lorentzian) Regge action, as opposed to $(+\imath)$ times the (Lorentzian) Regge action. We have also Euclidean data, where the Regge exponent gives  minus the Euclidean action, and another copy giving plus the Euclidean action.  The global change of sign can be interpreted to arise from a change of orientation. Such a change of orientation can be also done locally for each 4-simplex. It might therefore be possible that the complex Regge action already includes all possible choices of orientations for the 4-simplices in a given triangulation. In spin foams one does sum over all theses different choices of orientations.

We have also investigated in more detail the analytical extension of the complex Regge action for an example triangulation discussed in Section \ref{SecEx}. This example suggest that circling branch points, which result from configurations with null triangles or null tetrahedra, leads to a change in the deficit angles by multiples of $2\pi n$, with $n\in \mathbb{Z}$, which is consistent with the behaviour of the angle functions, when circling configurations with null edges. There is another remarkable connection to spin foams \cite{PRinSF}: there, instead of integrating over length variables one sums over discrete values for the areas. But, using Poisson resummation, one can rewrite the sum into a sum of integrals over now continuous area parameters. The new sums are over $n \in \mathbb{Z}$-labels,  which are added as $2\pi n$ terms to the deficit angles.

 The branch cuts arise for Lorentzian data, which violate hinge causality. We have seen that in Regge calculus, and therefore also related approaches such as spin foams, violations of causality conditions as formulated in \cite{LollJordan} and revisited in Section \ref{CCond}, are rather generic. Even triangulations describing simple cosmological models include such causality violations \cite{DGS}. These are, in fact, related to Courant-Friedrichs-Lewy instabilities. Related effects have been also discussed in the context of loop quantum cosmology \cite{Bojowald}. 

For a certain type of causal irregularities, namely violations of hinge causality, one encounters imaginary terms in the Regge action.  These imaginary terms lead to either a suppression or enhancement of these configurations. Which one occurs seems to be \textit{a priori} a choice in the construction of the Regge action.  But either choice leads to the enhancement of some type of causality violations,  which can easily lead to non-sensible results \cite{EffSF3}.

Here we found that violations of hinge causality lead to branch cuts, with one side of the branch cut leading to a suppression and the other side to an enhancement. 

One thus faces the question whether to include such configurations violating hinge causality into the Lorentzian contour or not. If one does include these configurations, how should one navigate the branch cuts? 

We proposed and illustrated here that the answer to the second question can be provided by applying Picard-Lefschetz theory \cite{PLTheory} for the computation of the (path) integral. Picard-Lefschetz theory allows to identify a contour, given in general by a certain combination of Lefschetz-thimbles, along which the integral is `most' convergent. We have seen in our examples that if one wants to deform the original contour to this combination of Lefschetz-thimbles, one is also given a prescription of how to navigate the branch cuts. In our examples, this has lead to a suppression of the causally irregular configurations. For configurations admitting saddle points on regular Lorentzian data, we have also identified the general reason for finding a suppression for these causally irregular configurations. 

We also investigated the possibility of excluding configurations violating hinge causality. Using the integration contour defined by the Lefschetz thimble, we found that this amounts to adding or leaving out an integration over an arc, that goes around the irregular configurations. We found that for the examples with saddle points along Lorentzian data the differences are small as compared to the over-all partition function. In the case where we have an Euclidean saddle point, the contribution from the Lefschetz-thimble, which includes the irregular configurations, is exponentially suppressed (e.g. as a function of the boundary scale factor). The additional contribution, that is needed to compute the path integral without irregular configurations, is quite small, but does become the dominating one for sufficiently large boundary scale factor, where the contribution from the Lefschetz-thimble is much smaller than one.

This additional contribution, which arises if one does exclude configurations violating hinge causality, does not have a counterpart in the continuum theory. Indeed, for the ball model, one does obtain a better match to the continuum if one does include the causally irregular configurations. 

We expect that excluding causally irregular configurations, can become  computationally quite cumbersome. This is due to the fact that the violation of these conditions cannot be checked locally on the building blocks, but requires \textit{e.g.} all the building blocks glued around a hinge, or glued around a vertex. Checking in addition non-local causality conditions, such as the existence of closed time-like curves, would require even more effort. 

We therefore see the possibility of including causality violating configurations, together with a mechanism that leads to a suppression of such configurations, as promising.

One of our examples, used for the construction of the no-boundary wave function, realizes a type of causality violation different from hinge causality, which we discussed above, namely a violation of vertex causality.  Vertex causality violations do not lead to the appearance of imaginary terms in the Regge action \textit{a priori}. But for our example we find that we do not have saddle points for Lorentzian data, but only for Euclidean data. Using Picard-Lefschetz-theory we  find an exponential suppression, in agreement with the continuum discussion \cite{TurokEtAl}.

There is a remarkably  similar result in spin foams, see also the discussion in \cite{EffSF3}, which  suggest that we should  allow vertex causality violations: The spin foam amplitude for one 4-polytope (typically a 4-simplex) can be reformulated in terms of an integral over $\text{SU(2)}$ group elements. The amplitude is typically  not suppressed if for the Lorentzian theory, the boundary data satisfy the Lorentzian triangle inequalities. But for configurations which define  Euclidean boundary data, one does find an amplitude that is exponentially suppressed with the Regge action \cite{BarrettFoxon,HanLiu}. 

The boundary data for the 4-polytope discussed in Section \ref{SecEx} also violate the Lorentzian triangle inequalities. But they do realize an Euclidean flat 4-polytope. In Section \ref{SecEx} we considered a refinement of the 4-polytope, which admits a Lorentzian realization -- but violates vertex causality. Now, to realize triangulation independence --- which can be considered as a form of diffeomorphism invariance \cite{DittrichDiff} --- one would like to have the same amplitude for the original polytope and the subdivided polytope. This might be indeed achievable in spin foams, where Euclidean boundary data for the not-subdivided polytope in the Lorentzian theory, lead to an exponentially suppressed amplitude.  

On the other hand, to achieve such a triangulation independence for the Regge path integral, we would have to either include 4-simplices which satisfy the Euclidean triangle inequalities with exponentially suppressed amplitudes, or exclude Lorentzian configurations which include subdivided 4-polytopes, with boundary data that can only be realized in Euclidean space. This latter option would again amount to complicated highly non-local conditions, whereas the first option would remove a condition, namely the part of the generalized triangle inequality, that imposes positive squared volume for the 4-simplices.

In this work we considered quite simple examples, which already allowed us to draw very interesting conclusions. It will be very interesting to see whether the suppression of causality violations holds also for (much) more complicated triangulations, and to investigate further the difference between including and excluding causality violating configurations.  The discussion around subdivided 4-polytopes has shown that demanding triangulation independence can already lead to a preference between these options.

The cosmological model considered here offers already a number of interesting extensions: We can consider the path integral for several time steps. The corresponding continuum system has (time) diffeomorphism symmetry, but this symmetry is broken by the discretization \cite{DittrichDiff}. Choosing more and more time steps for fixed boundary conditions, one expects to regain diffeomorphism invariance \cite{BahrDitt09,BDS}, and thus a divergence if one performs the integration for all bulk variables. Demanding diffeomorphism invariance can fix the path integral measure \cite{BDS,PIMeasure}. It might also help to answer the question of which (irregular) configurations to include, as one could consider to gauge fix either the $s_l$ variables or (a part of the) $s_h$ variables, and both procedures should lead to the same result. Regaining (time) diffeomorphism invariance would also  allow to derive a Hamiltonian description of the theory \cite{DittHoehn}, where one can also study the impact of excluding or including causally irregular configurations.


\appendix

\section{On the analytical extensions of $\theta^\pm$}\label{AppA}

Here we detail the analytical extensions of the dihedral angels $\theta^\pm$ for the various cases of the vectors $a,b$ lying in the different quadrants, depicted in Figure \ref{Fig1}.

\begin{itemize}
\item For $a$ and $b$ from quadrant I we parametrize the vectors as $a=(\sinh\alpha,\cosh\alpha),b=(\sinh\beta, \cosh \beta)$. The square roots in the denominator in (\ref{eqtheo1}) are fine, but we have a branch cut at  $\phi=\pm \pi$ from the square root in the numerator. This case avoids the  log branch cut. 

The analytical extensions are therefore given by (for $\phi \in (-2\pi,2\pi]$)
\ba
\theta^\pm&=&
-\imath \log\left( \frac{e^{\imath\phi} \sinh \alpha \sinh\beta +\cosh\alpha \cosh\beta \mp\imath   e^{\imath \phi/2}|\sinh(\alpha-\beta)| }{\sqrt{\cosh^2\!\alpha+ e^{\imath\phi} \sinh^2\!\alpha}\,\,\sqrt{\cosh^2\!\beta+ e^{\imath\phi} \sinh^2\!\beta}}\right) \q .
\ea

\item For $a$ from quadrant I and $b$ from quadrant II, we use $a=(\sinh\alpha,\cosh\alpha),\,b=(\cosh\beta, \sinh \beta)$: Apart from the branch cut in the numerator for $\phi=\pm \pi$, we have also a branch cut at $\phi=\pm \pi$ from one of the square roots in the denominator. This determines $\sqrt{\cdot}_+$ for $\theta^+$ and $\sqrt{\cdot}_-$ for $\theta^-$. We do not encounter the branch cut from the logarithm.

Thus we obtain (for $\phi \in (-2\pi,2\pi]$)
\ba
\theta^\pm&=&
-\imath \log\left( \frac{e^{\imath\phi} \sinh \alpha \cosh\beta +\cosh\alpha \sinh\beta \mp\imath   e^{\imath \phi/2}\cosh(\alpha-\beta) }{\sqrt{\cosh^2\!\alpha+ e^{\imath\phi} \sinh^2\!\alpha}\,\, e^{\imath \phi/2} \sqrt{\cosh^2\!\beta+ e^{-\imath\phi} \sinh^2\!\beta}}\right)  \q .
\ea

\item For $a$ from quadrant II and $b$ from quadrant II, we write $a=(\cosh\alpha,\sinh\alpha),\,b=(\cosh\beta, \sinh \beta)$. We have only a branch cut from the numerator and obtain the analytical extension
\ba
\theta^\pm&=&
-\imath \log\left( \frac{e^{-\imath\phi} \sinh \alpha \sinh\beta +\cosh\alpha \cosh\beta \mp\imath   e^{-\imath \phi/2}|\sinh(\alpha-\beta)| }{\sqrt{\cosh^2\!\alpha+ e^{-\imath\phi} \sinh^2\!\alpha}\,\,\sqrt{\cosh^2\!\beta+ e^{-\imath\phi} \sinh^2\!\beta}}\right) \q .
\ea

\item
For $a=(\sinh\alpha,\cosh\alpha)$ being in quadrant I and $b=(\sinh\beta,-\cosh\beta)$  being in quadrant III we can assume (due to having a convex wedge) $\alpha+\beta\leq 0$.
 We have, in addition to the usual branch cut of the numerator a branch cut of the logarithm. Analytical extension gives
\ba
\theta^+&=&
\begin{cases}
-\imath \log_-\left( \frac{e^{\imath\phi} \sinh \alpha \sinh\beta -\cosh\alpha \cosh\beta -\imath   e^{\imath \phi/2}\sinh(\alpha+\beta) }{\sqrt{\cosh^2\!\alpha+ e^{\imath\phi} \sinh^2\!\alpha}\,\,\sqrt{\cosh^2\!\beta+ e^{\imath\phi} \sinh^2\!\beta}}\right) \;\q\q \q \text{for} \,\, \phi\in (-\pi,+\pi) \\
-\imath \log_+\left( \frac{e^{\imath\phi} \sinh \alpha \sinh\beta -\cosh\alpha \cosh\beta -\imath   e^{\imath \phi/2}\sinh(\alpha+\beta) }{\sqrt{\cosh^2\!\alpha+ e^{\imath\phi} \sinh^2\!\alpha}\,\,\sqrt{\cosh^2\!\beta+ e^{\imath\phi} \sinh^2\!\beta}}\right)  -2\pi \q \text{for} \,\, \phi\in (-2\pi,-\pi] \cup [+\pi,+2\pi] 
\end{cases} \nn\\
\theta^-&=&
\begin{cases}
-\imath \log_+ \left(\frac{e^{\imath\phi} \sinh \alpha \sinh\beta -\cosh\alpha \cosh\beta +\imath   e^{\imath \phi/2}\sinh(\alpha+\beta) }{\sqrt{\cosh^2\!\alpha+ e^{\imath\phi} \sinh^2\!\alpha}\,\,\sqrt{\cosh^2\!\beta+ e^{\imath\phi} \sinh^2\!\beta}} \right)\;\q\q \q \text{for} \,\, \phi\in [-\pi,+\pi] \\
-\imath \log_- \left(\frac{e^{\imath\phi} \sinh \alpha \sinh\beta -\cosh\alpha \cosh\beta +\imath   e^{\imath \phi/2}\sinh(\alpha+\beta) }{\sqrt{\cosh^2\!\alpha+ e^{\imath\phi} \sinh^2\!\alpha}\,\,\sqrt{\cosh^2\!\beta+ e^{\imath\phi} \sinh^2\!\beta}} \right) +2\pi \q \text{for} \,\, \phi\in (-2\pi,-\pi) \cup (+\pi,+2\pi] \q .\q\q
\end{cases} 
\ea

\item For $a$ from quadrant II and $b$ from quadrant IV, $a=(\cosh\alpha,\sinh\alpha),\,b=(-\cosh\beta, \sinh \beta)$ we can again assume $\alpha+\beta\geq 0$.  This case behaves very similar to the quadrant I -- quadrant III case. 

The analytical extensions are given by
\ba
\!\!\!
\theta^+&=&
\begin{cases}
-\imath \log_-\left( \frac{e^{-\imath\phi} \sinh \alpha \sinh\beta -\cosh\alpha \cosh\beta -\imath   e^{-\imath \phi/2}\sinh(\alpha+\beta) }{\sqrt{\cosh^2\!\alpha+ e^{-\imath\phi} \sinh^2\!\alpha}\,\,\sqrt{\cosh^2\!\beta+ e^{-\imath\phi} \sinh^2\!\beta}}\right) \;\q\q \q \text{for} \,\, \phi\in (-\pi,+\pi) \\
-\imath \log_+\left( \frac{e^{-\imath\phi} \sinh \alpha \sinh\beta -\cosh\alpha \cosh\beta -\imath   e^{-\imath \phi/2}\sinh(\alpha+\beta) }{\sqrt{\cosh^2\!\alpha+ e^{-\imath\phi} \sinh^2\!\alpha}\,\,\sqrt{\cosh^2\!\beta+ e^{-\imath\phi} \sinh^2\!\beta}}\right)  -2\pi \q \text{for} \,\, \phi\in (-2\pi,-\pi] \cup [+\pi,+2\pi] 
\end{cases}\nn\\
\!\!
\theta^-&=&
\begin{cases}
-\imath \log_+ \left(\frac{e^{-\imath\phi} \sinh \alpha \sinh\beta -\cosh\alpha \cosh\beta +\imath   e^{-\imath \phi/2}\sinh(\alpha+\beta) }{\sqrt{\cosh^2\!\alpha+ e^{-\imath\phi} \sinh^2\!\alpha}\,\,\sqrt{\cosh^2\!\beta+ e^{-\imath\phi} \sinh^2\!\beta}} \right)\;\q\q \q \text{for} \,\, \phi\in [-\pi,+\pi] \\
-\imath \log_- \left(\frac{e^{-\imath\phi} \sinh \alpha \sinh\beta -\cosh\alpha \cosh\beta +\imath   e^{-\imath \phi/2}\sinh(\alpha+\beta) }{\sqrt{\cosh^2\!\alpha+ e^{-\imath\phi} \sinh^2\!\alpha}\,\,\sqrt{\cosh^2\!\beta+ e^{-\imath\phi} \sinh^2\!\beta}} \right) +2\pi \q \text{for} \,\, \phi\in (-2\pi,-\pi) \cup (+\pi,+2\pi] \, .\q\q
\end{cases} 
\ea

\end{itemize}

\section{ Projecting a simplex onto a wedge orthogonal to a hinge }\label{AppProj}

 A  $n$-simplex in a flat spacetime\footnote{This includes both Euclidean and Minkowski spacetimes.} consist of the convex hull of $n+1$ vertex points  $\{ v_0 ,v_1 \cdots ,v_n \}$. The flat metric allows us to define the length square for the (geodesic) line segments between each pair of points. 
 For a non-degenerate simplex the $(n+1)$ points are affinely independent and therefore for a fixed vertex $v_k$ the set of $n$ vectors $x_{i(k)}:= v_i - v_k, i\neq k$ representing the edges of the form $v_iv_k$ are linearly independent.  

By convention, we choose $k=0$ as the vertex $v_0$ at the origin, so that the basis edge vectors are $ x_i = v_i - v_0$. The independent edge vectors can be used to define a quadratic form  
 \be 
 g_{ij} = x_{i}\star x_{j}  
 \ee
with $0< i,j \leq n$, which defines the components of a metric for the simplex, where $\star$ is the generalized inner product defined by
\be
x_{i} \star x_{j} := x_{i}^0 x_{j}^0 e^{\imath \phi} + \sum_{k=1}^{n-1} x_{i}^k x_{j}^k \q .
\ee

The metric components $ g_{ij}$  can also be computed in terms of the squared edge lengths   
\be \label{metricC}
 g_{ij} = \tfrac12 \left( s_{0i} + s_{0j} -s_{ij} \right)
\ee
for $i,j \neq 0$ and $s_{ii}=0$. Here $s_{ij} = (v_i-v_j) \star (v_i-v_j)$ is the square edge length between vertices $i,j$. The metric $g_{ij}$ defines an $n\times n$ matrix  which is a Gram matrix $G$ for the basis vectors $ \{ x_{i} \}_{i=1}^n$. 

The determinant of the Gram matrix gives $\det G = (n!)^2 \, \mathbb V$  where $\mathbb V$ is the volume squared of the simplex. Let us denote by $\mathbb V_i$ the squared volume  of the $(n-1)$-sub-simplex (face $f_i$) by removing vertex $v_i$ and $\mathbb V_{ij}$ the square volume of the $(n-2)$-sub-simplex (hinge $h_{ij}$) by removing vertices $v_i$ and $v_j$. The volume square of a sub-simplex is also proportional to the determinant of the corresponding sub-matrix. The minor of the $(i,j)$-th entry of $G$ is given by
\be
{\rm M}_{ij} = \begin{cases} ((n-1)!)^2 \, \mathbb  V_i & \text{for } i = j \\ (n!)^2\, \frac{\partial \mathbb V}{\partial s_{ij}} & \text{for } i \neq j  \end{cases}
\ee
and defines the inner product between the normal vectors to the faces $f_i,f_j$ of the simplex.

For $n=2$, the triangle with basis edge vectors $a,b$ and square edge lengths $s_{a},s_{b},s_{c}$ is associated to a  Gram matrix
\be
G = \begin{pmatrix} a \star a && a \star b \\ a \star b && b \star b \end{pmatrix}  = \begin{pmatrix} s_{a} & \tfrac12 \left( s_{a} + s_{b} -s_{c}  \right) \\ \tfrac12 \left( s_{a} + s_{b} -s_{c}  \right) & s_{b} \end{pmatrix}  \q .
\ee
The components of the this matrix have been used to derive the formula for the generalized dihedral angles between vectors $a,b$. 

In $n$ dimensions, the dihedral angles are based at the $(n-2)$-sub-simplices (hinges) of the $n$-simplex. A hinge $h_{ij}$ is defined by the vertex set $\{ v_k\}_{k\neq i,j}$. The dihedral angle at the hinge $h_{ij}$ can be computed by projecting out the hinge which in turn projects the simplex onto a 2D plane. The faces $f_i,f_j$ are projected onto edges, which span a triangle $t_{ij}$. To compute the dihedral angle  we can then apply formula \ref{thetaTria}.

The projecting out of the hinge can be performed as follows: For a given hinge $h_{ij}$, we can always arrange the order of the vertices so that the Gram matrix is of the form 
\be \label{MetG}
G = \begin{pmatrix} H & B \\  B^T& \bar H \end{pmatrix}   = \begin{pmatrix} H & B_i & B_j \\  B^T_i & \bar H_{ii} & \bar H_{ij} \\ B^T_j & \bar H_{ij} & \bar H_{jj} \end{pmatrix}  
\ee
where $H$ contains the metric components of  the $(n-2)$-sub-simplex defining the hinge and $\bar H$ is a $2\times 2$ matrix of the remaining components opposite the hinge and $i,j$ are the labels of the hinge $h_{ij}$.  The metric in the plane onto which one projects can then be expressed as the Gram matrix for the triangle $t_{ij}$, which is given by
\be\label{projMetric}
\tilde G = \bar H - B^T ( H^{-1} ) B  =  \begin{pmatrix}  \tilde G_{ii} & \tilde G_{ij} \\  \tilde G_{ij} & \tilde G_{jj} \end{pmatrix}  \q .
\ee
The minor of $G$ associated to removing the $(n-1)$-th row and $n$-th column from \eqref{MetG} is given by 
\be
{\rm M}_{ij} = \det \begin{pmatrix} H & B_j \\  B^T_i& \bar H_{ij} \end{pmatrix}  = \det H \left( \bar H_{ij} - B^T_{i} H^{-1} B_j \right) = \tilde G_{ij} \, \det H \, ,
\ee
where in the last equality we used the components of $\tilde G$ in \eqref{projMetric}. The matrix $H$ is the metric associated to the hinge $h_{ij}$ and hence we have $\det H =  ((n-2)!)^2 \mathbb V_{ij}$.  Hence, we get the metric of the projected triangle to be
\be
\tilde G = \frac{(n-1)^2}{\mathbb V_{ij}} \begin{pmatrix} \mathbb V_{i} & n^2 \frac{\partial \mathbb V}{\partial s_{ij}} \\  n^2 \frac{\partial \mathbb V}{\partial s_{ij}} &\mathbb V_{j} \end{pmatrix}  \q.
\ee
The (star) inner product of the edges $a,b$ of the triangle $t_{ij}$ are therefore 
\be
a \star a =(n-1)^2 \frac{ \mathbb V_{i}}{\mathbb V_{ij}} := s_a, \q b \star b =(n-1)^2 \frac{ \mathbb V_{j}}{\mathbb V_{ij}} := s_b, \q a \star b = n^2(n-1)^2 \frac{\partial \mathbb V}{\partial s_{ij}}\; .
\ee 
The Euclidean signature case and Lorentzian signature case \cite{DittrichFreidelSpeziale,EffSF3} are reproduced for $\phi=0$ and $\phi=\pi$ respectively.

\section{Action and analytical continuations for the ball model}\label{AppC}

Here we will consider the analytical extension of the Regge exponent 
\ba\label{C1}
 \tilde \Lambda W^\pm(s_h)&=& 720 \sqrt{\!\!{}_{{}_\pm} \mathbb{A}_{\rm blk}} \,\, \delta^\pm_{\rm blk} +1200 \sqrt{\!\!{}_{{}_\pm} \mathbb{A}_{\rm bdry}} \, \,\delta^\pm_{\rm bdry} -600 \sqrt{\!\!{}_{{}_\pm}\mathbb{V}_\sigma}    \q .
\ea
for the ball model. The Regge exponent (\ref{C1})  can be expressed in terms of areas, deficit angles and four-volumes, which are given by
\ba\label{B1}
\sqrt{\!\!{}_{{}_\pm} \, \mathbb{A}_{\rm blk}} &=&\frac{\sqrt{s_l}}{4\sqrt{2}}\, \sqrt{\!\!{}_{{}_\pm} \,s_l+8 s_h} \q , \nn\\
\delta^\pm_{\rm blk}&=& 2\pi \mp 5\imath \log_\mp
\left(
 \frac{-s_l+8s_h \mp\imath 8(s_l +8s_h)\sqrt{\!\!{}_{{}_\pm}\,\,  \frac{s_h}{s_l+8s_h} } }{s_l+24 s_h} 
 \right)         \q , \nn\\
\sqrt{\!\!{}_{{}_\pm} \, \mathbb{A}_{\rm bdry}} &=& \frac{\sqrt{3}}{4} s_l                    \q ,  \nn\\
\delta^\pm_{\rm bdry}&=& \pi \mp2 \imath \log_\mp
\left(
\frac{ 
\sqrt{s_l} \mp\imath 2\sqrt{6}\sqrt{ \!\!{}_{{}_\pm}\,\,  s_h}
}
{ 
\sqrt{\!\!{}_{{}_\pm} \,\,s_l+24 s_h}
}
\right)      \q , \nn\\
\sqrt{\!\!{}_{{}_\pm}\mathbb{V}_\sigma}  &=& \frac{\sqrt{2}}{48} s^{3/2}_l \sqrt{   \!\!{}_{{}_\pm}\,\,      s_h}        \q .
\ea
Here we already used a version for the deficit angles, which has branch cuts only on the negative real axis. 
We express $s_h$ in polar coordinates $s_h=r_h\exp(\imath \phi)$ and define an analytical continuation of the functions (\ref{B1}) from the region $(0,\pi)$ to the extended range $(-2\pi,2\pi)$ for $W^+$, and from the region $(-\pi,0)$ to the extended range $(-2\pi,2\pi)$ for $W^-$. For the analytical continuation we cross $\phi=\pi$ respectively $\phi=-\pi$  for $r_h>\tfrac{1}{8}s_l$.  This defines $W^\pm_a(r_h,\phi)$ with $\phi \in (-2\pi,2\pi)$ and $r_h>0$. We have $W^+_a=W^-_a$. $W^\pm_a$ is analytical except for branch cuts at $\phi=\pm \pi$, that extend from $r_h=0$ to $r_h=\tfrac{1}{8}$. For $W_a^\pm$ we replace the geometric functions specified in (\ref{B1}) by
\ba\label{B2}
\sqrt{\!\!{}_{{}_\pm} \, \mathbb{A}_{\rm blk}} &\rightarrow&\frac{\sqrt{s_l}}{4\sqrt{2}}\,e^{\imath \phi/2} \sqrt{ s_l  e^{-\imath \phi}  + 8 r_h}\q , \nn\\
\delta^\pm_{\rm blk}&\rightarrow& 2\pi - 5\imath \log
\left(
 \frac{-s_l+8 r_he^{\imath \phi} - \imath 8(s_l +8 r_he^{\imath \phi}) \sqrt{ \frac{  r_h}{s_l e^{-\imath \phi}+8 r_h}}} {s_l+24 r_he^{\imath \phi}}
 \right)         \q , \nn\\
\sqrt{\!\!{}_{{}_\pm} \, \mathbb{A}_{\rm bdry}} &\rightarrow& \frac{\sqrt{3}}{4} s_l                    \q ,  \nn\\
\delta^\pm_{\rm bdry}&\rightarrow& \pi -2 \imath \log
\left(
\frac{ 
\sqrt{s_l} -\imath 2\sqrt{6} \,e^{\imath \phi/2} \sqrt{ r_h}
}
{ 
e^{\imath \phi/2} \sqrt{ s_l  e^{-\imath \phi}  + 24 r_h}
}
\right)      \q , \nn\\
\sqrt{\!\!{}_{{}_\pm}\mathbb{V}_\sigma}  &\rightarrow& \frac{\sqrt{2}}{48} s^{3/2}_l\,  e^{\imath \phi/2} \sqrt{ r_h}      \q .
\ea
Here we used that for $s_h<-\tfrac{1}{8}s_l$ we only encounter branch cuts for most of the square roots, and we applied for these the analytical extension
\ba\label{B3}
\sqrt{\!\!{}_{{}_\pm} \,s_l+8 s_h} &\rightarrow& e^{\imath \phi/2} \sqrt{ s_l  e^{-\imath \phi}  + 8 r_h}  \q ,\nn\\
\sqrt{\!\!{}_{{}_\pm} \,\,s_l+24 s_h} &\rightarrow& e^{\imath \phi/2} \sqrt{ s_l  e^{-\imath \phi}  + 24 r_h}  \q ,\nn\\
\sqrt{   \!\!{}_{{}_\pm}\,\,      s_h}  & \rightarrow& e^{\imath \phi/2} \sqrt{ r_h} \q .\nn\\
\ea

Crossing the branch cuts $\phi=\pm \pi$ at  $\tfrac{1}{24}s_l < r_h < \tfrac{1}{8}s_l$ we avoid the branch cut of $\sqrt{\!\!{}_{{}_\pm} \,s_l+8 s_h}$. On the other hand 
\ba\label{B4}
\sqrt{\!\!{}_{{}_\pm}\,\,  \frac{s_h}{s_l+8s_h} } \rightarrow e^{\imath \phi_b/2} \sqrt{ \frac{  r_h}{s_l +8 r_he^{\imath \phi_b}}}  
\ea
has a branch cut for $-\tfrac{1}{8}s_l\leq s_h <0$ and is extended as indicated in (\ref{B4}). In summary $W_b(r_h,\phi_b)$ is defined by
\ba\label{B5}
\sqrt{\!\!{}_{{}_\pm} \, \mathbb{A}_{\rm blk}} &\rightarrow&\frac{\sqrt{s_l}}{4\sqrt{2}}\, \sqrt{ s_l  + 8 r_he^{\imath \phi_b}  } \q , \nn\\
\delta^\pm_{\rm blk}&\rightarrow& 2\pi - 5\imath \log
\left(
 \frac{-s_l+8 r_he^{\imath \phi_b} - \imath 8(s_l +8 r_he^{\imath \phi_b}) e^{\imath \phi_b/2} \sqrt{ \frac{  r_h}{s_l +8 r_he^{\imath \phi_b}}}   } {s_l+24 r_he^{\imath \phi_b}}
 \right)         \q , \nn\\
\sqrt{\!\!{}_{{}_\pm} \, \mathbb{A}_{\rm bdry}} &\rightarrow& \frac{\sqrt{3}}{4} s_l                    \q ,  \nn\\
\delta^\pm_{\rm bdry}&\rightarrow& \pi -2 \imath \log
\left(
\frac{ 
\sqrt{s_l} -\imath 2\sqrt{6} \,e^{\imath \phi_b/2} \sqrt{ r_h}
}
{ 
e^{\imath \phi_b/2} \sqrt{ s_l  e^{-\imath \phi_b}  + 24 r_h}
}
\right)      \q , \nn\\
\sqrt{\!\!{}_{{}_\pm}\mathbb{V}_\sigma}  &\rightarrow& \frac{\sqrt{2}}{48} s^{3/2}_l\,  e^{\imath \phi_b/2} \sqrt{ r_h}      \q .
\ea

For $r_h<\tfrac{1}{24}$ we only have a branch cut for the square roots $\sqrt{s_h}$ and  $\sqrt{s_h/(s_l+8s_h)}$. But we also encounter a branch cut for the logarithm in $\delta^\pm_{\rm blk}$, and need to change to $\log(\cdot)-2\pi \imath$ for $|\phi|\geq \pi$. That is $W^\pm_c$ is defined by
\ba\label{B6}
\sqrt{\!\!{}_{{}_\pm} \, \mathbb{A}_{\rm blk}} &\rightarrow&\frac{\sqrt{s_l}}{4\sqrt{2}}\, \sqrt{ s_l  + 8 r_he^{\imath \phi_c}  } \q , \nn\\
\delta^\pm_{\rm blk}&\rightarrow&
\begin{cases}
 +2\pi - 5\imath \log
\left(
 \frac{-s_l+8 r_he^{\imath \phi_c} - \imath 8(s_l +8 r_he^{\imath \phi_c}) e^{\imath \phi_c/2} \sqrt{ \frac{  r_h}{s_l +8 r_he^{\imath \phi_c}}}   } {s_l+24 r_he^{\imath \phi_c}}
 \right)    \q\q \text{for} \, |\phi_c|<\pi \\
  -8\pi - 5\imath \log
\left(
 \frac{-s_l+8 r_he^{\imath \phi_c} - \imath 8(s_l +8 r_he^{\imath \phi_c}) e^{\imath \phi_c/2} \sqrt{ \frac{  r_h}{s_l +8 r_he^{\imath \phi_c}}}   } {s_l+24 r_he^{\imath \phi_c}}
 \right)  \q\q \text{for} \, |\phi_c|\geq\pi
 \end{cases}     \q , \nn\\
\sqrt{\!\!{}_{{}_\pm} \, \mathbb{A}_{\rm bdry}} &\rightarrow& \frac{\sqrt{3}}{4} s_l                    \q ,  \nn\\
\delta^\pm_{\rm bdry}&\rightarrow& \pi -2 \imath \log
\left(
\frac{ 
\sqrt{s_l} -\imath 2\sqrt{6} \,e^{\imath \phi_c/2} \sqrt{ r_h}
}
{ 
 \sqrt{ s_l   + 24 r_h e^{\imath \phi_c} }
}
\right)      \q , \nn\\
\sqrt{\!\!{}_{{}_\pm}\mathbb{V}_\sigma}  &\rightarrow& \frac{\sqrt{2}}{48} s^{3/2}_l\,  e^{\imath \phi_c/2} \sqrt{ r_h}      \q .
\ea

We can easily construct further analytical continuations, that cross the branch cuts e.g. of $W_b$ for $r_h>s_l/8$. Let us for instance consider a path that starts in $\phi\in(-\pi,\pi)$ and circles either the branch point $(r_h=s_l/8,\phi=\pi)$ or the branch point $(r_h=s_l/24,\phi=\pi)$ or both. The resulting Regge exponent differs from $W_a$ only in multiples of $\pi$ added or subtracted to $\delta^\pm_{\rm blk}$ and $\delta^\pm_{\rm bdry}$. This can be parametrized as 
\ba
\delta^\pm_{\rm blk}&\rightarrow& 2\pi - 5\imath \log
\left(
 \frac{-s_l+8 r_he^{\imath \phi} - \imath 8(s_l +8 r_he^{\imath \phi}) \sqrt{ \frac{  r_h}{s_l e^{-\imath \phi}+8 r_h}}} {s_l+24 r_he^{\imath \phi}}
 \right)   +10n\pi -4m\pi     \q , \nn\\
\delta^\pm_{\rm bdry}&\rightarrow& \pi -2 \imath \log
\left(
\frac{ 
\sqrt{s_l} -\imath 2\sqrt{6} \,e^{\imath \phi/2} \sqrt{ r_h}
}
{ 
e^{\imath \phi/2} \sqrt{ s_l  e^{-\imath \phi}  + 24 r_h}
}
\right)   +2k\pi    \q ,
\ea
where $m=0,1$ and $k,n \in \mathbb{Z}$. 

For instance, circling $(r_h=s_l/8,\phi=\pi)$ once (crossing $\phi=\pi$ first at some $s_l/24<r_h<sl/8$) we obtain $(m,n,k)=(1,0,0)$. Circling twice we go back to $(0,0,0)$. 

Circling both $(r_h=s_l/8,\phi=\pi)$  and $(r_h=s_l/24,\phi=\pi)$ once  (crossing $\phi=\pi$ first at some $r_h<s_l/24$) we obtain $(m,n,k)=(1,1,1)$. A second circling gives $(0,0,2)$, and thus a third $(1,1,3)$. 

Circling $(r_h=s_l/24,\phi=\pi)$ once  (crossing $\phi=\pi$ first at some $r_h<s_l/24$) we obtain $(m,n,k)=(0,-1,1)$. Circling twice leads to  $(m,n,k)=(0,-2,2)$, and so on.

\section{Action for the shell model}\label{AppD}

The Regge exponent for the shell model is very similar to the one for the ball model. It is given by
\ba\label{D1}
 \tilde \Lambda W^\pm(s_h)&=& 720 \sqrt{\!\!{}_{{}_\pm} \mathbb{A}_{\rm blk}} \,\, \delta^\pm_{\rm blk} +1200 \left(\sqrt{\!\!{}_{{}_\pm} \mathbb{A}_{\rm bdry_1}} \, \,\delta^\pm_{\rm bdry_1}  + \sqrt{\!\!{}_{{}_\pm} \mathbb{A}_{\rm bdry_2}} \, \,\delta^\pm_{\rm bdry_2}  \right)   -600 \sqrt{\!\!{}_{{}_\pm}\mathbb{V}_{\text{4-frust}}}    \q .
\ea
where
\ba\label{D2}
\sqrt{\!\!{}_{{}_\pm} \, \mathbb{A}_{\rm blk}} &=&\frac{\sqrt{s_{l_1}}+ \sqrt{s_{l_2}  }}{4\sqrt{2}}\, \sqrt{\!\!{}_{{}_\pm} \,(\sqrt{s_{l_2}}-\sqrt{s_{l_1}})^2+8 s_h} \q , \nn\\
\delta^\pm_{\rm blk}&=& 2\pi \mp5\imath \log_\mp
\left(
 \frac{-(\sqrt{s_{l_2}}-\sqrt{s_{l_1}})^2+8s_h \mp\imath 8((\sqrt{s_{l_2}}-\sqrt{s_{l_1}})^2 +8s_h)\sqrt{\!\!{}_{{}_\pm}\,\,  \frac{s_h}{(\sqrt{s_{l_2}}-\sqrt{s_{l_1}})^2+8s_h} } }{(\sqrt{s_{l_2}}-\sqrt{s_{l_1}})^2+24 s_h} 
 \right)         \q , \nn\\
\sqrt{\!\!{}_{{}_\pm} \, \mathbb{A}_{\rm bdry_1}} &=& \frac{\sqrt{3}}{4} s_{l_1}                    \q ,  \q\q
\sqrt{\!\!{}_{{}_\pm} \, \mathbb{A}_{\rm bdry_2}} \,=\, \frac{\sqrt{3}}{4} s_{l_2}   \q ,  \nn\\
\delta^\pm_{\rm bdry_1}&=& \pi \mp2 \imath \log_\mp
\left(
\frac{ 
(\sqrt{s_{l_1}}-\sqrt{s_{l_2}}) \mp\imath 2\sqrt{6}\sqrt{ \!\!{}_{{}_\pm}\,\,  s_h}
}
{ 
\sqrt{\!\!{}_{{}_\pm} \,\,(\sqrt{s_{l_2}}-\sqrt{s_{l_1}})^2+24 s_h}
}
\right)       \, , \q
\delta^\pm_{\rm bdry_2}\,=\, \pi \mp2 \imath \log_\mp
\left(
\frac{ 
(\sqrt{s_{l_2}}-\sqrt{s_{l_1}}) \mp\imath 2\sqrt{6}\sqrt{ \!\!{}_{{}_\pm}\,\,  s_h}
}
{ 
\sqrt{\!\!{}_{{}_\pm} \,\,(\sqrt{s_{l_2}}-\sqrt{s_{l_1}})^2+24 s_h}
}
\right)       \q , \nn\\
\sqrt{\!\!{}_{{}_\pm}\mathbb{V}_{\text{4-frust}} } &=& \frac{\sqrt{2}}{48}  (\sqrt{s_{l_1}}+\sqrt{s_{l_2}}) (s_{l_2}+s_{l_1})    \sqrt{   \!\!{}_{{}_\pm}\,\,      s_h}        \q .
\ea
The analytical extensions can be defined in the same way as for the ball model in Appendix \ref{AppC}.

 \vspace{3mm}

\begin{acknowledgments}
 
The authors thank Ding Jia for collaboration at the start of this project and extensive discussions. BD thanks Renate Loll for interesting discussions. JPA is thankful to Fedro Guillén for discussions and supported by a NSERC grant awarded to BD. SKA is funded by the Deutsche Forschungsgemeinschaft (DFG, German Research Foundation) - Projektnummer/project-number 422809950. This project was also supported by FQXi FFF Grant number FQXi-MGB-2012 from the Foundational Question Institute and the Fetzer Franklin Fund, a donor advised fund of Silicon Valley Community Foundation. Research at Perimeter Institute is supported in part by the Government of Canada through the Department of Innovation, Science and Economic Development Canada and by the Province of Ontario through the Ministry of Colleges and Universities.
\end{acknowledgments}

\providecommand{\href}[2]{#2}
\begingroup
\endgroup

\end{document}